\documentclass[preprint]{aastex}

\shorttitle{D/H in Complex~C}
\shortauthors{Sembach et al.}
\slugcomment{{\it To appear in ApJ Supplement Series, vol. 150 (February 2004)}}
\begin{document}

\newcommand{\kms}{\,km\,s$^{-1}$}     

\title{The Deuterium-to-Hydrogen Ratio
in a Low-Metallicity Cloud Falling onto the Milky Way}

\author{K.R.~Sembach\altaffilmark{1}, 
 	B.P.~Wakker\altaffilmark{2},
	T.M.~Tripp\altaffilmark{3,4},
	P.~Richter\altaffilmark{5}, 
	J.W.~Kruk\altaffilmark{6},
	W.P.~Blair\altaffilmark{6},
	H.W.~Moos\altaffilmark{6},
	B.D.~Savage\altaffilmark{2},
	J.M.~Shull\altaffilmark{7},
	D.G.~York\altaffilmark{8},
	G.~Sonneborn\altaffilmark{9},
	G.~H\'ebrard\altaffilmark{10},
	R.~Ferlet\altaffilmark{10},
	A.~Vidal-Madjar\altaffilmark{10},
	S.D.~Friedman\altaffilmark{1},
	E.B.~Jenkins\altaffilmark{3}}
\altaffiltext{1}{Space Telescope Science Institute, 3700 San Martin Dr., 
	Baltimore, MD  21218.}
\altaffiltext{2}{Department of Astronomy, University of Wisconsin-Madison,
	475 N. Charter Street, Madison, WI  53706.}
\altaffiltext{3}{Princeton University Observatory, Princeton, NJ 08544.}
\altaffiltext{4}{Current address: Department of Astronomy, University
	of Massachusetts, Amherst, MA 01003.}
\altaffiltext{5}{Osservatorio Astrofisico di Arcetri, Largo 
        E. Fermi 5, 50125 Florence, Italy.}
\altaffiltext{6}{Department of Physics and Astronomy, Johns Hopkins
	University, Baltimore, MD  21218.}
\altaffiltext{7}{Center for Astrophysics and Space Astronomy, Department of 
	Astrophysical and Planetary Sciences, University of Colorado, 389-UCB, 
	Boulder, CO  80309.}
\altaffiltext{8}{Department of Astronomy and Astrophysics and the Enrico Fermi 
	Institute, University of Chicago, Chicago, IL  60637.}
\altaffiltext{9}{Laboratory for Astronomy and Solar Physics, NASA Goddard
	Space Flight Center, Code 681, Greenbelt, MD 20771.}
\altaffiltext{10}{Institut d'Astrophysique de Paris, 98 bis Boulevard Arago, 
	F-75014, Paris, France.}

\begin{abstract}
Using {\it Far Ultraviolet Spectroscopic Explorer} (FUSE)
and {\it Hubble Space Telescope} (HST)
observations of the QSO PG\,1259+593,
we detect \ion{D}{1} Lyman-series absorption in high 
velocity cloud Complex~C, a low-metallicity gas cloud falling onto the Milky 
Way.  This is the first detection of atomic deuterium in the local universe 
in a location other than the nearby regions of the Galactic disk. 
We construct a 
velocity model for the sight line based on the numerous \ion{O}{1}
absorption lines detected in the ultraviolet spectra.  
We identify 8 absorption-line 
components, two of which are associated with the high velocity gas in
Complex~C at $\approx -128$ and $\approx -112$ \kms.  
A new Westerbork Synthesis Radio Telescope (WSRT) 
 interferometer map of the \ion{H}{1} 21\,cm emission 
toward PG\,1259+593 indicates that the sight line passes through a 
compact concentration of neutral gas in Complex~C.  We use the WSRT data
together with single-dish data from the Effelsberg 100-meter radio telescope 
to estimate the \ion{H}{1} column density of the high velocity
gas and to constrain 
the velocity extents of the \ion{H}{1} Lyman-series
absorption components observed by FUSE.  We find
N(\ion{H}{1}) = $(9.0\pm1.0)\times10^{19}$ cm$^{-2}$,
N(\ion{D}{1}) = $(2.0\pm0.6)\times10^{15}$ cm$^{-2}$, and 
N(\ion{O}{1}) = $(7.2\pm2.1)\times10^{15}$ cm$^{-2}$ for the Complex~C
gas (68\% confidence intervals). 
The corresponding light-element abundance 
ratios are D/H = $(2.2\pm0.7)\times10^{-5}$, O/H = $(8.0\pm2.5)\times10^{-5}$,
and D/O = $0.28\pm0.12$.  
The metallicity of Complex~C gas toward PG\,1259+593 is approximately
1/6 solar, as inferred from the oxygen abundance 
[O/H] = $-0.79\pm^{0.12}_{0.16}$.
While we cannot rule out a value of D/H similar to that 
found for the local ISM (i.e., D/H $\sim 1.5\times10^{-5}$), we can 
confidently exclude values as low as those 
determined recently for extended sight lines in the Galactic disk 
(D/H~$< 1\times10^{-5}$).  Combined with the sub-solar 
metallicity estimate and the low nitrogen abundance, this conclusion lends 
support to the hypothesis 
that Complex~C  is located outside the 
Milky Way, rather than inside in material 
recirculated between the Galactic disk and halo.
The value of D/H for Complex~C is consistent with 
the primordial abundance of deuterium inferred from recent {\it Wilkinson
Microwave Anisotropy Probe} (WMAP) observations
of the cosmic microwave background and simple chemical evolution
models that predict the amount of deuterium astration as a function of
metallicity. 
\end{abstract}

\keywords{cosmology: observations 
-- Galaxy: evolution -- ISM: abundances -- ISM: clouds
-- nucleosynthesis -- quasars: individual (PG\,1259+593)}

\section{Introduction}

Observations of the abundance of deuterium relative to hydrogen (D/H)
in different environments provide insight into the evolution of the
light elements in the universe.  With the excellent concordance in the 
estimates of the cosmic baryon density from 
measurements of D/H in low-metallicity 
quasar absorption-line systems and measurements of the cosmic 
microwave background, the cosmic baryon density and the 
primordial value of D/H are now tightly constrained (Burles, Nolett,
\& Turner 2001; O'Meara et al.\
2001; Spergel et al.\ 2003).  Thus, it should be possible to test
chemical evolution models by examining the progression of D/H with 
time (see Lemoine et al.\ 1999 and 
Olive, Steigman, \& Walker 2000 for recent discussions).  The abundance
of deuterium is expected to decrease with time since there are no
known sources of deuterium capable of increasing the cosmic
abundance significantly (Epstein, Lattimer, \& Schramm 1976; see also
Prodanovi\'c \& Fields 2003).
Many chemical evolution models predict moderate 
levels of deuterium destruction by stellar nucleosynthesis, typically less
than a factor of 3--5 (Clayton 1985; Edmunds 1994;
Steigman \& Tosi 1995; Tosi et al.\ 1998).
More recent models suggest that slightly lower levels of astration are 
also possible (e.g., Chiappini, Renda, \& Matteucci 2002).

Unfortunately, measurements of D/H are particularly difficult, and there 
have been relatively few opportunities to directly measure the
detailed changes in the abundance of deuterium as a function of metallicity
or time.  A key piece of 
information missing in discussions of the evolution of the light element
abundances with time is the behavior of D/H in environments with 
metallicities between those of the high-redshift systems (typically 
$Z\lesssim0.01Z_\odot$) and those of gas in the local neighborhood of 
the Sun (typically $Z \sim Z_\odot$).

There are several reasons why it it is important to 
determine the D/H ratio in a wide variety of galactic and extragalactic
environments.   
First, there have been few high-precision estimates of D/H at moderate to
high 
redshifts ($z \sim 2-4$), where the amount of stellar processing of deuterium 
is presumably low, as evidenced by low metallicity (e.g., O'Meara et al.\
2001; Pettini \& Bowen 2001 and references therein). The measurements 
that have been made appear to yield conflicting values for the primordial
abundance of deuterium, with the observed values
depending on the type of system observed 
(e.g., Lyman-limit or damped Ly$\alpha$ systems 
-- see Pettini \& Bowen 2001).  Second,
estimates of D/H in many locations yield a more global perspective of 
the chemical history of gas at different epochs than is possible from a 
few isolated measurements.  Chemical evolution models seeking to describe 
the general evolution of the light-element abundances need a large
sample of measurements to avoid systematic problems encountered by 
relying upon data for only a few types of environments.
Third, although measurements of deuterium in
nearby gas clouds  imply
a relatively constant value of D/H within the local interstellar medium 
 (ISM; Moos et al.\ 2002 and references therein),  
substantial variations may exist in D/H and D/O over  
distances of only a few hundred parsecs (Jenkins et al.\ 1999; Sonneborn
et al.\ 2000; Hoopes et al.\ 2003; H\'ebrard \& Moos 2003).   
If a sufficiently large number of high-precision D/H and D/O
 measurements can be made in a diverse set of 
nearby environments, it may be possible to understand the exact causes of this 
variability and the degree to which galactic chemical evolution and
accretion of intragroup gas clouds  influence
the scatter in the observed ratios, both locally and at high redshift.  
This goal is a major science driver for the {\it Far Ultraviolet 
Spectroscopic Explorer} (FUSE) mission (Moos et al.\ 2000). 

To bridge the gap in D/H between low and high metallicity environments,
we have obtained an extensive set of FUSE, {\it Hubble 
Space Telescope} (HST), and interferometric \ion{H}{1}
21\,cm observations of the quasar PG\,1259+593 behind 
high velocity cloud (HVC) Complex~C.  The HVC is located
at least 3.5 kpc from the Galactic plane (Wakker 2001), well
beyond all Milky Way clouds with current D/H determinations.  
Unlike previous investigations of D/H and D/O 
in either high-redshift clouds or the local ISM, 
we know which gaseous system is responsible for 
the high-velocity \ion{D}{1} Lyman-series
absorption observed toward PG\,1259+593.  A global description of the gas
in Complex~C is available from both emission and absorption-line 
measurements.  The neutral
gas in Complex~C has been mapped extensively in \ion{H}{1} 21\,cm emission
(see Wakker 2001 and references therein) and low-ionization absorption
(e.g., Wakker et al.\ 1999; Richter et al.\ 2001b; Collins, Shull, \& Giroux
 2003).
The ionized gas in Complex~C has been investigated in absorption 
by Sembach et al.\ (2003) and Fox et al.\ (2003), and in emission 
by Tufte, Reynolds, \& Haffner (1998) and Wakker et al.\ (1999).
Complex~C is an excellent site to determine D/H for  comparisons
 with the high-redshift values because it is chemically young
(Richter et al.\ 2001b; Collins et al. 2003;
Tripp et al.\ 2003) and has a metallicity (10--25\% solar) lower
than that of the general ISM of the Milky Way and higher than that of 
intergalactic clouds at high redshifts.

In this paper we describe these new measurements and the resulting 
D/H and D/O ratios in Complex~C.   In \S2
we describe the FUSE and HST Space Telescope Imaging Spectrograph 
(HST/STIS) absorption-line observations and the \ion{H}{1} 21\,cm 
interferometer observations.  Section~3 contains a 
short summary of the properties of the PG\,1259+593 sight line.  In \S4
we outline the methods and general assumptions used to determine
 the column densities of 
\ion{H}{1}, \ion{D}{1}, and \ion{O}{1} in Complex~C.
We determine the \ion{H}{1} column density from 
interferometric \ion{H}{1} 21\,cm emission data and use the 
FUSE and HST/STIS ultraviolet absorption-line data to determine the 
\ion{O}{1} and \ion{D}{1} column densities.
Sections 5, 6, and 7 contain descriptions of these determinations 
for oxygen, hydrogen, and deuterium, respectively, and provide
estimates of the various errors. 
The column densities and error ranges are summarized and discussed in
 \S8, and comments on future progress appear in \S9.
We summarize the results of the study in \S10.

\section{Observations}

\subsection{Far Ultraviolet Spectroscopic Explorer Observations}
We observed PG\,1259+593 on nine occasions 
with FUSE between 2000~February~25 and 
2001~March~28 for a total (orbital day + night) 
integration time of $\sim600$ kiloseconds (ks)
spread over 236 individual exposures. 
Approximately 350 ks of data were obtained during orbital night.  An
observation log for the nine visits is provided in Table~1.
Data were obtained through the large (LWRS; $30\arcsec\times30\arcsec$) 
apertures in all four FUSE channels (LiF1, LiF2, SiC1, SiC2) with varying 
degrees of success.   PG\,1259+593 was always 
well-centered in the LWRS aperture of 
the LiF1 channel
used for guiding, but thermal  effects caused the light of PG\,1259+593 to 
drift around inside (and at times
even partially outside) the remaining three apertures.
Detector
high voltages were also at reduced levels during some of the exposures 
as a result of operational difficulties (see Table~1 notes). 
We obtained all of the data 
in time-tagged photon-address mode to allow for data 
screening and time-dependent corrections necessary to fully calibrate the 
data.  
The raw FUSE data for PG\,1259+593
can be found in the Multi-Mission Archive at the Space Telescope
Science Institute under the observation identifications P1080101--P1080109.

We processed the data with a customized version of the FUSE pipeline 
software ({\tt CALFUSE} v2.2.2), which is publicly available from the 
FUSE Project at the Johns 
Hopkins University.\footnotemark
\footnotetext{See {\tt http:$\slash\slash$fuse.pha.jhu.edu.}}  
The 
details of this processing follow the same general principles outlined by 
Sembach et al.\ (2001b). However, since PG\,1259+593 is a faint object
($F_\lambda \sim 2\times10^{-14}$ erg cm$^{-2}$ s$^{-1}$ \AA$^{-1}$
between 920 and 1200\,\AA),
we modified these procedures in the following manner to process the 
data.  First, we screened
the raw photon lists in every exposure for Earth limb avoidance, South 
Atlantic Anomaly passage, pulse height 
distribution constraints, and 
particle event 
bursts (see Sahnow et al.\ 2000).  We chose the pulse height restrictions 
after inspection of the pulse height profiles to reduce background events 
while minimizing the number of source events discarded;
pulse height data numbers from 4 to 24 were allowed.  Event bursts were
common in these observations, particularly those obtained in the March 2001
time period, so we carefully checked the cleaned lists to be sure that 
no obvious signal from these events remained after screening.

Next, we combined the screened exposure lists for each channel within
each observation using the default spectral registration provided by the 
software.  This default assumption is necessary since there is not enough
signal in the individual exposures to reliably cross-correlate the positions
of narrow spectral features; the typical exposure time of $\sim2-3$ ks
per exposure yielded an average of $\sim2-3$ counts per resolution element in 
the highest sensitivity (LiF1) channel and $\sim1-2$ counts per resolution
element in the SiC channels.
 The summed channel lists for each observation
were corrected for geometric distortions, Doppler shifts, thermally-induced 
grating motions, astigmatism, detector backgrounds, and  scattered light 
following the standard procedures (Sahnow et al.\ 2000).  The extracted 
spectra for each channel in each observation were then flux calibrated
and wavelength 
calibrated.  We registered the spectra from each of the nine
observations to a common wavelength scale by noting the integral pixel shifts
required to align the same narrow absorption 
features observed in each spectrum.  These shifts were typically 
less than 3 detector FUSE pixels and always less than 15 pixels.  We then 
co-added the aligned spectra for all 9 observations to produce a final 
composite spectrum for each channel.  

We registered the composite channel spectra to a common 
heliocentric wavelength scale by performing a channel-to-channel
registration similar to that performed at the observation level and then
setting the zero point of the wavelength scale to that provided by the 
STIS spectra (discussed below).  For the FUSE-to-STIS registration, we
compared the velocities of lines of the same species observed by each 
instrument (e.g., \ion{C}{2} $\lambda\lambda1036.337, 1334.532$; \ion{Si}{2}
$\lambda\lambda 1020.699, 1304.370$; 
\ion{Fe}{2} $\lambda\lambda1144.938, 1608.451$). The \ion{C}{2} 
and \ion{Fe}{2} comparisons used lines with approximately the same 
line strengths, $f\lambda$, to ensure that the comparisons were not 
influenced by asymmetric profiles that saturate in one line but not the 
other.   Additional checks were
performed using cross-element comparisons (e.g., \ion{Si}{2} $\lambda1020.699$
with \ion{S}{2} $\lambda\lambda1250.584, 1253.811$, etc.).
We then corrected the 
FUSE spectra to the Local Standard of Rest (LSR) reference frame. For the 
PG\,1259+593 sight line, the correction for standard solar motion
$v_{\rm LSR} = v_{\rm helio}
+10.5$ \kms, assuming a solar speed of 19.5 \kms\ in the direction 
$l_{std}=56\degr$,
$b_{std} = 23\degr$ (Delhaye 1965; Lang 1980; see also Kerr \& 
Lynden-Bell 1986).  This correction is within $\approx 1.6$ \kms\ of the 
LSR reduction based on a solar speed of 16.5 \kms\ in the direction
$l = 53\degr, b = 25\degr$ (Mihalas \& Binney 1981).
Unless stated otherwise, all
velocities quoted in this paper are in the LSR reference frame.  The 
FUSE spectra for PG\,1259+593 have a nominal zero-point velocity uncertainty 
of $\pm5$ \kms\ ($1\sigma$) after these calibrations have been applied.  

The FUSE data are oversampled in the spectral domain.  Therefore, 
after calibration we binned the spectra to a spectral bin size
of 4 pixels, or $\sim0.025$\,\AA\ ($\sim7.5$ \kms).  This binning
provides approximately 3 samples per spectral resolution element of 
22--25 \kms.  The night-only data used in this study have continuum 
signal-to-noise ratios $S/N \sim 25$ and 16 
per spectral resolution element at 1030\,\AA\
in the LiF1 and LiF2 channels, respectively, and $S/N \sim 10$ and 12 at 
950\,\AA\ in the SiC1 and SiC2 channels, respectively.
 
The  version of  {\tt CALFUSE} (v2.2.2) used for this study
has significant improvements over earlier versions used in previous
studies, most notably improved wavelength solutions and background
corrections tailored for day+night or night-only extractions.
Portions of the fully reduced night-only 
SiC2 and SiC1 spectra in the 915--955\,\AA\ 
wavelength range are shown in Figure~1.  The locations of the Lyman 
series \ion{H}{1} and \ion{D}{1} lines are indicated in the top panel
along with the locations of prominent interstellar 
\ion{O}{1} and \ion{N}{1} lines.
Terrestrial airglow emission features, which are present in all of the 
\ion{H}{1} lines observed, are marked below the top spectrum with crossed
circle symbols.  Terrestrial
\ion{O}{1} and \ion{N}{1} emissions are undetectable in the night-only
data shown.  The background subtraction for both channels results in 
nearly zero residual flux in the cores of the \ion{H}{1}
lines as expected, 
except at velocities affected by airglow contamination. (A similar
conclusion holds for the very strong \ion{C}{2} $\lambda1036.337$ line
observed in all four channels.)  The slight rise in the sub-Lyman-limit
regime ($\lambda < 912$\,\AA) in the SiC1 channel
may be the result of the radiative recombination continuum emission from
atomic oxygen near 911\,\AA\ (see Feldman et al.\ 1992; L\'opez-Moreno
et al.\ 2001)
or improper background subtraction at these very short wavelengths.
The SiC2 channel extends
only to 916.6\,\AA, so no independent check of the SiC1 behavior 
at $\lambda  < 916$\,\AA\ is  
possible.  

In the discussions that follow, we restrict our analysis to the 
orbital night-only data to minimize the impact of terrestrial \ion{H}{1} and 
\ion{O}{1} airglow emissions on the absorption lines of interest in this 
study.  In particular, we concentrate on data from the two SiC channels
 covering the FUSE bandpass below 
1000\,\AA, but we also use the independent 
data from the two LiF channels to check the quality of the data at
longer wavelengths. The checks allowed by multiple channel observations
of the same spectral region are important for assessing noise in the 
data. When we derive H, D, and O abundances, we use both the SiC1 and SiC2
data.
Additional illustrations of FUSE spectra of PG\,1259+593 can be found 
in Richter et al.\ (2001b) and Wakker et al.\ (2003).

\subsection{Space Telescope Imaging Spectrograph Observations}

PG1259+593 was observed several times with HST/STIS as part of 
Guest Observer program GO-8695 on 2001~January~17-19
and 2001~December~19.  An observation log is provided in Table~2.
All observations used the 
intermediate-resolution echelle mode (E140M) and 
the $0\farcs 2 \times 0\farcs 06$
slit  to minimize the power in the wings of the
spectral line-spread function (see Figure~13.90 in the STIS
Instrument Handbook, Leitherer et al.\ 2002). This instrument
mode provides a resolution of $\sim7$~\kms\ (FWHM) per
two-pixel resolution element and covers the 1150--1729\,\AA\ wavelength
band with
only five small gaps between echelle orders at $\lambda >$ 1634\,\AA.
For further details on the design and performance of
STIS, see Woodgate et al.\ (1998) and Kimble et al.\ (1998). 

We reduced the data for each observation in the manner described by Tripp et 
al.\ (2001), including application of the two-dimensional scattered light 
correction developed by the STIS Team (Landsman \& Bowers 1997; 
Bowers et al.\ 1998).
We weighted the extracted 
flux-calibrated spectra for each observation  by
their inverse variances 
and added them together to produce a composite spectrum.
The continuum signal-to-noise 
ratio per resolution element in the final co-added spectrum ranges from 
$\sim$7 to 17 and peaks near 1400\,\AA.  We corrected the STIS data to the 
LSR reference frame by adding +10.5 \kms\ to the nominal heliocentric 
velocity scale provided by the standard processing.  The zero-point accuracy 
of the STIS velocity scale for these data is approximately 
$\pm1$ \kms.

Examples of continuum normalized profiles for the absorption lines of 
several low ionization species are shown in Figure~2.  Additional 
examples of absorption lines in these STIS spectra of PG\,1259+593 can be 
found in Richter et al.\ (2003) and Collins et al. (2003).

\subsection{\ion{H}{1} 21\,cm Observations}

High velocity clouds often show structure at angular scales down to at
least $\sim$1\arcmin\ (e.g., Wakker \& Schwarz 1991), particularly in
cloud cores. The PG\,1259+593 sight line samples the Complex~C core 
named CIII (Giovanelli, Verschuur, \& Cram 1973; Wakker 2001). 
The Leiden-Dwingeloo Survey (LDS, Hartmann \& Burton 1997) shows that
this core has an extent of about $1\degr\times2\degr$ 
at a column density of $3\times10^{19}$ cm$^{-2}$, with PG\,1259+593 lying 
at its edge.  To determine a more accurate estimate of the amount of 
high velocity \ion{H}{1} in this region of the sky, we observed CIII with the 
Westerbork Synthesis Radio Telescope (WSRT).  Since this interferometer 
filters out the large-scale structure, we supplemented the WSRT data in the 
direction of PG\,1259+593 with an Effelsberg single-dish observation having a
9\farcm7 beam. This latter observation is described by Wakker et al.\
(2001).

The half-power beam width (HPBW) of the WSRT primary beam is 
35\arcmin, which allows an area with a diameter of about 44\arcmin\ to be 
mapped.
We therefore observed CIII with
 a mosaic of $2\times4$ pointings spaced 27\arcmin\ apart, resulting in 
mapping over an area of $(22\arcmin+27\arcmin+22\arcmin)\times(22\arcmin+3\times27\arcmin+22\arcmin)=71\arcmin\times125\arcmin$ arc minutes.
 To increase 
the sensitivity, we included an extra pointing in the direction of 
PG\,1259+593 itself. For each of these pointings, we obtained full 
uv-plane coverage in $4\times12$ hours,
using shortest spacings of 36, 54, 72, and 90\,m. The correlator was set to
cover the LSR velocity range between $-333$ and +197 \kms\ with 2.1 \kms\
velocity resolution after on-line Hanning smoothing was applied. The WSRT 
observations were completed in late April 2001.

R.\ Braun of the Netherlands Foundation for Research in Astronomy
performed the calibration of the observations. The
calibrated data were mapped using uniform weighting, and included a
Gaussian taper in the uv-plane such that the final synthesized
beam is $1\arcmin\times
1\arcmin$. Next, a map of the continuum was created for each pointing by 
adding the channels without \ion{H}{1} emission. After subtracting the 
continuum, 
the individual pointings were cleaned using the {\tt Multi-Resolution 
Clean} algorithm (Wakker \& Schwarz
1988). To increase the accuracy of this step, the areas containing signal were
first delineated, taking into account the overlap between pointings. The
resulting cleaned maps were then mosaiced together. The fully processed
data have a final root-mean-square (RMS) residual of
$\sim3.2$ mJy beam$^{-1}$, or $\sim0.5$ K.  

Figure~3 shows a grey-scale map of N(\ion{H}{1}) for Complex~C core CIII
based on the WSRT data (integrated
between $-$148 and $-$109 \kms). 
PG\,1259+593 lies in the brightest concentration in the field, which
coincides with the brightest spot in the LDS data. Although it seems unusual
that the extragalactic background source lies toward the center of a 
cloud core,
this is not an artifact of the mosaic process;
if the pointing centered on PG\,1259+593 is
removed, the core is still clearly seen at the half-power point of each of the 
four surrounding pointings, and the final \ion{H}{1} map looks similar, though
noisier.  We discuss the \ion{H}{1} data and the derived column density
for the high velocity gas in \S6.1.

\section{The PG\,1259+593 Sight Line}

PG\,1259+593 lies in the direction $l = 121\fdg09,~b = +57\fdg80$ behind
HVC Complex~C, which spans Galactic longitudes 
 between $l\sim30\degr$ and $l \sim 150\degr$ in the northern
Galactic hemisphere.  Maps of the \ion{H}{1} 21\,cm emission in
Complex~C can be found in Wakker (2001) and Sembach et al.\
(2003); these maps show larger regions than the area around the sight line
covered in Figure~3.
Complex~C has a mass M~$>1.2\times10^6$ M$_\odot$
and a distance $d > 5$ kpc (or altitude z $>$ 3.5 kpc)
(van~Woerden et al.\ 1999).  Wakker (2001) suggests a  distance limit  
$d > 6.1$ kpc and a mass
M $>3\times10^6$ M$_\odot$.  Various determinations of the 
metallicity of the high velocity gas yield $Z/Z_\odot\sim0.1-0.3$ on a linear
scale where
$Z = Z_\odot$ is solar (Wakker et al.\ 1999; Gibson et al.\ 2001; 
Collins et al.\
2003; Tripp et al.\ 2003).  
Previously, Richter et al.\ (2001b) found $Z/Z_\odot 
= 0.093\pm^{0.125}_{0.047}$ 
for Complex~C based on a subset of the oxygen absorption lines 
toward PG\,1259+593 considered here.

In the direction of PG\,1259+593, interstellar gas within 10 kpc of the 
Galactic plane that is participating in differential
Galactic rotation has velocities $v_{LSR} \approx -30$ to 0 \kms.
There is also a large
intermediate-velocity cloud known as the Intermediate-Velocity Arch in
this general direction (Kuntz \& Danly 1996).
Complex~C has a velocity 
$v_{LSR} \approx -157$ to $-100$ \kms, which makes it possible to 
distinguish between the absorption produced by the HVC and the
lower velocity foreground absorption produced by the nearby 
ISM ($v_{LSR} \sim -5$~\kms) 
and the Intermediate Velocity Arch (IV~Arch; 
$v_{LSR} \sim -55$ \kms). 
The three principal groups of gas along the sight line
are reasonably well separated in velocity in the \ion{H}{1} 21\,cm emission
and weak ultraviolet absorption-line profiles shown in Figure~2.  
In the strongest absorption lines shown (\ion{Si}{2} $\lambda1526.707$, 
\ion{O}{1} $\lambda1302.168$, \ion{C}{2} $\lambda1334.532$),
the ISM and IV~Arch components blend together.
It is important to note that the  21\,cm emission associated with
Complex~C is stronger than that from the IV~Arch or the Milky Way 
ISM in this direction, a point we will return to in the derivation of 
the Complex~C \ion{H}{1} Lyman series velocity structure in \S6.
Other sight lines through 
Complex~C usually exhibit much weaker HVC 21\,cm emission and/or have a 
much more complicated velocity structure that hinders a clean separation of 
the Complex~C gas from the ISM and intermediate velocity gas. 
For examples of the 21\,cm emission toward  other sight lines passing
through Complex~C, see
Sembach et al.\ (2003) and Wakker et al.\ (2001).  
The ``enhanced'' 21\,cm emission
toward PG\,1259+593 is undoubtedly due to the fact that the 
sight line is located directly behind the Complex C core CIII (see 
\S2.3).   

In addition to neutral gas, Complex~C contains ionized gas that can be 
traced through its H$\alpha$ emission (Tufte et al. 
1998; Wakker et al.\ 1999), \ion{O}{6} absorption 
(Murphy et al.\ 2000; Sembach et al.\ 2000, 2003), 
and \ion{C}{4} and \ion{Si}{4}
absorption (Fox et al.\ 2003; see also \S6 below).  We must consider the 
impact of these
ionized regions on our derivation of D/H in Complex~C 
since they may contain trace amounts of \ion{H}{1} detectable in 
the \ion{H}{1} Lyman series lines (see \S6).

\section{General Methodology}
\subsection{Overview}

Even though the PG\,1259+593 sight line has many desirable attributes that
facilitate a detection of deuterium  
in Complex~C, the conversion of this detection into a reliable estimate 
for the deuterium abundance depends upon many factors.
It is necessary to take a methodical approach to determining the
 D/H and D/O ratios in the high velocity gas.  
The sight line is more complicated than short sight lines through
the local ISM used to determine the local value of D/H 
(e.g., Moos et al.\ 2002 and references therein; H\'ebrard \& Moos 2003).
It is also more complex than the high-redshift sight lines, which can be 
selected for simple intergalactic absorption
velocity structure and typically contain one or 
two predominant absorption components (e.g., O'Meara et al.\ 2001).  
The velocity structure of the PG\,1259+593
sight line does not allow for a precise measure of the \ion{H}{1} column 
density in Complex~C solely through analysis of the \ion{H}{1} Lyman-series
absorption due to blending of the high velocity \ion{H}{1} absorption 
with the ISM and IV~Arch components.  However, it is possible to derive an accurate \ion{H}{1}
column density from 21\,cm emission, provided that the angular resolution
is sufficiently high to assure that small-scale structure within the 
radio beam is not biasing the result.  It is also possible to draw upon
the information in the \ion{O}{1} absorption
profiles for the sight line and to constrain the range of possible 
Lyman-series absorption parameters for both \ion{D}{1} and \ion{H}{1}. 

Our adopted method for determining the D/H and D/O ratios in Complex~C
can be broken down into several key steps:  
1)~determination of an accurate  \ion{O}{1} column density for Complex~C 
and a model for the velocity structure of the neutral gas along the 
PG\,1259+593 sight line 
using the numerous \ion{O}{1} absorption
lines in the FUSE and STIS spectra,  2)~application of
the \ion{O}{1} velocity model to the \ion{H}{1} Lyman-series lines
in the FUSE data to create an \ion{H}{1} model that reproduces the 
observed \ion{H}{1} absorption and 21\,cm emission profiles,
3)~determination of
the \ion{H}{1} column density in Complex~C  from the interferometric \ion{H}{1}
21\,cm data 
described in \S2.3,  4)~determination of the amount of \ion{D}{1} 
required to account for the Complex~C absorption in the 
negative velocity wings of the \ion{H}{1} Lyman-series lines, and
5)~consideration of the uncertainties associated with steps (1) -- (4).

\subsection{Methods}
We use a variety of spectral line analysis techniques to determine the 
velocity structure and column density of each species studied.  Our primary 
tool for quantifying the properties of the absorption lines in the 
FUSE and HST/STIS spectra is a suite of software for fitting line profiles,
written specifically for this study in the Interactive Data Language 
(IDL).
The software constructs synthetic line profiles that can be compared to the
observed data after convolution with an instrumental line spread function.  
It allows complex absorption lines to be modeled as a superposition of 
components with Maxwellian velocity distributions, each of which can be 
described as Voigt functions with appropriate natural damping constants.
Each component has 
a central velocity ($\langle v_i \rangle$), line width (b$_i$), and 
column density (N$_i$), with the best-fit parameters found through a 
$\chi^2$-minimization of the differences between the model profiles and the 
data.  To explore the sensitivity of our adopted results to various input 
parameters/assumptions and boundary conditions, 
we perform some of these profile fitting analyses 
multiple times with other profile fitting codes used in
FUSE analyses of other sight lines (e.g., the {\tt Owens.f} code -- see
H\'ebrard et al 2002; Lemoine et al.\ 2002).  In all cases checked, 
the column densities are consistent to within the $1\sigma$ errors, 
confirming that the formal uncertainties have been estimated properly.
For the \ion{O}{1} and \ion{D}{1} lines we also use curve-of-growth 
analyses to estimate column densities and to check the reliability of the 
profile fitting results.  
We describe these different analyses in the 
following discussions of the absorption produced by each species 
(\ion{H}{1}, \ion{D}{1}, \ion{O}{1}).

\subsection{Input Data and Reference Abundances}

The input atomic data parameters for this study are known with high enough 
accuracy that they do not present a significant source of systematic
uncertainty in our final results.  We adopt wavelengths,  oscillator
strengths ($f$-values), and radiation damping constants
 from the atomic data compilations of Morton (1991, 2003). For \ion{O}{1},
the original sources of the $f$-values are Zeippen, Seaton, \& Morton (1977),
Bi\'emont \& Zeippen (1992), and Tachiev \& Froese Fischer (2002).  The 
source of $f$-values for \ion{H}{1} and \ion{D}{1} is Pal'chikov (1998).
We use multiple
transitions of \ion{H}{1}, \ion{D}{1}, and \ion{O}{1} in our line analyses.  
Thus, typical uncertainties
of $10-20$\% in the $f$-values for individual lines are 
mitigated by considering several transitions simultaneously in the
column density determinations. 
 To identify  molecular hydrogen lines and
assess their possible contamination of the atomic absorption features, we use 
the H$_2$ line lists of Abgrall et al.\ (1993a,b).

Throughout this work,  we adopt a solar reference abundance
(O/H)$_\odot = 
4.90\times10^{-4}$ from Allende Prieto, Lambert, \& Asplund 
(2001).  This reference solar oxygen abundance is in good agreement with the 
value of $4.5\times10^{-4}$ derived from analyses of the solar [\ion{O}{1}],
\ion{O}{1}, and OH line shapes
and asymmetries implied by new 3-D hydrodynamical model atmospheres
(Asplund 2003).  It is in better agreement with 
the nearby average ISM oxygen abundance of 
(O/H)$_{\rm ISM} = 3.4\times10^{-4}$ (Meyer 2001; 
Meyer, Jura, \& Cardelli 1998)
than earlier determinations based on 1-D model atmospheres.
It also
agrees well with recent determinations of \ion{H}{2} region oxygen
abundances (see Pilyugin, Ferrini, \& Shkvarun 2003 and references 
therein).  We will adopt this solar reference abundance in our discussions 
of the metallicities of the various absorbers along the sight line.
Using (O/H)$_\odot = 5.45\times10^{-4}$ from Holweger (2001)
or (O/H)$_\odot = 7.41\times10^{-4}$  from Grevesse \& Noels
(1993) yields values of [O/H] that are
a factor of 0.05 dex and 0.18 dex lower, respectively, than those adopted
in this study.

\subsection{Simplifying Assumptions}
\subsubsection{Number of Components in the Model}
Throughout this work, we adopt the minimum number of velocity 
components required
to produce acceptable fits to the observed absorption lines in the 
FUSE and HST/STIS observations.  Our choice of 
component structure is guided in part by the high-resolution HST/STIS data 
available for several metal-line species (e.g., \ion{O}{1}, \ion{Si}{2}, 
\ion{S}{2}, \ion{Fe}{2}).  In all,  8 components are
 required
to fit the \ion{O}{1}, \ion{H}{1}, and \ion{D}{1} profiles.    The 
attributes of these components will be discussed for each species in
\S\S5--7.  A standard
F-test indicates that adding additional components does not significantly 
improve the 
quality of the fit to the observed data, with the possible exception of 
an additional weak \ion{H}{1} absorption feature discussed in \S7.1.  Adopting
 fewer components does not 
provide an acceptable fit to both the \ion{O}{1} and \ion{H}{1} 
absorption-line data.  Some
of the identified components,  especially those at low velocities,
are likely to consist of multiple sub-components 
that cannot be resolved at the resolution of the FUSE and HST/STIS data.

\subsubsection{Velocity Structure Assumptions in the Model}

We 
require the \ion{O}{1}, \ion{H}{1}, and \ion{D}{1} lines to have a similar 
velocity structure, but we do not simultaneously fit these species 
with other
metal-line species (e.g., \ion{N}{1}, \ion{Si}{2}, \ion{S}{2}, \ion{Fe}{2}) 
as has
been done in some previous D/H analyses of other sight lines.  To 
do so would impose additional constraints on the \ion{O}{1}, \ion{H}{1},
and \ion{D}{1}  lines that 
are not only unwarranted but also potentially misleading for
this sight line.
The absorption profiles of other species may be influenced 
significantly by the presence of ionized gas in the different 
components along the sight line. Unlike
sight lines to nearby stars where the coupling of many species makes 
sense because the velocity structure of the 
ISM is simple and the neutral and ionized species
have very similar velocity structure 
(e.g., Vidal-Madjar et al.\ 1998; H\'ebrard et al.\ 2002), 
the PG\,1259+593 sight line contains 
a variety of absorbing regions with a range of physical properties.
Linking the \ion{O}{1} velocity structure directly to the profile 
shapes of species that
trace both neutral and ionized gas could result in an   
inferred velocity extent of the neutral Complex~C absorption that is  
sensitive to the profile shapes of singly ionized species 
in Complex~C or in the IV~Arch gas at nearby velocities.
A secondary concern with tying the velocity structure of \ion{O}{1} 
to ions such as \ion{Si}{2} or \ion{Fe}{2} is the inclusion of 
refractory elements into dust.  Differential depletion of elements into
dust grains or selective shock-disruption of grains
can alter the relative gas-phase abundances of ISM clouds 
encountered along the sight line, which in turn can translate into different
absorption-line velocity structures for different species. 
We use all of the information available to construct
an initial guess at the velocity structure of the sight line,  but
take the conservative approach of coupling  the velocity
structure of only those species (i.e., \ion{H}{1}, \ion{D}{1}, \ion{O}{1})
for which ionization  and
dust depletion effects are unlikely to differ significantly.  Information for 
the singly ionized species can be found in recent studies by 
Richter et al.\ (2001b) and Collins et al.\ (2003).

\subsubsection{The FUSE Line Spread Function}
We assume that the FUSE line spread function (LSF) has a Gaussian
shape and a constant width (FWHM) of 0.075\,\AA\ ($\approx12$ FUSE detector 
pixels) throughout the FUSE bandpass.  
This width corresponds to a velocity width of FWHM $\approx20-25$ \kms, depending
upon wavelength.  This is slightly broader than the nominal value of 20 \kms\
often assumed in FUSE studies (see Sahnow et al.\ 2000).  It is reasonable
to expect the LSF for the combined 
PG\,1259+593 data to be degraded slightly compared to the 
nominal LSF for single FUSE exposures because the number of 
individual exposures in the combined set of observations is  large and
registration of the individual exposures is certain to introduce some 
minor spectral degradation, particularly in the SiC channels.
  Given the velocity structure of the sight line, it is 
difficult to find narrow isolated lines to determine the LSF width directly.
The narrowest lines for which this is possible in the FUSE spectrum of 
PG\,1259+593 have observed (instrumentally convolved)
widths 
of $\approx0.08$\,\AA\ (LiF1A detector segment) 
and $\approx 0.10$\,\AA\ (SiC2A detector segment).  The measured
widths are broader than
the adopted LSF width but are consistent with the expected line widths based
on the velocity structure observable at higher resolution in the STIS 
data shown in Figure~2.

In previous investigations of
 D/H along kinematically simple sight lines, a multiple-component 
LSF has sometimes been used  to fit the data.  For example, Kruk et al.
(2002) and Wood et al. (2002) used a two-component LSF with a narrow component
(FWHM = 9 pixels) containing $\sim 70-75$\% of the total LSF area and a 
broad component (FWHM $= 17-24$ pixels) containing $\sim 25-30$\% of the 
LSF area.  
When summed, the double-Gaussian functions studied have a roughly Gaussian 
shape with a width similar to that adopted in this study.
The resulting column 
densities of \ion{D}{1} and \ion{O}{1} obtained with a double-component LSF
have been  similar to those 
obtained with single-component LSFs (Kruk et al.\ 
2002; Wood et al.\ 2002).  The reasons for this are straight-forward:
the nearby absorption is simple, and the effects of a broad LSF component are
most pronounced for strongly saturated lines requiring 
very accurate estimates of the 
optical depth at low residual flux levels.  Toward PG\,1259+593, the 
velocity structure is much more complex.  Since we must consider the 
case where weak absorption is in close proximity to strong absorption,
it is not so easy to dismiss the effects of a non-Gaussian LSF.  
For our PG\,1259+593 FUSE
dataset, which is the combination of many individual exposures,
the central-limit theorem implies that the LSF of the combined data should
also closely represent a single Gaussian function.
After modeling the sight line and considering the impact of a more complex
LSF, we find that the addition of a second 
LSF component with broad wings does not change our estimates of the 
\ion{H}{1}, \ion{D}{1}, and \ion{O}{1} 
line strengths. Other systematic uncertainties dominate the 
potential
uncertainties caused by the adoption of a single-component LSF rather than
a two-component LSF of the type used in  
previous FUSE studies.

We performed a consistency check on our LSF estimation by
comparing the shape of the \ion{Fe}{2} $\lambda1144.938$ line in the 
FUSE LiF2A spectrum with that of the \ion{Fe}{2} $\lambda1608.451$ line
in the HST/STIS spectrum. The two lines have intrinsic strengths ($f\lambda$)
within $\sim20\%$ of each other, so direct comparisons of the velocity 
structure and component absorption depths 
could be made by relating the FUSE spectrum to various smoothed versions 
of the higher resolution 
STIS spectrum.  We found that a good approximation
to the FUSE \ion{Fe}{2} profile was obtained by convolving the STIS spectrum 
with a Gaussian function having FWHM $= 20\pm4$ \kms. 
(At 1145\,\AA, our adopted instrumental
width of 0.075\,\AA\ corresponds to FWHM = 19.6 \kms.) Smaller smoothing 
widths left structure in the smoothed 
STIS spectrum that was not obvious in the 
FUSE data, while larger smoothing widths underestimated the line depths 
and overestimated the line widths of the FUSE absorption.
This comparison 
does not reveal the exact shape of the FUSE LSF, but it is
sufficient to bracket the LSF width and demonstrate that the adopted 
Gaussian LSF width is well-constrained.

\subsubsection{Estimation of the Quasar Continuum}

Another simplifying assumption we make is that there are no 
undulations  in the ultraviolet continuum of 
PG\,1259+593 on scales of less than a few \AA ngstroms.  
This assumption is made in essentially all studies of 
quasar absorption line spectra.  Compared to the spectra of normal and 
degenerate hot stars, power-law quasar continua are exceedingly smooth over 
small wavelength intervals; see, for example, the FUSE spectrum of the 
quasar 3C\,273 (Sembach et al.\ 2001b). 
The PG\,1259+593 spectrum contains no 
signatures of a stellar population contribution to the continuum light, which
can be difficult to model for some AGNs, particularly Type~II Seyferts.
We model the PG\,1259+593 continuum in the vicinity of the interstellar 
lines as a smoothly varying function of wavelength.  The low-order 
Legendre polynomial
fits adopted for the SiC channel data below 960\,\AA\ are shown in
Figure~1.  The quasar continuum is relatively flat
(i.e., d$F_\lambda$/d$\lambda \approx 0$) over the large wavelength interval
considered ($\sim 900-1100$\,\AA).
Uncertainties associated with the placement of these continua are small
compared to the other sources of uncertainty discussed below.  Continuum
placement is especially important for the weak lines of \ion{D}{1}, for
which we have estimated upper limits including this systematic 
uncertainty.  We used the methods outlined by Sembach \& Savage
(1992) to determine the continuum levels and to extract the continuum 
placement uncertainties.

\subsubsection{Possible Contamination from Absorption Lines of Other Species}
There is little molecular hydrogen along the 
PG\,1259+593 sight line, as evidenced by the paucity of H$_2$ R(0) and R(1)
Lyman-band absorption
features in the FUSE spectra. Richter et al.\ (2001a) estimate
\(\sum_{J=0}^{1} \log {\rm N_J(H_2)}\) $< 13.96$ ($3\sigma$) in Complex~C.
 There may be a small amount of H$_2$ absorption 
present in the strongest lines of the $J=0-3$ rotational levels
at the velocities of the Milky Way ISM ($v_{\rm LSR} \approx -5$ \kms);
we estimate \(\sum_{J=0}^{3} \log {\rm N_J(H_2)}\) $\approx 14.7$.  
A small amount of H$_2$ may
also be present in the IV~Arch [\,\(\sum_{J=0}^{3} \log {\rm N_J(H_2)}\) 
= $14.0\pm^{0.21}_{0.44}$\,].
We examined each \ion{O}{1} and \ion{D}{1} line used in this study
for possible interstellar H$_2$ lines that might be confused with the 
atomic absorption in Complex~C at $-160 \le v_{LSR} \le -110$ \kms.
We found that the predicted interstellar
H$_2$ lines in the $J=0-3$ rotational levels 
with the total column densities quoted above produce negligible
absorption in the vicinity of these lines.  H$_2$ absorption by the 
foreground ISM can safely be ignored in our analysis
of the Complex~C absorption features.
Some H$_2$ lines arising in the Milky Way ISM occur
at wavelengths corresponding to the low-velocity portions of the \ion{H}{1}
lines, but these features do not affect the inferred strengths or 
shapes of the strongly saturated \ion{H}{1} lines.

There are no known absorption 
lines produced by the intergalactic medium (IGM) along the PG\,1259+593
sight line at the 
wavelengths of the \ion{O}{1} and \ion{D}{1} lines between 915\,\AA\ and 
950\,\AA.  The 
strongest IGM systems occur at redshifts $z = 0.04606$ and 0.21949
(Richter et al.\ 2003).  In
both cases the \ion{H}{1} Lyman-series 
lines in these systems occur at wavelengths
longward of the $910-950$\,\AA\ spectral region.  
The remaining 19 IGM systems with $z=0.00229-0.43148$
have \ion{H}{1} detectable only 
in the Ly$\alpha$\,$-$\,Ly$\gamma$ transitions (Richter et al.\ 2003), 
so there is no possibility
of contamination at shorter wavelengths by those systems.  
Of the possible metal-line species that are 
likely to be detectable, \ion{C}{3} $\lambda977.020$ tends to be the strongest
and most common; it too is safely longward of the absorption features
considered in this study.  Potential redshifted extreme ultraviolet species 
(e.g., \ion{O}{2} $\lambda834.466$, \ion{O}{3} $\lambda832.927$, \ion{O}{4} $\lambda787.711$) in the 
intergalactic absorbers along the sight line do not pose a problem either.

\section{Oxygen}
\subsection{\ion{O}{1} Velocity Structure}
The first step in our study of the D/H ratio in Complex~C is a determination 
of the \ion{O}{1} velocity structure of the gas along the entire sight line 
in the velocity range  $|v_{LSR}| \le 300$ \kms.  We use both the 
moderate-resolution FUSE observations of numerous \ion{O}{1} lines
at far-ultraviolet wavelengths, combined with the higher resolution HST/STIS 
echelle observations of the \ion{O}{1} $\lambda1302.168$ line, to examine
the optical depth of the absorption in detail.  (Note that the weak
intersystem \ion{O}{1} line at 1355.598\,\AA\ is not detected in the 
STIS spectrum.) Deriving the \ion{O}{1} 
velocity structure
 is critical because it provides an excellent initial 
approximation to the velocity structure of the \ion{D}{1}
and \ion{H}{1} lines and is required for an accurate 
metallicity ([O/H]) estimate for the Complex~C gas.

The ionization 
fraction of \ion{O}{1} is coupled to that of 
\ion{H}{1} (and thus \ion{D}{1}) through a strong charge exchange reaction
at typical interstellar densities (Field \& Steigman 1971):

\begin{equation}
{\rm O^+ (^4S_{3/2}) + H^0 \longrightarrow O^0 (^3P_2)+ H^+ + 0.02~eV}.
\end{equation}

\noindent
Detailed {\tt CLOUDY} photoionization models 
for a range of physical conditions spanning those likely to be encountered 
along the PG\,1259+593 sight line show that the ionization fractions of 
\ion{O}{1} and \ion{H}{1} track each other
closely.  For the specific case of the 3C\,351 sight line through
Complex~C, Tripp et al.\ (2003) find that the ratio of ionization fractions of
\ion{H}{1} and \ion{O}{1} is
unity to within a few percent for $-5.0 \le \log U \le -3.0$, where 
$U = n_\gamma/n_{\rm H}$ is the ratio of hydrogen-ionizing photon density to 
total hydrogen number density.  A valuable check that the ionization 
balance of \ion{O}{1} and \ion{H}{1} is not impacted by extreme conditions
is provided by the observed
ratio of \ion{Ar}{1} to \ion{O}{1} in Complex~C toward PG\,1259+593.
Collins et al.\ (2003) find that the 
two species have a ratio of column densities within a factor of a 
few of the expected value of Ar/O in a solar-abundance gas. 
Neither Ar nor O is severely depleted into dust grains, and  since both
are $\alpha$-elements, their chemical histories should be similar. 
Tripp et al. (2003) deduce that $\log U \sim -4.0$ based on the Collins 
et al. result (see their Figure~14).  Therefore, the ionization correction
for \ion{O}{1} should be small.

The present-day cosmic abundances of D and O are within a factor of 
$\sim20-50$ of each other, which makes the weak \ion{O}{1}
lines in the FUSE bandpass  an excellent guide 
to the velocity structure expected for the \ion{D}{1} lines.  The ratio of 
N$f\lambda$ for the \ion{O}{1} lines to the \ion{D}{1} lines considered 
in this study is roughly unity.  The strong
\ion{O}{1} $\lambda1302.168$ line in the HST/STIS data helps to constrain 
the amount and velocity structure of neutral gas contained in low column 
density components unobservable in the weaker \ion{O}{1}
transitions in the FUSE bandpass, which
is an especially important consideration in modeling the \ion{H}{1}
velocity structure at the velocities of the \ion{D}{1} lines.  An additional
incentive for determining the \ion{O}{1} velocity structure is that
some of the \ion{O}{1} lines overlap the \ion{D}{1} and 
\ion{H}{1} lines (see Figure~1).  It is therefore desirable to estimate the 
contributions of the \ion{O}{1} absorption
 to the blended (\ion{D}{1}+\ion{O}{1}) absorption features observed toward
PG\,1259+593.

We list the \ion{O}{1} lines considered in this study in Table~3.  The 
table entries include the wavelength, line strength ($\log f\lambda$),
instrument used for the measurements, and notes about the line and 
possible blends with other absorption features. We have adopted
$f$-values from Morton (1991, 2003), but using the $f$-values 
recommended by  Verner, Barthel, \& Tytler (1994) does
not change the predicted \ion{O}{1} absorption strengths appreciably.
The transitions considered span a factor of 276
in line strength ($f\lambda$) from $\lambda922.200$ to $\lambda1302.168$.
The lines used in our \ion{O}{1} analyses below are indicated in column
5 of Table~3.

To reproduce the observed velocity structure of the \ion{O}{1} lines,
we start with a model that contains components clearly present in the 
neutral gas.
The HST/STIS data shown in Figure~2  are the best guide available.
Three primary components at $v_{LSR}$ = --129 \kms\ (Complex~C), 
--55 \kms\ (IV~Arch),
 and --2 \kms\ (Milky Way ISM) are prominent
in the \ion{S}{2} $\lambda\lambda1250.584$, 1253.811, 1259.519 lines; these
primary components are also seen in the \ion{H}{1} 21\,cm  emission 
profiles shown at the 
top of the figure.  A second Complex~C component at --112 \kms\ is 
present in the \ion{O}{1} $\lambda1302.168$ line profile and in the strong
\ion{Si}{2}, \ion{Al}{2}, and \ion{Fe}{2} lines observed by STIS.  
An additional 
ISM component at +21 \kms\ is also visible in these same lines, and a weak
intermediate velocity component is present at +69 \kms\ in several 
lines (\ion{C}{2} $\lambda1334.532$, \ion{O}{1} $\lambda1302.168$, 
\ion{Al}{2} $\lambda1670.787$, \ion{Si}{2} $\lambda1260.422$, 
\ion{Si}{3} $\lambda1206.500$).
 These 6 components form the basis for our initial model --
a set of velocities, line widths, and estimated optical depths.  The 
resulting model was convolved with the FUSE and STIS LSFs, and the parameters
were varied to minimize the residuals between the model fits and the 
observed FUSE and STIS \ion{O}{1} data.  During this process, we found  
that two additional components at --82 and --29 \kms\ were needed 
to produce an acceptable fit to the ensemble of \ion{O}{1} 
lines. These components are included in the final model spectrum, which has 
the parameters listed in Table~4.

We show the results of the \ion{O}{1} fitting process in Figure~4 for a 
representative set of \ion{O}{1} lines with strengths $-1.0 \lesssim 
\log f\lambda \le 1.826$.  The STIS data for the $\lambda1302.168$ line
are shown at the bottom of the figure, and the FUSE data (orbital night-only)
are shown as either
solid (SiC2) or dashed (SiC1) lines.  The adopted  model, which is
overplotted on the data
as a heavy solid line, reproduces the observed absorption
features in both the weak and strong lines quite well.  The predicted 
strengths of those lines that are not shown are equally well 
represented.  This model solution is not necessarily unique, but it 
provides some fundamental information.  First, it indicates that the 
Complex~C absorption has two distinct components.  Earlier determinations of
N(\ion{O}{1}) using subsets of the present data
(Richter et al.\ 2001b; Collins et al.\ 2003) assumed single-component
representations for the absorption.  Second, the Complex~C absorption is 
distinct from the IV~Arch absorption.  Therefore, degeneracies in the 
parameters for the lower velocity gas are not critical for estimating 
N(\ion{O}{1}), although they are important for understanding the 
possible range of \ion{H}{1} (and thus \ion{D}{1}) velocity structure. 
Third, the model reproduces the 
absorption in both weak and strong \ion{O}{1} transitions, so 
it provides a reliable
estimate of the \ion{O}{1} absorption for those remaining \ion{O}{1}
lines that are blended with other absorption features (see Table~3).

In constructing the model shown in Figure~4, we found that the reduced
absorption strength at $v_{LSR} = 20.5$ \kms\ ($v_{helio} = 10$ \kms) in the 
HST/STIS \ion{O}{1} $\lambda1302.168$ profile  must be 
due to geocoronal \ion{O}{1} airglow emission.  This feature is indicated
by the cross-hatched region in the figure.  
The velocity of the  emission 
feature matches (to within 1 \kms) that of the geocoronal \ion{H}{1} 
Ly$\alpha$ in the same STIS spectrum. Additional weak airglow features of 
\ion{O}{1}$^*$ $\lambda1304.858$ and 
\ion{O}{1}$^{**}$ $\lambda1306.029$ at $v_{LSR} = 20.5$ \kms\ are also
present in the STIS 
spectrum at the 1--2$\sigma$ confidence level, lending further support to the
geocoronal identification of the feature.  The FUSE \ion{O}{1} lines, which are
weaker than  $\lambda1302.168$ and were observed during orbital night,
are not expected to contain geocoronal emission.
They show considerable absorption at the velocity of the emission feature, 
as do other metal lines in the 
STIS spectrum that are not substantial constituents of the Earth's atmosphere
(e.g., \ion{Si}{2} $\lambda\lambda1304.370, 1526.707$;
\ion{Fe}{2} $\lambda1608.451$).  Thus, our \ion{O}{1} model requires
an interstellar absorption component at this velocity.  
  The precise details of
how much interstellar \ion{O}{1} absorption occurs at the velocity of the 
atmospheric \ion{O}{1} $\lambda1302.168$ airglow feature do not affect the 
conclusions of this paper. 

\ion{P}{2} $\lambda1301.874$  ($\log f\lambda = 1.351$; Morton 1991)
 occurs at --68 \kms\ with respect
to \ion{O}{1} $\lambda1302.168$. 
Estimates of the line strengths of interstellar ($v \sim 0$ \kms) 
\ion{P}{2} 
$\lambda1152.818$  ($\log f\lambda = 2.435$, $W_\lambda = 97\pm16$\,m\AA) 
and \ion{P}{2} $\lambda1532.533$ 
($\log f\lambda = 1.067$, $W_\lambda < 50$\,m\AA\ [3$\sigma$]) imply a 
column density log N(\ion{P}{2}) $\approx$ 13.6 and a corresponding equivalent 
width of only $\approx 10$\,m\AA\ for the 1301.874\,\AA\ line.
Therefore, the interstellar 
\ion{P}{2} line
is buried in the strong \ion{O}{1} absorption shown in Figure~4 and contributes
negligibly to the observed \ion{O}{1} profile.    The same is true of the 
IV~Arch absorption.
\ion{P}{2} $\lambda1301.874$ absorption associated with the IV~Arch would
occur near an \ion{O}{1} $\lambda1302.168$ velocity of --123 \kms.  This
line is also expected to be weak ($<10$ m\AA) since the strength of the 
IV~Arch \ion{P}{2} $\lambda1152.818$ absorption is comparable to that of the 
ISM line.  As a consistency check, the IV~Arch \ion{P}{2} $\lambda1301.874$
absorption should have a strength
similar to
that of \ion{P}{2} $\lambda1301.874$ in Complex~C (Richter et al.\ 2001b),
which has a velocity of $\approx -197$ 
\kms\ on the \ion{O}{1} $\lambda1302.168$
velocity scale shown in Figure~4.  There is no detectable absorption
at this velocity over a 50 \kms\ wide interval
($W_\lambda < 21$\,m\AA, log N $< 13.91$ [3$\sigma$]).  
We conclude that the 
\ion{P}{2} $\lambda1301.874$
lines for the ISM, the IV~Arch, and Complex~C
do not contribute significantly to the observed \ion{O}{1}
$\lambda1302.168$ absorption shown in Figure~4.

\subsection{\ion{O}{1} Column Densities}

In Table~4 we list the velocities, line widths, and approximate \ion{O}{1} 
column densities for each of the components used in the fit to the \ion{O}{1} 
lines.  The errors on the column densities and widths are difficult to 
estimate precisely for some of the individual ISM components because of the 
complex velocity structure of the overlapping features.  However, the summed 
column densities for the three primary absorption groups (Complex~C,
IV~Arch, ISM) and the positive velocity IVC
are well-constrained by the observations.  The total column density and 
$1\sigma$ error
for each group are listed after the individual component values in Table~4.
We discuss the column densities in each component group below.

\subsubsection{Complex~C}
Most of the \ion{O}{1} in Complex~C is contained in the component at 
--129.5 \kms\ (component \#1).  This component dominates the absorption produced by Complex~C
in the weak \ion{O}{1} lines in the FUSE bandpass and is the main 
constituent of the $\lambda1302.168$ absorption observed with HST/STIS.
The weaker component at --112.5 \kms\ (component \#2) 
accounts for the absorption in 
the lower velocity Complex~C absorption in the $\lambda1302.168$ line and
$\lambda1039.230$ lines, but does not 
contribute significantly
to the weaker \ion{O}{1} absorption features observed.  The primary constraints
on the weaker feature are set by the shape of the $\lambda1302.168$ line since the 
absorption produces only a small inflection in the $\lambda1039.230$ line.
The widths of both Complex~C components are tightly constrained by the steep 
absorption walls of the $\lambda1302.168$ line.

The large range of \ion{O}{1} line strengths available constrains the 
total column density of the Complex~C absorption.  We find a total Complex~C
\ion{O}{1} column density of

\begin{center}
N(\ion{O}{1}) = $(7.2 \pm 2.1) \times10^{15}$ cm$^{-2}$  (68\% confidence) \\
N(\ion{O}{1}) = $(7.2 \pm 4.2) \times 10^{15}$ cm$^{-2}$  (95\% confidence) \\
\end{center}

\noindent
where the errors are dominated by systematic effects associated with 
the saturation correction necessary for the higher velocity
component containing most of the column density.  The statistical error
is negligible ($\lesssim5$\%) compared to the systematic error since many 
lines are used to determine the quality of the fit.  The level of inferred 
saturation in the strong component must be judged primarily by the strengths
of the lines observed in the FUSE bandpass because the $\lambda1302.168$ 
line has a very high optical depth.  Fortunately, the width of the stronger
component is well-constrained (b = $6.0\pm1.0$ \kms).  The small amount
of \ion{O}{1} in the lower velocity Complex~C component contributes at most
$\sim10$\% to the total \ion{O}{1} column density in Complex~C.  However, this 
component must be included in the profile fits 
because it affects the inferred b-value for the strong component;
single-component fits to the 
ensemble of \ion{O}{1} lines shown in Figure~4
systematically underestimate the true column density by $\sim20$\%.

Previous studies of the \ion{O}{1} abundance in Complex~C used
single-component curves of growth to estimate the \ion{O}{1} column
density and
found N(\ion{O}{1}) $\approx 5.7\times10^{15}$ cm$^{-2}$, which is 
within our $1\sigma$ confidence range but about 20\% less than our
preferred value (Richter et al.\ 2001b: log~N = $15.77\pm^{0.37}_{0.31}$;
Collins et al.\ 2003: log~N = $15.75\pm^{0.18}_{0.24}$).
Our best-fit
column density estimate, log N(\ion{O}{1}) = $15.86\pm0.15$, 
is slightly higher than the values
reported in these previous studies because we derive a 
smaller b-value for the main Complex~C component than the single-component
COG studies; statistically, all three values are within the quoted
errors on each measurement. The two previous studies adopted 
b $\approx$ 10 \kms\ for the Complex~C
absorption, compared to b $\approx 6$ \kms\ for the stronger component in 
this study, and neither made use of the detailed 
velocity structure revealed by the high-resolution STIS data even though the 
$\lambda1302.168$ line was included in the COG fits.
One expects a larger b-value for the single-component COG fits because the 
weaker ($-112.5$ \kms) Complex~C
component contributes to the total equivalent widths of the strong
lines used in the COG fits ($\lambda\lambda1302.168, 1039.230$).
In these cases, the b-value is an ``effective''
b-value for the combination of the two components weighted by the individual
line strengths, line widths, and velocity separation of the two components.
The Complex~C absorption requires a minimum of 
two components to fit the non-symmetrical 
velocity structure observed in the $\lambda1302.168$  line.  
Adding additional weak components does not improve the fit in a 
statistically meaningful way.

We checked the veracity of our \ion{O}{1} profile fit results by measuring the 
equivalent widths of the lines shown in Figure~4 and plotting these data
points on curves of growth.  The measurements are listed in Table~5,
and the curves of growth are shown in Figure~5.
We considered two cases: a single-component curve of growth (like the ones 
used in previous analyses) and a 
double-component curve of growth based on the fit parameters listed in 
Table~4.  For both curves of growth, we have plotted the corresponding \ion{O}{1}
$\lambda1302.168$ model absorption profile over the observed data
shown in the inset box in each panel of Figure~5.
The single-component fit yields log~N(\ion{O}{1}) = 
$15.66\pm^{0.09}_{0.06}$ and b~=~$9.7\pm0.7$ \kms.  This column density
is at the 2$\sigma$ 
lower limit
of the column density in our best fit model.  The double-component
curve of growth improves upon the single-component column density estimate by 
providing both a good fit to the observed equivalent widths {\it and} a 
better fit to the velocity structure 
of the $\lambda1302.168$ 
line profile, as can be seen from the insets in Figure~5.

The error derived for N(\ion{O}{1}) from the profile fitting 
process is shown graphically on the double-component curve of growth
illustrated in Figure~6.  The dashed curves represent the COG results
when the value of N(\ion{O}{1}) in the stronger Complex~C component is
changed by $4.2\times10^{15}$ cm$^{-2}$
($\pm2\sigma$).  The change in column density affects the
predicted strengths of the weak \ion{O}{1} lines to a much greater extent
than the strength of the $\lambda1302.168$ line.  The range of column density
values allowed is consistent with the range predicted from the profile
fits.

\subsubsection{IV~Arch and Galactic ISM}
The column densities derived for the IV~Arch and the ISM components 
toward PG\,1259+593 do not strongly affect the value of N(\ion{O}{1})
derived for Complex~C. 
The total \ion{O}{1} column density for the IV~Arch in our model
is log~N(\ion{O}{1})
= $16.07\pm0.10$, which is roughly a factor of 2 smaller than
the value of $16.34\pm^{0.35}_{0.27}$ derived by Richter et al.\ (2001b) from 
a single-component COG. The
two estimates are within the measurement uncertainties of each other,
and the exact value adopted depends weakly upon the lower velocity 
integration limit.
The two weak IV~Arch components at --82 and --29 \kms\ are not 
visible in the \ion{H}{1} 21\,cm emission profiles; the column densities
are an order of magnitude or more lower  than the column density in the 
principal component 
at --55 \kms.  However, these two components contribute to the \ion{O}{1}
(and presumably \ion{H}{1} and \ion{D}{1}) absorption.
Using log~N(\ion{H}{1}) = 19.48
derived from the Effelsberg data shown at the top of 
Figure~2 (see also Wakker et al.\
2003), we find 
[O/H]$_{\rm IV\,Arch}$ = log~[N(\ion{O}{1})/N(\ion{H}{1})] --
log~(O/H)$_\odot \approx -0.10$.

The total \ion{O}{1} column density derived for the ISM components 
with $v_{LSR}$ = --2.5 and +21 \kms\ is 
log~N = $16.11\pm0.10$, which is 
in reasonable agreement with what would be expected for gas with solar 
abundances and an \ion{H}{1} column density log N(\ion{H}{1}) = 19.67
derived from the Effelsberg data shown in Figure~2.  We find 
(O/H)$_{\rm ISM} \approx 2.75\times10^{-4}$, or
[O/H]$_{\rm ISM}$ $\approx -0.25$, for the gas between --30 and +30
\kms.  

\section{Hydrogen}
\subsection{\ion{H}{1} 21\,cm Emission Data and the \ion{H}{1} 
Column Density of Complex~~C}
To derive the best value for N(\ion{H}{1}) in the direction of PG\,1259+593, we
combined the WSRT map shown in Figure~3 
with the single-dish Effelsberg observation centered on the QSO (see \S2.3). 
The method we followed is fully described by Schwarz \& Wakker (1991) and 
Wakker, Oosterloo, \& Putman (2002). 
In summary, we first ``observed'' the WSRT map with a 
9\farcm7 beam.  The resulting spectrum was subtracted from the Effelsberg 
spectrum, and the full-resolution WSRT data were added back in. To each of 
these spectra we fit a single Gaussian emission feature
to estimate the \ion{H}{1} column 
density in Complex~C. Figure~7 shows the results of these procedures.  
From the fit
shown in the figure,
we estimate N(\ion{H}{1}) = $(8.9\pm0.7) \times10^{19}$ cm$^{-2}$ 
in Complex~C in the direction of PG\,1259+593.  The single-component 
fit to the Complex~C emission has FWHM =  $16\pm2$ \kms\ (b = $9.6\pm1.2$
\kms).
A straight integration of the \ion{H}{1}
profile from --145 to --109 \kms\ yields N(\ion{H}{1}) =
$(9.0\pm0.6)\times10^{19}$ cm$^{-2}$, where the statistical
error in this case is given by propagating the RMS noise per channel
(0.32 Jy beam$^{-1}$) over 28 channels of width 1.288 \kms\ per channel.
Note that N(\ion{H}{1}) is nearly identical to that derived 
from the Effelsberg data because N(\ion{H}{1})$_{_{\rm recovered}} \sim$ 
N(\ion{H}{1})$_{_{\rm  interferometer}}$.
The higher resolution WSRT data reveal
that the profile is about 20\% narrower than previously measured, showing 
that some of the width of the single-dish profile is due to a small velocity 
gradient in the 9\farcm7 Effelsberg  single-dish beam.

Sources of systematic errors associated 
with our determination of N(\ion{H}{1}) in Complex~C 
include cleaning and calibration of the data,
scaling of the single-dish data, the shape of the single-dish beam, 
the ``negative bowl'' in the interferometer maps, and the fine-scale
\ion{H}{1} structure within the beam.
Adding the uncertainties associated with the first four of 
these errors  together yields an ``instrumental'' systematic
error of $0.7\times10^{19}$ cm$^{-2}$ if all
act in the same sense; this uncertainty is comparable to the 
statistical error quoted above.  
Although the WSRT data are not able to provide information about
gas structures on scales smaller than the 1\arcmin\ effective beam size
($\approx1.5$ pc at a distance of 5 kpc),
it is possible to assess the magnitude of the uncertainties caused
by \ion{H}{1} fine structure.  We calculated the fluctuations in the 
\ion{H}{1} column density in box sizes of $3\times3$ pixels 
($60\arcmin\times60\arcmin$), $5\times5$ pixels ($100\arcmin\times110\arcmin$),
and  $7\times7$ pixels ($140\arcmin\times154\arcmin$).  We find averages
and variations of $(9.0\pm0.5)\times10^{19}$ cm$^{-2}$, 
$(10.0\pm1.6)\times10^{19}$ cm$^{-2}$, and $(11.0\pm2.4)\times10^{19}$ 
cm$^{-2}$,
respectively.  The minimum and maximum values in the $3\times3$ pixel box
are $8.4\times10^{19}$ cm$^{-2}$ and $10.0\times10^{19}$ cm$^{-2}$, 
respectively.  Some of these small-scale structure effects are already
covered indirectly in the ``instrumental'' systematic error estimate, but we 
note that there could still be additional uncertainties contributed by 
fluctuations at small scales that we cannot assess.
A total systematic error of $\sim1.1\times10^{19}$ cm$^{-2}$ accounts 
for most of the systematic errors listed above, including small-scale 
structure within the $3\times3$ pixel box, 
even if the majority of these uncertainties 
affect the column density measurement in the same sense.   We adopt this 
value as our 95\% confidence uncertainty
on the systematic error, and a value of $0.8\times10^{19}$ cm$^{-2}$
as our 68\% confidence uncertainty.

Adding the corresponding systematic and statistical errors in quadrature to
produce a total error  yields a final \ion{H}{1} column density 
estimate based on the 21\,cm emission data of
\begin{center}
N(\ion{H}{1}) = $(9.0\pm1.0)\times10^{19}$ cm$^{-2}$  (68\% confidence) \\
N(\ion{H}{1}) = $(9.0\pm1.6)\times10^{19}$ cm$^{-2}$  (95\% confidence).
\end{center}
\noindent We adopt this column density and its associated errors in our 
discussion of the \ion{H}{1} Lyman-series profiles, and we use the 21\,cm
data as a constraint on the velocity structure of the neutral gas along
the sight line.

Other sight lines show higher velocity 21\,cm emission in Complex~C
than the PG\,1259+593 sight line (e.g., the Mrk~876 sight line; Murphy et al.\
2000).  
Therefore, it is useful to consider whether
the PG\,1259+593 sight line might also contain some lower column density 
\ion{H}{1} at higher negative velocities that might be confused with  
\ion{D}{1} in the Lyman-series lines.  The 21\,cm emission data 
set a limit of N(\ion{H}{1}) $<2.0 \times10^{18}$ cm$^{-2}$ at velocities
 $-275 \le v_{LSR} \le -175$ \kms, the velocity range over which
 low column density \ion{H}{1} could masquerade as \ion{D}{1} absorption.  
In constructing
this limit, we have used the high-S/N side-lobe-cleaned 21\,cm spectra
shown in Figure~2 from
the NRAO 140-foot telescope (Murphy, Sembach, \& Lockman, private
communication). These
data set a more stringent constraint on the amount of very high velocity gas 
than either the WSRT or Effelsberg data.

A limit on the amount of \ion{H}{1} that could be present at very high
 velocities
can also be set by the lack of absorption in the \ion{O}{1}
$\lambda1302.168$ and \ion{C}{2} $\lambda1334.532$ profiles.
We find 3$\sigma$ equivalent width limits of 30\,m\AA\ for both lines
over this 100 \kms\ velocity interval.
These limits correspond to N(\ion{O}{1}) 
$< 4.1\times10^{13}$ cm$^{-2}$ and N(\ion{C}{2}) 
$< 1.5\times10^{13}$ cm$^{-2}$, assuming the lines lie on the linear part
of the curve of growth.  For 
 gas with solar abundances, (O/H)$_\odot = 4.90\times10^{-4}$ (Allende
Prieto et al.\ 2001)
and (C/H)$_\odot  = 2.45\times10^{-4}$ (Allende Prieto, Lambert, \& 
Asplund 2002),   the 
metal-line limits translate into \ion{H}{1} limits of $<8.4\times10^{16} (Z/Z_\odot)^{-1}$
cm$^{-2}$ (from \ion{O}{1}) and $<6.1\times10^{16} (Z/Z_\odot)^{-1}$ cm$^{-2}$ 
(from \ion{C}{2}).  For $Z/Z_\odot\sim0.1$, these limits are only a factor of 
a few times more stringent than those provided by the \ion{H}{1} 21\,cm 
emission data.  Consideration
of narrower velocity intervals would produce smaller upper limits. 
Unfortunately, neither the 21\,cm emission nor the metal-line absorption
provide strict enough limits to rule out the possibility that some 
of the absorption attributed to \ion{D}{1}  may be high-velocity, low-density
\ion{H}{1} (see \S7.1).

While it is not possible to distinguish between \ion{D}{1} and 
interloping \ion{H}{1} in the Lyman-series lines observed by FUSE, an estimate
of the amount of neutral gas at high positive velocities can be made.
The lack 
of \ion{H}{1} absorption  at $+100 < v_{\rm LSR} < +300$ \kms\ in the 
Ly$\epsilon$ line sets a limit of 
log~N(\ion{H}{1}) $< 14.55$ $(3\sigma)$ on high positive velocity gas along 
this sight line.  This is the strongest 
Lyman-series line for which this velocity interval is not confused by 
overlapping absorption features;  Ly$\alpha$ and Ly$\beta$ have damping 
wings from lower velocity components, Ly$\gamma$ has a nearby intervening
intergalactic absorption line at these velocities, and Ly$\delta$ 
has nearby interstellar \ion{O}{1} absorption at these velocities. 

\subsection{Lyman-Series \ion{H}{1} Velocity Structure}
An important step in our estimation of the deuterium abundance
and the D/H ratio in Complex~C is the conversion of the \ion{O}{1} velocity
structure into the \ion{H}{1} velocity structure.  Once this conversion is 
completed, the D/H ratio in the  Complex~C components 
can be estimated by minimizing 
the residual absorption in the negative velocity wings of the 
\ion{H}{1} Lyman-series
lines.  To provide the most accurate \ion{H}{1} velocity structure determination, 
we placed several constraints on the \ion{H}{1} model profiles
constructed with the \ion{O}{1} template, as follows:

\begin{enumerate}
\item{We required that the  column densities of the three  
principal component groups (Complex~C, IV~Arch, and ISM) have total \ion{H}{1} 
column densities determined from the \ion{H}{1} 21\,cm emission data.  
The \ion{H}{1} column density of a fourth component group, the positive 
IVC, was calculated from N(\ion{O}{1}) assuming a solar abundance of oxygen.
These column densities are summarized in Table~6.
Within these 
component groups, we distributed the \ion{H}{1} column
densities in the proportions derived for the \ion{O}{1} components
(see Table~4).  For 
example, in Complex~C, we required that the --127.4 \kms\ component
contain $\sim97$\% of the total \ion{H}{1} column density.  We 
call the column density fraction in the Complex~C components $x_C$, and set 
$x_{C_1} = 0.97$ and $x_{C_2} = (1-x_{C_1}) = 0.03$.} 

\item{We required that the adopted \ion{H}{1} model reproduce the 
available 21\,cm emission profiles.  For Complex~C, this required
a match to the interferometric data described in \S2.3.  For the 
remaining components at $v_{LSR} > -100$ \kms, this required a match to the 
single-dish (9\farcm7 beam) Effelsberg data. Model parameters
constrained by this criterion can be found in the notes for Table~6.}

\item{We required the eight \ion{H}{1} components to have velocities 
within a few \kms\ of those of the \ion{O}{1} components.  It was necessary to 
deviate from the \ion{O}{1} velocities slightly for only two of the 
eight components.  For the highest negative velocity component in Complex~C
(component \#1),
the difference was 2 \kms; this deviation is well within the uncertainty of 
the \ion{O}{1} model velocity but was necessary to reproduce the \ion{H}{1}
21\,cm emission profile.   For the ISM feature at --6.5 \kms
(component \#6), the
deviation was 4 \kms.  In this case, the precise central velocity is not 
well constrained since additional sub-components (which do not affect
our analysis) may be present.  The \ion{O}{1} and \ion{H}{1} velocities of
all other components in the model were the same.}

\item{With a few notable exceptions discussed below, we adopted the \ion{O}{1} 
b-values for the \ion{H}{1} components under
the assumption that turbulent broadening dominates the line widths.  
In most cases the \ion{H}{1}
components are strongly saturated and
overlapping, so a precise estimate of the individual component widths was not 
possible.  However, the total widths of the component groups was well
constrained by both the \ion{O}{1} absorption-line data and the \ion{H}{1}
21\,cm emission data.
For the Complex~C and positive IVC components, the widths could be constrained 
by the steep absorption walls of the higher order Lyman-series lines.
For these cases, the differences in the \ion{H}{1} and \ion{O}{1} b-values
can be understood if there is a thermal contribution to the derived
line widths.}

\item{We fixed the D/H ratio in all components except the two Complex~C 
components at the local ISM value of D/H = $1.5\times10^{-5}$ (see Moos et 
al.\ 2002 and references therein).  The exact choice of D/H for the
ISM and intermediate-velocity components does not influence the results of 
this study in any way, as long as the value chosen is not absurdly large.  
Our study of the \ion{H}{1} lines does not provide any meaningful 
constraints on those D/H values;
by adjusting the \ion{H}{1} parameters slightly, values of D/H as large as 
those toward $\gamma^2$ Vel 
(D/H = $2.2\times10^{-5}$;
Sonneborn et al.\ 2000) could be accommodated easily, as could 
values as low as  D/H = 0.  For the two Complex~C components, the D/H ratio
was allowed to vary under the constraint that D/H was equal in the two 
components.}

\end{enumerate}

Terrestrial \ion{H}{1} airglow is a major contaminant of the observed
\ion{H}{1} Lyman-series 
profiles.  The width of the airglow emission lines is approximately
100 \kms\ since the airglow  fills the FUSE LWRS apertures.  Fortunately, the 
intensity of the emission is reduced substantially in the night-only data 
considered here, and the velocity of the \ion{H}{1} airglow 
emission is sufficiently low that 
it does not occur at the velocities expected for the \ion{D}{1} absorption
in Complex~C (roughly  --250 to --180 \kms\
with respect to the rest velocity of the \ion{H}{1} lines).
Furthermore, the airglow strength diminishes in the higher-order 
lines in the Lyman series.  Since the airglow contamination
occurs at velocities where the \ion{H}{1} lines are expected to be strongly 
saturated, the airglow does not compromise the \ion{H}{1} column 
density and velocity analyses.

In Figure~8 we show the absorption produced by the \ion{H}{1} model
assuming (D/H)$_{\rm Complex~C}$~=~0.  The wavelengths of the eight
\ion{H}{1} and \ion{D}{1} components are marked above each spectrum, and a
 velocity scale for the \ion{H}{1} lines is given at the top
of each panel.  The data are plotted as histogrammed lines. 
The model fit shown with the heavy dashed lines 
in Figure~8 contains \ion{H}{1}, \ion{D}{1}, and 
\ion{O}{1}; no lines of other species are present in these panels.  
The contribution of the \ion{O}{1} absorption to 
the model is shown by the smooth solid line.  For the Ly$\delta$ and Ly$\theta$
lines, \ion{O}{1} absorption occurs at wavelengths near the \ion{H}{1}
absorption but does not
overlap the \ion{H}{1} and \ion{D}{1} lines.  For the Ly$\epsilon$
line, \ion{O}{1} absorption overlaps the low velocity portions of the 
profile but does not occur at the wavelengths expected for \ion{D}{1} 
absorption in Complex~C.
The residual absorption in the negative velocity wings of the 
Ly$\delta$ and Ly$\epsilon$ profiles in Figure~8 is due to the absence 
of \ion{D}{1} at Complex~C velocities in the model illustrated. 

 We have isolated the \ion{H}{1}
absorption due to different component groups in Figure~9, where we 
show the Complex~C, IV~Arch, and ISM/positive IVC absorption in the 
\ion{H}{1} Ly$\theta$ line.  Note that the ISM and IV~Arch produce
no absorption at the expected velocities of the \ion{D}{1} lines in 
Complex~C.
The steep
absorption wings in the higher order
Lyman-series lines set limits on the widths of 
components 1 and 2 (Complex~C), and component 8 (the positive IVC). 
These widths and 
their associated uncertainties are listed in Table~6.  We adopt a line width 
b = $9.6\pm1.2$ \kms\ for component \#1 based on considerations of the 
\ion{H}{1} 21\,cm emission.

We investigated the effects of adding to our model a broad \ion{H}{1} feature
associated with the hot gas traced by the \ion{O}{6} absorption in
Complex~C.  Sembach et al.\ (2003) find log~N(\ion{O}{6}) $\approx 13.72\pm0.13$
 with 
a central absorption velocity of --110 \kms.  The \ion{H}{1} column
density associated with this hot gas is 

\begin{equation}
{\rm N(H\,I)} = \frac{\rm N( O\,VI)}{(Z/Z_\odot)~(O/H)_\odot}~\frac{\rm f_{H\,I}}{\rm f_{O\,VI}},
\end{equation}

\noindent 
where f$_{\rm H\,I}$ and f$_{\rm O\,VI}$ are the ionization fractions
of \ion{H}{1} and \ion{O}{6}, respectively.

For gas at $T=2.8\times10^5$\,K, the temperature at which \ion{O}{6}
peaks in abundance in collisional ionization equilibrium, 
f$_{\rm H\,I} = 1.65\times10^{-6}$
and f$_{\rm O\,VI} = 0.22$ (Sutherland \& Dopita 1993).  
The solar abundance of oxygen is (O/H)$_\odot = 4.90\times10^{-4}$ 
(Allende Prieto et al.\ 2001).
Thus, for 
$Z/Z_\odot = 0.1$, we find N(\ion{H}{1}) $\approx 8\times10^{12}$ cm$^{-2}$.  At 
$T = 2.8\times10^5$\,K, the \ion{H}{1} lines arising in the hot gas would have 
b $\approx 68$ \kms\ and would produce negligible absorption at the 
velocities of the \ion{D}{1} lines in Complex~C.  (The maximum \ion{H}{1}
absorption depth would be $\approx1\%$ of the continuum level in the Ly$\beta$
line, and $\sim0.01$\% in the higher order Lyman-series lines we are 
interested in here.)
The assumption of ionization equilibrium in Eq. (2) is probably overly
simplistic, and more realistic estimates would treat the (presently 
unstudied) non-equilibrium cooling of the gas.
The \ion{H}{1} estimates increase if the gas is not in collisional ionization
equilibrium and is at lower temperatures than implied by the presence of 
\ion{O}{6}.  However, for the \ion{H}{1} and \ion{D}{1} lines of 
interest in this study, the expected \ion{H}{1} absorption depths at the 
velocities of the higher order \ion{D}{1} absorption lines are still less
 than a few percent of the local continuum, even if the value 
of f$_{\rm H\,I}$ / f$_{\rm O\,VI}$
in the above equation is increased by a factor of 50--100. A similar 
conclusion is reached for the \ion{H}{1} associated with the 
\ion{O}{6} absorption produced by the Galactic disk and halo along the 
sight line.  Even though the Galactic \ion{O}{6} column density is higher
than it is in Complex~C [log~N(\ion{O}{6}) = $14.22\pm0.05$; 
Savage et al.\ (2003)],
the velocity is lower and the impact on the \ion{D}{1} absorption in
Complex~C is negligible.  We conclude that the presence of \ion{H}{1}
associated with the hot gas traced by \ion{O}{6} in
Complex~C does not affect the results of our study.

A check on the overall validity of the 
velocity structure along the sight line is provided by the 
\ion{H}{1} and \ion{D}{1} Ly$\alpha$ and Ly$\beta$ lines shown in Figure~10.
These lines were not included explicitly in the derivation of the 
\ion{H}{1} and \ion{D}{1} models.  The ability of the model to reproduce
the damping wings
of the Ly$\alpha$ and Ly$\beta$ absorption features  so reliably in Figure~10
indicates that the overall column density estimates for the two dominant 
components along the sight line (Complex~C and the ISM) are reasonable.

A second check on our \ion{H}{1} model comes from the 21\,cm emission
data shown in Figure~11, where we have plotted a composite profile using
the interferometric results for Complex~C ($-150 < v_{\rm LSR} < -100$ \kms),
and the single-dish Effelsberg data for the IV~Arch and ISM
emission features at $v_{\rm LSR} > -100$ \kms.  
We converted the \ion{H}{1} absorption 
model into an emission model under the assumption that the gas is 
optically thin, such that 
N(\ion{H}{1}) $= \int{{\rm N}(v) dv} = 1.823\times10^{18} \int{T_b(v) dv}$
(Spitzer 1978).  The agreement of the model and the 21\,cm data in Figure~11
is excellent,
as required by the constraints imposed above.  The weak \ion{H}{1} 
absorption components (\#5 and \#7)
at --29 \kms\ and +21 \kms\ do not produce significant 21\,cm emission.

We note that the interferometric data for the lower velocity \ion{H}{1} 
features toward PG\,1259+593
 are consistent with the single-dish measurements, but are much
noisier.  The interferometer data show that there is relatively little 
small-scale structure in the IV~Arch and the low-velocity gas. 
The interferometer
observation automatically filters out the larger scale ($\gtrsim20\arcmin$)
 features,
leaving  little signal  in the ISM or IV~Arch features and 
 implying that arcminute-scale 
variations are small, no more than 10--20\% of the total column density.
This can be explained by the fact that the IV~Arch and the low-velocity clouds
along the PG\,1259+593 sight line are relatively close and do not exhibit any
bright compact cloud cores.

\section{Deuterium}

\subsection{Complex~C Profile Fitting Results}

Using the information derived above for \ion{O}{1} and \ion{H}{1}, it is 
possible to bracket the allowed range of values for N(\ion{D}{1}) in 
Complex~C.  We considered several approaches for estimating the \ion{D}{1}
column density.  Since most of the \ion{D}{1} lines are weak, the strongest
constraints on N(\ion{D}{1}) come from the parameters assumed for the 
stronger Complex~C feature (component \#1).  The weaker feature (component \#2)
does not contribute 
significantly to the observed \ion{D}{1} absorption features unless the 
D/H ratio in this component is an order of magnitude or more higher than the
D/H ratio in the stronger component.  For simplicity, we assume that the 
D/H ratio is the same in both components, and focus our attention on the 
--127.4 \kms\ component.

The widths of the \ion{O}{1} and \ion{H}{1} lines in the stronger component
bracket the width expected for \ion{D}{1} (i.e., b = 6.0--9.6 \kms).  
If we assume that the differences in the b-values between \ion{O}{1} and 
\ion{H}{1} are due to a combination of turbulent and thermal broadening of 
the lines, then we can calculate the component widths 
assuming that b$^2$ = b$_{\rm turb}^2$ + b$_{\rm therm}^2$.  Solving this 
equation for the observed \ion{O}{1} and \ion{H}{1} b-values
(see Tables~4 and 6) simultaneously 
leads to  b$_{\rm turb} = 5.7\pm1.1$ \kms\ and $T = \phn3600\pm1700$\,K.  Using
these parameters implies b $\approx 8\pm1$ \kms\ for \ion{D}{1}.

The model found by minimizing the residuals of the fit 
for the \ion{H}{1} and \ion{D}{1} lines has N(\ion{D}{1}) = 
$2.35\times10^{15}$ cm$^{-2}$ and 
(D/H)$_{_{\rm Complex~C}} = 2.6\times10^{-5}$. This result is insensitive to 
the width
chosen for the weaker Complex~C component, which was assumed to be 
15 \kms; values of b for component~2
spanning the widths implied by the \ion{O}{1} and 
\ion{H}{1} lines (b = $10-24$ \kms) are plausible and indistinguishable.
Parameters for the \ion{D}{1} components in this 
model can be found in Table~7.  For the ISM
and IVC lines, we have adopted the component widths and velocities
found for \ion{H}{1}, and we have set the D/H ratio
to D/H = $1.5\times10^{-5}$.

Model results for the \ion{H}{1}, \ion{D}{1}, and \ion{O}{1}
absorption along the sight line are shown in Figures~12 and 13 for the 
FUSE SiC2 and SiC1 data, respectively. The absorption features
produced by the 
Lyman-series lines of Ly$\delta$--Ly$\kappa$ are shown as histogrammed lines.
The total (\ion{H}{1} + \ion{D}{1} + \ion{O}{1}) 
absorption model is shown as thick dashed curves.
The smooth solid line isolates the contribution to the fit produced by
\ion{O}{1} absorption along the sight line. The velocity scale applies to 
the \ion{H}{1} line shown in each panel.
 In this model, (D/H)$_{\rm Complex\,C} = 2.6\times10^{-5}$.
The composite model fits the data reasonably well, with the noticeable
exception of a small underestimate in the Ly$\delta$ and Ly$\epsilon$ lines
near $v_{\rm LSR} = -190$ \kms\ on the \ion{H}{1} velocity 
scale ($\approx -108$ \kms\ on a \ion{D}{1} velocity scale).  The 
underestimated absorption at these velocities, which is highlighted with
arrows in the top panels of Figures~12 and 13, is important because it 
impacts the quality of the fit for the higher-order lines.  The 
model is driven toward higher values of D/H to fit 
the Ly$\delta$ and Ly$\epsilon$ lines at the expense of slightly 
over-producing the absorption in the weaker lines. 
Before adopting a final column density and error for N(\ion{D}{1}),
we  consider some
reasons for this underestimate.

First, we considered the possibility that the underestimated absorption 
may be due to how the total \ion{D}{1} column density in Complex~C is
distributed between the two components. 
In our model we have assumed that the 
column densities of \ion{O}{1}, \ion{H}{1}, and \ion{D}{1} are divided 
among the two components such that $x_{C_1} = 0.97$ and $x_{C_2} = 0.03$.
Holding other parameters fixed,  we allowed the ratios of N(\ion{D}{1})
and N(\ion{H}{1}) in the two components to vary.   
Values of $x_{C_1}$ as low as 0.94 improved the fit to the data.  Values
of $x_{C_1} < 0.94$ produced noticeable discrepancies with the negative
velocity absorption walls of the higher order \ion{H}{1} lines, in the 
sense that too much absorption was predicted.  Lowering the b-value of 
component 2 to alleviate this concern reintroduced the discrepancy 
in the Ly$\delta$ and Ly$\epsilon$ lines.  The changes considered affected
both N(\ion{D}{1}) and N(\ion{H}{1}) by the same percentage since the 
D/H ratio was assumed constant in Complex~C.  Thus, while the fit could
be improved slightly, such changes had no effect on (D/H)$_{_{\rm Complex~C}}$.

We also considered whether a difference in the D/H ratio in the two Complex~C
components might lead to an underestimate of the absorption at lower
velocities.  To account for the extra absorption
by increasing the D/H ratio in the weaker feature (component \#2 at 
--112.5 \kms), requires
a value of D/H  that is at least a factor of 8--10 times
greater than D/H in component~1.  Such a discrepancy is inconsistent with 
Big Bang nucleosynthesis estimates of the primordial value of 
D/H, (D/H)$_p \approx (2.6-3.0)\times10^{-5}$
(Nollett \& Burles 2000;
Burles et al. 2001; Spergel et al. 2003), and the expected variations in D/H 
caused by deuterium astration.

Next, we explored the possibility that the 
residual absorption could be due to  
an underestimate in the width of the \ion{D}{1} and \ion{H}{1} lines in the 
main Complex~C component at --127.4 \kms.  To fit the absorption well 
would require
a width b$_{\rm HI} \sim 14$ \kms, which is inconsistent with the width of the 
\ion{H}{1} 21\,cm emission shown in Figure~11.  It is possible for the 
\ion{H}{1} absorption width to be larger than that found for the 21\,cm 
emission if the wings of the absorption consist primarily of low column
density gas that is below the detection limit of the radio observations.
In this case, the \ion{D}{1} line width could also be increased, resulting
in a slightly better fit to the data.  The resulting D/H value 
is similar 
to that found for the single-component curve of growth analysis with 
b$_{\rm DI} \approx 14$ described
in the next section.

The underestimated absorption 
may be due to a small amount of fixed-pattern noise in the 
data for one or both of the SiC channels.  Note that the underestimate is 
slightly different for the lines shown in Figure~12 for the SiC2 channel 
and those shown in Figure~13 for the SiC1 channel.  If the absorption is 
actually fixed-pattern noise, the fit may change the derived 
value of N(\ion{D}{1})  since the \ion{D}{1} profile fit is sensitive to 
absorption at this velocity.  

We considered whether adding an additional component to the 
absorption model to account for the residual absorption affected the 
fit results.  This feature could be either a \ion{D}{1} component 
corresponding to a stronger \ion{H}{1} feature near --108 \kms, or
a weak interloping \ion{H}{1} component near --190 \kms.  To account for the 
absorption with
additional \ion{D}{1} requires an \ion{H}{1} component near --108 \kms\
with  N(\ion{H}{1}) $\approx (2-5)\times10^{19}$ cm$^{-2}$ if D/H = 
$(1.5-3.0)\times10^{-5}$.
While this additional \ion{H}{1} component would remain hidden in the 
\ion{H}{1} Lyman-series lines, it would lead to a serious conflict with both 
the observed \ion{H}{1} 21\,cm emission and the \ion{O}{1} absorption at this 
velocity. 

An \ion{H}{1} feature at --190 \kms\ 
with N(\ion{H}{1}) $\approx 3\times10^{14}$ cm$^{-2}$
and b$_{\rm HI} \approx 10-15$ \kms\ can account for the absorption 
and remove the noted discrepancy.
The weak high-velocity \ion{H}{1} interloper would be 
undetectable through either its 21\,cm emission or corresponding
metal-line absorption (see \S6.1).  The associated \ion{D}{1} absorption
 would not be 
detectable at higher velocities.  In considering this
possibility, we note that the Ly$\alpha$ and Ly$\beta$ lines do not
help constrain whether the absorption could be \ion{H}{1}
because in both cases the features occur in the strong cores of the 
overall sight line absorption.  Ly$\gamma$ is also not restrictive 
since strong \ion{O}{1} absorption blends with \ion{D}{1} and \ion{H}{1}
absorption at this velocity.  If we add a weak \ion{H}{1} feature with these
properties to our model and recalculate the fit, we find that the 
preferred value of N(\ion{D}{1}) in Complex~C drops from $2.35\times10^{15}$
cm$^{-2}$ to $1.62\times10^{15}$ cm$^{-2}$, and (D/H)$_{_{\rm Complex~C}} $
drops from $2.6\times10^{-5}$ to $1.8\times10^{-5}$.
We reproduce Figures~12 and 13 in Figures~14 and 15 with the additional
weak \ion{H}{1} feature added to the model (i.e., D/H = $1.8\times10^{-5}$).   
The greater the 
contribution of this feature, the smaller the 
inferred D/H ratio in Complex~C
becomes.  Many additional parameters
could be tuned to improve the fits shown in Figures~14 and 15, but the 
degeneracies in the possible solutions preclude a definitive statement 
about the exact magnitude of the absorption
caused by this weak feature.
The discrepancies in the fit caused by the presence or absence of 
this feature lead to an uncertainty 
of  $\approx \pm0.4\times10^{15}$ cm$^{-2}$ in the \ion{D}{1} 
column density. 

It is important to note that \ion{H}{1} 21\,cm emission with 
N(\ion{H}{1}) $\sim 10^{19}$ cm$^{-2}$ is seen near --190 \kms\ along
a high-velocity ridge running through Complex~C (see Figure~6 in Wakker 2001
and Figure~13 in Sembach et al. 2003).
PG\,1259+593 lies off of this ridge by several degrees, and even though there
are no metals detected in our absorption-line spectra of PG\,1259+593,
there is still a reasonable possibility that there may be low column
density \ion{H}{1} near --190 \kms\ in this direction. 
In our final estimation of the deuterium abundance in 
Complex~C, we will account for the possible range of values 
of (D/H)$_{\rm Complex~C}$ resulting from the inclusion or exclusion of 
a weak \ion{H}{1} feature.

\subsection{Curve of Growth Results for Complex~C}

As an additional consistency check on our profile fitting results for the
\ion{D}{1} column density in Complex~C, we constructed curves of growth for 
the Complex~C absorption using the equivalent widths of the \ion{D}{1}
lines measured between $-160 \le v_{\rm LSR} \le -110$ \kms.  To make the 
\ion{D}{1} equivalent width measurements listed in Table~8, 
 we removed the \ion{H}{1} and \ion{O}{1}
signatures in the data by dividing the FUSE spectra by the 
adopted \ion{O}{1} and \ion{H}{1} models with the parameters listed in
Tables~4 and~6.  Next, we measured the 
\ion{D}{1} absorption in the 50 \kms\ interval in each channel
 between --160 and --110 \kms\ (--242 to --192 \kms\ on the \ion{H}{1} 
velocity scale shown in Figures 12--15). This velocity range, 
which is indicated 
by the horizontal bar above the Ly$\epsilon$ spectra in Figures~12 and 13, 
includes the Complex~C \ion{D}{1}
absorption recovered by our model result but does not include the small
amount of residual absorption at lower velocities discussed in
the preceding section.  

The results of this process are shown in Figure~16,
where we plot the data along with both a single-component curve of growth
and the double-component curve of growth
based upon the \ion{D}{1} fit parameters listed in Table~7.  
The best-fit single-component curve of growth yields
log~N(\ion{D}{1}) = $15.22\pm^{0.17}_{0.12}$ (1$\sigma$) 
and b = $14.4\pm^{11.4}_{4.0}$ \kms\ (1$\sigma$).  This column density
is less than that implied by the profile fitting process (at least 
for the case where there is no \ion{H}{1} interloper) and the optimum
b-value is higher than the \ion{H}{1} b-value for the dominant component. 
However, it is still within the possible ranges allowed by the 
uncertainties in the profile fits.
The single-component COG b-value  is determined
in large part by the measurements for the Ly$\delta$ line in the two
channels.  Since the measurements for this line differ between the two
channels by an amount similar to the $1\sigma$ measurement uncertainties,
the resulting b-value is particularly uncertain.   For comparison,
the dashed curves in the top panel of Figure~16 show the single-component 
COGs for b-values of 10 and 20 \kms,  assuming the same value of 
N(\ion{D}{1}).

If we take the additional residual absorption redward of the Complex~C 
absorption into account and assume that it is all due to \ion{D}{1}, 
the column density from the single-component
 curve of growth method becomes 
log~N(\ion{D}{1}) 
= $15.33\pm^{0.13}_{0.11}$ and b = $26\pm^{27}_{08}$ \kms.  We conclude 
that the quality of the fit does not improve, which is consistent with our
contention in the previous section that this residual absorption is
unlikely to be due to \ion{D}{1}.

The double-component COG  shown in the bottom panel of Figure~16
 passes through the 
SiC1 Ly$\delta$ measurement but underestimates the SiC2 Ly$\delta$
measurement slightly, as expected from the profile fit 
results shown in Figure~12 and 13.  The SiC2 equivalent width measurement
includes some residual absorption located on the sides of
the Complex~C absorption that is not fit well by the model.  This 
discrepancy between the Ly$\delta$ measurements for the two channels may
be due to fixed-pattern noise in the data (see discussion of errors below).

The curve-of-growth
results illustrate an important point in the 
consideration of the Complex~C \ion{D}{1} column density.  The 
predicted strengths of the lines are sensitive to the value of 
b$_{\rm DI}$ chosen for the stronger component because the Ly$\delta$ 
line does not lie on the linear portion of the curve of growth.  Unfortunately,
omitting the Ly$\delta$ measurements is a poor alternative to including them 
since the remaining detections are weak and have relatively large
errors.  Thus, it is important to consider the systematic error associated 
with a range of possible b-values.

\subsection{Sources of Systematic Errors}

In addition to our concerns about the ability of the profile fitting
model to reproduce
the small amount of residual absorption in the 
\ion{H}{1} + \ion{D}{1} absorption profiles, 
there are other sources of error that must be considered in making an 
estimation of N(\ion{D}{1}) in Complex~C.
The dominant systematic errors include:

\begin{enumerate}

\item{{\it Inability to 
distinguish between small column density differences due to the unresolved
nature of the strongest Complex~C component}.  This is probably the most 
troublesome problem since it necessarily guarantees that the $\chi^2$ 
minimum in N--b parameter space is shallow.  This occurs because the 
main \ion{D}{1} component in Complex~C has a width 
(FWHM $\approx 15$ \kms) that is less 
than the  width of the FUSE LSF (FWHM $\approx20-25$ \kms).  
Using the \ion{O}{1}
and \ion{H}{1} widths as guides, we estimate that the error arising 
from the uncertainty in b when holding other parameters fixed 
is $\approx 0.4\times10^{15}$ cm$^{-2}$.}

\item{{\it The assumed velocity structure for the sight line, particularly the 
Complex~C and intermediate velocity components}.  The error associated
with the assumed velocity structure is difficult to quantify.  One important
example, that of the possible existence of very high velocity \ion{H}{1}
absorption, was provided above and must be considered in the error
budget (see point 4 below). 
It is also possible that the velocity structure for the stronger
Complex~C component consists of several narrow absorption features closely
spaced in velocity. 
High velocity clouds typically consist primarily of warm gas and do not 
often exhibit evidence for cold components (Wakker \& van~Woerden 1997). 
Most HVCs show no evidence of molecular gas, including Complex~C
(Richter et al.\ 2001a).  In one region of the Magellanic Stream
containing dust in which H$_2$ has been detected in 
absorption by FUSE (Sembach et al.\ 2001a), the cloud structure consists
of a compact cloud core that is surrounded by warmer gas. The high-resolution
STIS data for Complex~C do not show any evidence of \ion{C}{1} absorption
that might be associated with cold components in the PG\,1259+593 direction.
Therefore, if narrow components do exist, there is probably no single cold
component that dominates the absorption.  In this case, the single-component
approximation usually results in column densities accurate to better than
20\% (Jenkins 1986; Savage \& Sembach 1991), or about $0.5\times10^{15}$
cm$^{-2}$ for the \ion{D}{1} Complex~C absorption toward PG\,1259+593.
An additional source of uncertainty related to the Complex~C velocity 
structure results from the component width and strength trade-offs that can be 
made between the strong and weak components (components \#1 and \#2).  
We find that these typically affect the total
column density at a level less than $0.2\times10^{15}$ cm$^{-2}$.}

\item{{\it Fixed-pattern
noise that may be present in these FUSE data}.  Small differences
are visible in the SiC2 and SiC1 data for some of the profiles considered in 
this work.  For example, the SiC1 data for the Ly$\delta$ line 
recover fully to the continuum blueward of the Complex~C absorption, while 
the SiC2 data do not (see the top panels of Figures~12 and 13).  
There are also slight discrepancies in the wings of the 
two Ly$\delta$ lines, which lead to the differences in \ion{D}{1} equivalent 
widths listed in Table~8.  If we consider the difference in column densities
resulting from the omission of the SiC data for one Ly$\delta$ line or the 
other as a measure of the error caused by fixed-pattern noise in these
lines, we find an uncertainty of $\approx 0.5\times10^{15}$ cm$^{-2}$ in 
the \ion{D}{1} column density.}

\item{{\it The possibility that the absorption we identify as  
\ion{D}{1} in Complex~C  might be 
high velocity \ion{H}{1}.}  We considered the case of the weak residual
absorption above.  However, our analysis of the metal lines and the 
inferred limits for the amount of \ion{H}{1} at high velocities cannot
formally exclude the possibility that some of the other absorption is 
high velocity \ion{H}{1}.  If {\it all}
of the observed absorption were \ion{H}{1} and not \ion{D}{1}, the
D/H ratio in Complex~C would be strange indeed!
We acknowledge the possibility that some of the absorption assumed to be 
due to \ion{D}{1} in Complex~C could be due to weak, high-velocity \ion{H}{1}
features.  However, without further evidence of this, 
we assume that the absorption observed is caused by \ion{D}{1}, with the 
caveat that the \ion{D}{1} column density decreases by $\sim0.8\times10^{15}$
cm$^{-2}$ if a weak \ion{H}{1}
feature is included to explain the residual absorption near the \ion{D}{1}
absorption.}

\end{enumerate}

\subsection{Complex~C Column Density}

We adopt a \ion{D}{1} column density for Complex~C that is an average of
the values of N(\ion{D}{1}) for the  case where the residual 
absorption is left untreated in our model (no \ion{H}{1} interloper,
N(\ion{D}{1}) = $2.4\times10^{15}$ cm$^{-2}$,
as shown in Figures~12 and 13) and 
the case where the residual absorption is assumed to be due to an 
\ion{H}{1} interloper (N(\ion{D}{1}) = $1.6\times10^{15}$ cm$^{-2}$,
as shown in Figures~14 and 15).  To derive an error, we add in quadrature
the uncertainties associated with this averaging and the systematic 
errors described in \S7.3.  We treat the individual uncertainties as 
2$\sigma$ estimates and combine them in quadrature because the
errors do not all add with the same sign.  We find  

\begin{center}
N(\ion{D}{1}) = $(2.0\pm0.6)\times10^{15}$ cm$^{-2}$  (68\% confidence) \\
N(\ion{D}{1}) = $(2.0\pm0.9)\times10^{15}$ cm$^{-2}$  
(95\% confidence),
\end{center}

\noindent
where we have scaled the $2\sigma$ uncertainty by the confidence 
interval to produce the $1\sigma$ uncertainty.  The upper $2\sigma$ limit 
produces a noticeably unsatisfactory fit to the data.

\section{Discussion}
We summarize the Complex~C column densities of \ion{H}{1}, \ion{D}{1},
and \ion{O}{1} in Table~9, together with the $1\sigma$ (68\% confidence
interval) and $2\sigma$ (95\% confidence interval) error estimates based
on the descriptions of errors given above.  Systematic errors dominate the
uncertainties for all three species.  We also list the  D/H, D/O, and
O/H column density ratios, along with the propagated errors from the column 
density determinations.   
A summary of the light element abundance ratios for Complex~C and other 
environments is given in Table~10, along with references for the 
ratios. The environments considered in this table
include the local ISM, the ISM of the Galactic disk at distances of 200--1000
pc from the Sun, and high-redshift 
absorption line systems (both Lyman-limit systems and damped Ly$\alpha$
systems).  We list both D/H and D/O when estimates of [O/H] are 
available.  
These comparisons
are not meant to be exhaustive summaries of all the available data in the 
literature; rather they are synopses of several studies given to provide 
insight into how the Complex~C values compare to typical values found 
elsewhere. The data in Table~10 are shown graphically in Figure~17, where we 
plot
D/H as a function of metallicity determined through measurements
of [O/H], or [Si/H] if [O/H] is not available.  For the Lyman-limit systems
where [Si/H] is used, the metallicity may be somewhat uncertain because 
large ionization corrections are necessary to convert estimates of 
N(\ion{Si}{2})/N(\ion{H}{1}) into estimates of N(Si)/N(H).  
Various other metallicity 
indicators are available, but [O/H] tends to be the best for 
the reasons described in \S5 (see also the detailed discussion presented
by Timmes et al.\ 1997).
Our estimate of D/H in 
Complex~C rules out values of D/H greater than $3.3\times10^{-5}$ or 
less than $1.1\times10^{-5}$ with a reasonable degree of confidence
($\sim95$\%).  Higher values over-produce the amount of \ion{D}{1} 
absorption in the higher order Lyman series lines, and lower values
under-produce the amount of absorption expected in the Ly$\delta$ line.
Both limits have interesting implications. 

Most models of Galactic chemical evolution predict that astration of 
deuterium should result in a factor of $\sim1.5-3$ decline in the cosmic
abundance of deuterium from the Big Bang to the present time 
(Edmunds 1994; Tosi et al.\ 1998; Chiappini et al. 2002), 
though higher levels of astration may be possible
if special conditions, such as a Galactic wind, are invoked to counter
over-production of elements heavier than helium (see, e.g., Scully et al.\
1997).  For a simple closed-box model with the 
assumption of instantaneous recycling, a deuterium astration efficiency of 
60\%, and standard oxygen yields, Pagel (1997) finds that the there should
be only a few percent reduction of deuterium relative to its primordial
abundance if [O/H] $\lesssim -1$, as found for Complex~C.  In
Figure 17, we show the results 
for a simple chemical evolution model from Fields et al.\ (2001)
having (D/H)$_p$ and little destruction of deuterium at low metallicities from 
Population~III stars.  More complicated chemical 
evolution models requiring bimodal episodes of star formation to explain
the scatter in D/H values measured at high redshift can be found in that work.
Our $2\sigma$
upper limit of D/H $< 3.3\times10^{-5}$ suggests  that the value of D/H
produced by Big Bang nucleosynthesis cannot be too much larger than
this upper limit unless the assumptions about the destruction of deuterium
and production of oxygen are invalid. Similar 
arguments have been made to infer a primordial value of 
(D/H)$_p \lesssim 3\times10^{-5}$ from
determinations of D/H in high-redshift, low-metallicity systems (Tytler et 
al.\ 2000; O'Meara et 
al.\ 2001; Kirkman et al.\ 2003).  
Values of (D/H)$_p \lesssim 3\times10^{-5}$ imply 
values of the baryon density, $\Omega_bh^2$, and the baryon-to-photon
ratio, $\eta$, 
in concordance with estimates derived from cosmic microwave background (CMB)
measurements.  Using the standard Big Bang nucleosynthesis predictions for the
abundance of deuterium (Burles et al. 2001), a value of 
(D/H)$_p \lesssim 3\times10^{-5}$ implies $\Omega_bh^2 \gtrsim 0.02$
and $\eta \gtrsim 5.6\times10^{-10}$.  The CMB estimates cluster near 
values of $\Omega_bh^2 \approx 0.022-0.024$ 
(Netterfield et al.\ 2002; Pryke et al.\ 2002;
Spergel et al.\ 2003), and imply 
(D/H)$_p = (2.62\pm^{0.19}_{0.20})\times10^{-5}$ (Spergel et al. 2003;
see also Steigman 2003).  The region between the dashed horizontal lines
in Figure~17 is the $\pm2\sigma$ range of (D/H)$_p$
allowed by the recent {\it Wilkinson Microwave Anisotropy Probe} (WMAP) 
observations.

With the PG\,1259+593 data, we can confidently 
rule out very high D/H values for Complex~C, like those claimed for some 
intergalactic systems [e.g., D/H $= (2.0\pm0.5)\times10^{-4}$ at 
$z\approx0.7$ toward PG\,1718+4807; 
Webb et al.\ 1997].  This is important because the Complex~C
measurement is the only other measurement of D/H outside the 
Milky Way at $z < 1.0$.
Such high values are also inconsistent 
with the CMB measurements, and recently the value for the 
$z=0.7$ absorber toward PG\,1718+4807 has been challenged on other
grounds (e.g., Kirkman et al.\ 2001 -- but see Crighton et al.\ 2003).

On the low end of the D/H range, we cannot exclude the possibility that
D/H in Complex~C is similar to D/H in the local ISM 
(D/H $\sim1.5\times10^{-5}$, Moos et al. 2002;  Linsky 2003).  However,
we  can  rule out D/H values
as low as those found for some Galactic disk sight lines extending beyond
the local ISM (i.e., at $d \gtrsim 100$ pc).  
Values of D/H less than $1.0\times10^{-5}$ have been
determined for two extended sight lines in the Galactic disk using FUSE
data.  Hoopes et al.\ (2003) find D/H = $(0.85\pm^{0.34}_{0.34})\times10^{-5}$
 ($2\sigma$) toward HD\,195965 ($d \sim 800$ pc), and 
D/H = $(0.78\pm^{0.52}_{0.25})\times10^{-5}$ ($2\sigma$) toward 
HD\,191877 ($d \sim 2200$ pc).  Using IMAPS data, Jenkins et al.\ (1999)
find D/H = $(0.74\pm^{0.19}_{0.13})\times10^{-5}$ (90\% confidence) 
toward $\delta$\,Ori\,A ($d \sim 500$~pc); a similar result was 
found by Laurent, Vidal-Madjar, \& York (1979) using Copernicus data.  
The low values found
for these sight lines indicate that more astration of deuterium may have 
occurred in these regions than in the local ISM or in Complex~C. 
The local star formation
histories of the regions explored toward $\delta$\,Ori\,A and HD\,195965
are roughly consistent with this idea 
(see Hoopes et al.\ 2003 and Jenkins et al.\ 1999), but the number of 
sight lines is still too small to draw general conclusions.  Eventually,
with a large enough number of sight lines, it may be possible to determine
whether refinements to  chemical evolution models are needed.

The value of D/O = $0.28\pm0.12$ in Complex~C is very different from the D/O
ratio in the local ISM or the disk of the Milky Way.  
H\'ebrard \& Moos (2003) find D/O = $0.038\pm0.002$ for white dwarf
and subdwarf sight 
lines within $\sim150$ pc of the Sun.  Most of this difference
can be accounted for by the different metallicities of the local
ISM and Complex~C.  Note that even without the \ion{H}{1} measurement,
the D/O ratio strongly suggests that the metallicity in Complex~C is lower 
than solar.   The Complex~C result is more similar to the value
of D/O = $0.37\pm0.03$ found for the $z=3.025$ absorber toward
Q\,0347-3819 (Levshakov et al. 2002) than it is to the value for the Galactic
disk.  However, values of D/O for  two other high-redshift systems 
are an order of magnitude or more larger than the Complex~C 
value.  We note that determinations of N(\ion{O}{1}) in the high-redshift
systems are challenging since the weaker \ion{O}{1} lines occur within
the \ion{H}{1} Ly$\alpha$ forest.

As estimates of D/H in the high-redshift systems have fallen in 
recent years, a few local measurements with high values of D/H 
have received more attention.  For example, the value of 
D/H = $(2.18\pm^{0.36}_{0.31}) \times10^{-5}$  for the $\gamma^2$ Vel
sight line (90\% confidence; Sonneborn et al.\ 2000) 
is indistinguishable from the 
values for the high-redshift ($z>2$) systems toward 
HS\,0105+1619 and Q\,1243+3407, and perhaps higher than
the value for Q\,2206--199 (see Table~10 and references therein).  
Preliminary FUSE results for 
the IX~Vel sight line, which lies $\sim1\degr$ from $\gamma^2$ Vel, 
indicate that the value of D/H for the first
$\sim100$ pc of the $\gamma^2$ Vel sight line is similar to that in the local
ISM (Blair et al.\ 2003).  
If this result is confirmed by upcoming FUSE observations, then the D/H ratio 
in the Vela region must be even higher than
the sight line average.  Note, however, that the $\zeta$~Pup sight line,
which is in the same region of the sky and extends further than the 
$\gamma^2$~Vel sight line has a D/H ratio similar to that of the local
ISM (see Table~17).
One possible explanation for 
a high value in the ISM toward $\gamma^2$ Vel is infall of ``primordial''
(metal-poor, D-rich) gas.  Clouds as large as Complex~C could contribute
significant amounts of deuterium to regions of the Galactic disk;
a mass of M(\ion{H}{1}) $\gtrsim(1.2-3.0)\times10^6$ M$_\odot$
and D/H = $2.2\times10^{-5}$ imply that 
Complex~C contains $\gtrsim(26-66)$ M$_\odot$ of deuterium.
The effects
of infalling low-metallicity gas have been considered by various authors 
in the context of deuterium astration and potential solutions to 
 the well-known ``G-dwarf'' problem 
(Tosi 1988; Matteucci \& Francois 1989; Tosi et al.\ 1998;
Wakker et al.\ 1999).   However, a serious problem with this interpretation 
for the $\gamma^2$~Vel region is that the gas along the sight line 
does not show a corresponding reduction in the ratios
of metals to hydrogen as would be expected if infall were the explanation
for the high D/H ratio. Furthermore, the constancy of 
the oxygen and krypton abundances within several hundred parsecs of the 
Sun (Meyer, Jura, \& Cardelli 1998; 
Cartledge et al.\ 2001)  supports the idea that the nearby
gas is reasonably well mixed and has similar heavy element abundances.  
Any explanation for the high D/H ratio in the Vela region
that incorporates infall must also include processing of the gas
to explain the relatively ``normal'' metal abundances in the region.

Other possibilities also exist for explaining the variations of D/H in
the Milky Way.  For example, D could be depleted onto dust grains (Jura
1982).  From simple thermodynamic considerations, Draine (2003) has
shown that differences in the binding energies of D and H on the
peripheries of polycyclic aromatic hydrocarbons (PAHs) could produce
extreme enrichment of deuterium in bound form, similar to what has been
found for some simple interstellar molecules.  He argues that it is
plausible that roughly 20\% of the H atoms bound to PAHs are replaced by
D.  If this is correct, he estimates that there are sufficient PAH
binding sites in a volume of normal interstellar material to account for
a decrease of free D atoms by one part in $10^5$ of H (i.e., roughly the
size of the reduction observed along some lines of sight).  In some
places, the atomic medium can revert to the intrinsic D/H when the PAHs
are destroyed by the recent passage of a shock, which could explain why
D/H seems to vary.  Reinforcing the concept that variable D abundances
might be explained by Draine's proposal is the recent detection of the
CD bond stretch features at 4.4 and 4.67 $\mu$m toward the Orion Bar
reported by Peeters (2002).  She found ${\rm D/H} = 0.17\pm 0.07$ in
bound form.  This detection, although not very far above the noise,
indicates that Draine's estimate of a 20\% replacement of H by D atoms
may be approximately correct.  For any gas with an overall metallicity
as low as that of Complex C, the concentration of PAHs would be so low
that there would be no measurable reduction in the concentration of free
deuterium atoms, which is consistent with our finding for the sight line
toward PG\,1259+593.  Finally, 
Mullan \& Linsky (1998) have proposed that deuterium production in stellar 
flares may cause some variations in D/H locally, but Prodanovi\'c \& 
Fields (2003) have recently reconsidered this issue and find that flare
production of deuterium is not a significant source of \ion{D}{1}  on galactic 
scales.  They concur with Mullan \& Linsky that local variations by 
stellar flare production cannot be ruled out.  
Dust depletion and stellar flare production of deuterium are 
unlikely to alter the present value of D/H in Complex~C substantially.
Complex~C contains no known stars, and analyses of its metal
content indicate that it contains little dust.

The metallicity of $Z/Z_\odot\approx0.17$ we derive for Complex~C 
through the ratio
of \ion{O}{1} and \ion{H}{1} is very similar to the recent determination of 
the metallicity in Complex~C toward 3C\,351 by Tripp et al.\ (2003).
They find [O/H] = $-0.76\pm^{0.23}_{0.21}$.  Both metallicity estimates
are slightly higher than the value of $Z/Z_\odot\sim0.1$ 
inferred for the Mrk~290 line of sight through Complex~C based on
measurements of \ion{S}{2} and \ion{H}{1} (Wakker et al.\ 1999).  The 
metallicity estimates may depend on environment and may change 
slightly as a function of position within Complex~C;  
Collins et al.\ 2003 find a range from 0.1 to 0.25 solar, but some of these
estimates are complicated by uncertain ionization corrections for \ion{S}{2}. 
One possible explanation for the slightly 
varying abundances in Complex~C might be that the gas is interacting with 
the Galactic corona or high Galactic halo, as indicated by measurements of 
highly ionized gas associated with the complex (Sembach et al.\ 2003; Fox et
al.\ 2003).  Mixing of the diffuse regions of Complex~C with the more metal 
abundant gas of the Galactic thick disk and halo
 would tend to increase the observed abundances in these regions, 
while the abundances in denser clumps of gas, as seen toward PG\,1259+593, 
should reflect more closely the original (pre-interaction) 
abundances in the cloud.

It is interesting that our best-fit value of D/H found for Complex~C is 
similar to that found for high-redshift damped Ly$\alpha$ systems (see
Pettini \& Bowen 2001 and Kirkman et al.\ 2003).  The \ion{H}{1} column 
density of N(\ion{H}{1}) $=9.0\times10^{19}$ cm$^{-2}$ 
is sufficient to produce damping wings on the \ion{H}{1} Ly$\alpha$ 
and Ly$\beta$ lines.  If observed in isolation from the Milky Way by
an outside observer, Complex~C  would produce  \ion{H}{1}
and \ion{D}{1} absorption similar
to that seen in these higher redshift systems.  Complex~C also has a low 
nitrogen abundance inferred
from the ratio of \ion{N}{1} to \ion{H}{1}, 
[N/H] $\approx -1.9$ toward PG\,1259+593 (Richter et al.\ 2001b;
Collins et al. 2003), about a factor of 10 lower than
[O/H].    (Initial work on the amount of \ion{N}{2} 
present indicates that photoionization 
of \ion{N}{1} in the \ion{H}{1} regions is not able to account for the 
deficit in the derived value of [N/H].) 
Commensurate values of [N/H] have been found for 
other Complex~C sight lines (Collins et al. 2003;
Tripp et al. 2003).
Some damped Ly$\alpha$ systems show similar deficits of nitrogen 
relative to oxygen, which several authors have explained as a nucleosynthetic
effect resulting from delayed release of nitrogen between massive
and intermediate mass stars or a top-heavy or truncated initial mass
function (Prochaska et al.\ 2002; Molaro 2003).  At the metallicity
of Complex~C, the low N/O ratio observed is probably more dependent 
upon the secondary nature of nitrogen production than these other processes
(see Figure~6 in Pettini et al. 2002).
The metallicity, abundance
pattern, D/H ratio, and lower distance limit all indicate that 
Complex~C is an external system (or the remains of an external system)
falling into with the Milky Way rather than
gas ejected from the Galactic disk.

\section{Future Progress}

Given the complex velocity structure of Galactic lines of sight, it may
be difficult to find other HVCs
for which an analysis similar to the
one presented here can be performed.  A similar conclusion was reached by 
Hoopes et al.\ (2003) for Galactic disk ISM along 
sight lines extending more than
$\sim1$ kpc.  We have inspected the FUSE spectra
of several other Complex~C sight lines, but none offer comparable 
prospects for a second determination of the D/H and D/O ratios in
Complex~C. Some brief comments on these sight lines are contained in
Table~11.  We are currently inspecting data for several other HVCs.

Reducing the error bars on the Complex~C D/H determination 
toward PG\,1259+593 is desirable.  A reduction in the error bars
from a $1\sigma$ uncertainty of $\sim30$\% to $\sim10\%$ would allow 
stronger statements to be made 
about the abundance of deuterium in this environment and the evolution of
deuterium as a function of metallicity.
It would be difficult to improve upon the current 
values for PG\,1259+593 with better FUSE data.  A large amount of additional 
observing time
($\sim2\times10^6$ seconds) 
would be needed to reduce the statistical errors by a factor of
$\sim2$.   This would improve the continuum
estimations and the \ion{O}{1} velocity model,  but it is unclear whether
 this would translate into a comparable reduction in the systematic errors
in the \ion{H}{1} and \ion{D}{1} models. 

A substantial improvement in the accuracy of the D/H determination in 
Complex~C could be made with higher spectral resolution far-ultraviolet
absorption-line data.
Even a factor of 1.5--2.0 increase in resolution to $R \sim 30,000-45,000$ 
(similar to that available with echelle spectrographs on large ground-based 
telescopes or with HST/STIS at ultraviolet wavelengths $>1150$\,\AA) would 
significantly enhance the discrimination between Complex~C 
\ion{D}{1} and \ion{H}{1} absorption in the Lyman-series transitions toward
PG\,1259+593.
This is currently beyond the capabilities of existing instrumentation, but 
perhaps someday it will be possible to 
observe PG\,1259+593 at higher spectral resolution in the far-ultraviolet
spectral region with a large space telescope equipped with an
efficient  high-resolution
spectrograph.  If and when that day comes, we recommend a re-analysis of 
the PG\,1259+593 sight line velocity structure and a redetermination of the 
range of D/H values allowed by the data.

\section{Summary}

We report the detection of \ion{D}{1} Lyman-series absorption in high 
velocity cloud Complex~C, a low-metallicity gas cloud falling onto the Milky 
Way.  This is the first detection of atomic deuterium in the local universe 
in a location other than the nearby regions of the Galactic disk.  The primary
results of our study are as follows:

\begin{enumerate}
\item{We combine data from the {\it Far Ultraviolet Spectroscopic Explorer},
{\it Hubble Space Telescope}, and Westerbork Synthesis Radio Telescope to 
derive the column densities of \ion{H}{1}, \ion{D}{1}, and \ion{O}{1}
in the direction of PG\,1259+593 ($l=120\fdg56, b = 58\fdg05$).}

\item{We use the FUSE and HST/STIS absorption-line data to construct a 
velocity model for the sight line based on the numerous \ion{O}{1}
absorption-lines detected.  We identify 8 absorption-line 
components common to all three species 
(\ion{H}{1}, \ion{D}{1}, and \ion{O}{1}) along the sight line.  
Two of these components, at 
$v_{\rm LSR}\approx -128$ and --112 \kms, are identified with Complex~C.
The remaining lower-velocity
components are identified with the Intermediate Velocity Arch, 
the Milky Way ISM, and a positive intermediate-velocity cloud in this
direction.}

\item{Our WSRT interferometer map of the \ion{H}{1} 21\,cm emission 
toward PG\,1259+593 indicates that the sight line passes through a 
compact concentration of neutral gas in Complex~C.  
We use the WSRT data having 
a resolution of $\sim1\arcmin$
together with single-dish data from the Effelsberg 100-meter
radio telescope to 
estimate the \ion{H}{1} column density in Complex~C and to constrain 
the velocity extents of the Complex~C \ion{H}{1} Lyman-series
absorption components observed by FUSE.}

\item{We use a combination of analysis techniques to estimate the column 
densities of \ion{H}{1}, \ion{D}{1}, and \ion{O}{1} in the absorbing regions
along the sight line.  We combine line-profile fitting analyses and 
curve-of-growth techniques with $\chi^2$ minimization codes to estimate
the most likely value and uncertainties on the column density of 
each species. The use of multiple analysis methods provides important
consistency checks on the results of the individual methods.  The good 
agreement found from independent analyses by members of our team confirm
that the formal errors are reliable representations of the uncertainties
involved in the analyses.}

\item{For each species, we describe the uncertainties involved in the 
estimation of the column density.  Systematic errors dominate the 
uncertainties, especially for \ion{D}{1}.  In estimating the
uncertainties on D/H, we have included the possibility that some of the 
absorption in the vicinity of the \ion{D}{1} lines might be caused 
by a weak, high-velocity \ion{H}{1} feature  near --190 \kms.}

\item{For the Complex~C absorption, we find 
N(\ion{H}{1}) = $(9.0\pm1.0)\times10^{19}$ cm$^{-2}$,
N(\ion{D}{1}) = $(2.0\pm0.6)\times10^{15}$ cm$^{-2}$, and 
N(\ion{O}{1}) = $(7.2\pm2.1)\times10^{15}$ cm$^{-2}$.  The uncertainties
quoted are $1\sigma$ errors (68\% confidence intervals);  two sigma error
estimates can be found in Table~9.
The corresponding light element abundance 
ratios are D/H = $(2.2\pm0.7)\times10^{-5}$, O/H = $(8.0\pm2.5)\times10^{-5}$,
and D/O = $0.28\pm0.12$.}

\item{ The metallicity of the Complex~C gas in this direction is approximately
1/6 solar, as inferred from the oxygen abundance of
[O/H] = $-0.79\pm^{0.12}_{0.16}$~($1\sigma$).  This metallicity is in 
excellent agreement with a recent high-quality 
determination of [O/H] in Complex~C toward 
3C\,351, which is located $\sim30\degr$ from PG\,1259+593.  Two other
Complex~C sight lines analyzed by Collins et al. (2003) yield similar
metallicities (Mrk~279: [O/H] = $-0.71\pm^{0.36}_{0.25}$;  Mrk~817:
[O/H = $-0.59\pm^{0.25}_{0.17}$). }

\item{While we cannot rule out a value of D/H for Complex~C similar to that 
found for the local ISM (i.e., D/H~$\sim1.5\times10^{-5}$), we can 
reasonably exclude values as low as those 
determined recently for extended sight lines in the Galactic disk 
(D/H~$< 1\times10^{-5}$), provided that the absorption observed is really
\ion{D}{1} and not very high velocity \ion{H}{1}.  This conclusion lends 
support to the hypothesis that Complex~C is a manifestation of gas 
originating outside the Milky Way, rather than material recirculated between 
the Galactic disk and halo.}

\item{A value of 
D/H $\approx 2.2\times10^{-5}$ for Complex~C is consistent with 
the primordial abundance of deuterium inferred from recent WMAP observations
of the cosmic microwave background, combined with simple chemical evolution
models that predict the amount of deuterium astration with time.  Reducing
the errors on the D/H estimate in Complex~C toward PG\,1259+593 will be 
difficult, but may eventually be possible with a large space telescope
equipped with a high-resolution ultraviolet spectrograph.}

\item{The number of sight lines for which \ion{D}{1} can be observed in 
Lyman-series absorption beyond a few hundred parsecs of the Sun appears
to be small because of \ion{H}{1} 
velocity crowding. We considered the data for several other sight lines 
through Complex~C and found that they are unsuitable for 
a reliable \ion{D}{1} column density determination.  We continue to 
examine FUSE data for \ion{D}{1} absorption in other HVCs.}

\end{enumerate}

\noindent
{\bf Acknowledgments.}
We thank Van Dixon for answering numerous questions about the {\tt CALFUSE}
pipeline software, and we thank Max Pettini and Jeff Linsky for useful
comments on the manuscript. This work is 
based on data obtained for the Guaranteed Time Team by the
NASA-CNES-CSA FUSE mission operated by the Johns Hopkins University.
We acknowledge the many contributions of the members of the FUSE D/H 
working group and the tireless efforts of the FUSE Operations Team.  
This work is also based in part upon observations made with the NASA/ESA 
Hubble Space Telescope, obtained at the Space Telescope Science Institute, 
which is operated by the Association of Universities for Research in 
Astronomy, Inc., under NASA contract NAS5-26555. 
The Westerbork Synthesis Radio Telescope is operated by the ASTRON (Netherlands
Foundation for Research in Astronomy) with support from the Netherlands
Foundation for Scientific Research (NWO).  The Effelsberg Telescope belongs to 
the Max Planck Institute for Radio Astronomy in Bonn, Germany. 
We are indebted to Robert 
Braun for assistance with the Westerbork observations.
Partial financial support has been provided by NASA contract NAS5-32985
and Long Term Space Astrophysics grants NAG5-3485 (K.R.S.) and NAG5-7262
(J.M.S.). B.P.W.\ was supported by NASA grants NAG5-9179, NAG5-9024, and 
NAG5-8967.  T.M.T.\ appreciates support for this work from NASA Long Term 
Space Astrophysics grant NAG5-11136 as well as NASA grant GO-08695.01-A
from the Space Telescope Science Institute.  P.R.\ was supported
by the Deutsche Forschungsgemeinschaft.
French co-authors were supported by CNES.  G.H.\ and 
R.F.\ acknowledge use of the {\tt Owens.f} software developed by Martin Lemoine
and the French FUSE Team.

\clearpage
\newpage
\begin{figure}[ht!]
\includegraphics{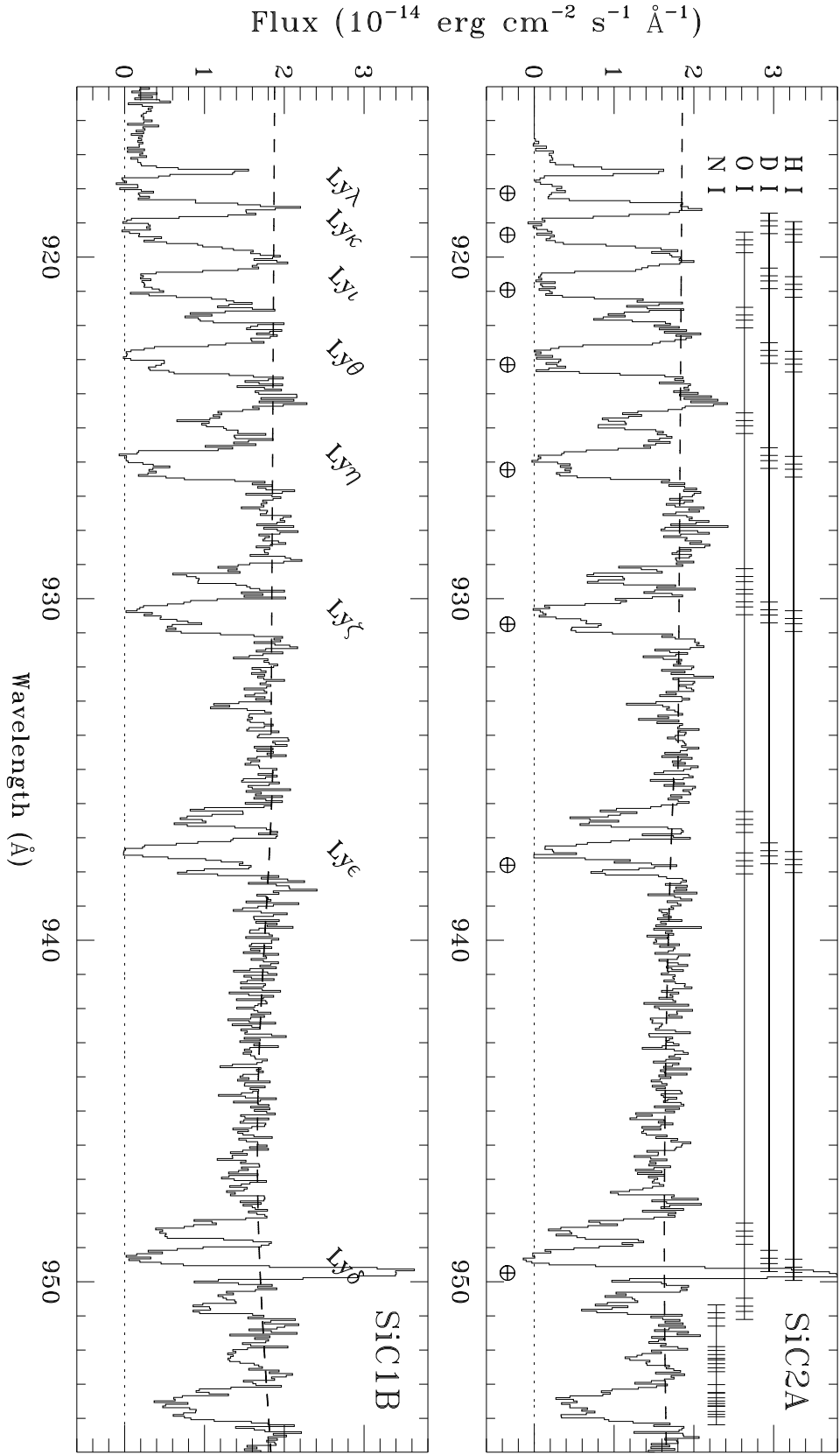}
\vspace{8.5in}

\caption{See caption on next page.}

\end{figure}

\clearpage
\newpage
\noindent
Fig. 1.---
Fully reduced FUSE spectra of PG\,1259+593 between 915 and 955\,\AA.
The two night-only 
datasets shown were obtained with the SiC1 and SiC2 channels.  The 
spectra are binned by 8 pixels (slightly less than a resolution element)
for illustration purposes only.
The locations of prominent \ion{H}{1}, \ion{D}{1}, \ion{N}{1}, and 
\ion{O}{1} lines are indicated at the top of the top panel.  Four components 
corresponding to absorption groups at --128 \kms\ (Complex~C), 
--54 \kms\ (IV~Arch), --5 \kms\ (ISM), and +69 \kms\ (Positive IVC)  
are shown 
for each transition.
The \ion{H}{1} Lyman-series lines are identified in the lower panel.
The heavy dashed lines indicate the continua adopted in our analysis of 
these
spectra.  The dotted line at the bottom of each panel marks the zero flux
level.
The wavelengths of terrestrial \ion{H}{1} airglow features in these 
night-only observations are denoted by crossed circles.  Note that 
the strong cores of the \ion{H}{1} lines reach zero residual intensity 
at wavelengths that are unaffected by these airglow features. The source
of the slight rise in the SiC1 spectrum below 917\,\AA\ is probably 
residual airglow at these wavelengths.  

\clearpage
\newpage
\begin{figure}[ht!]
\includegraphics{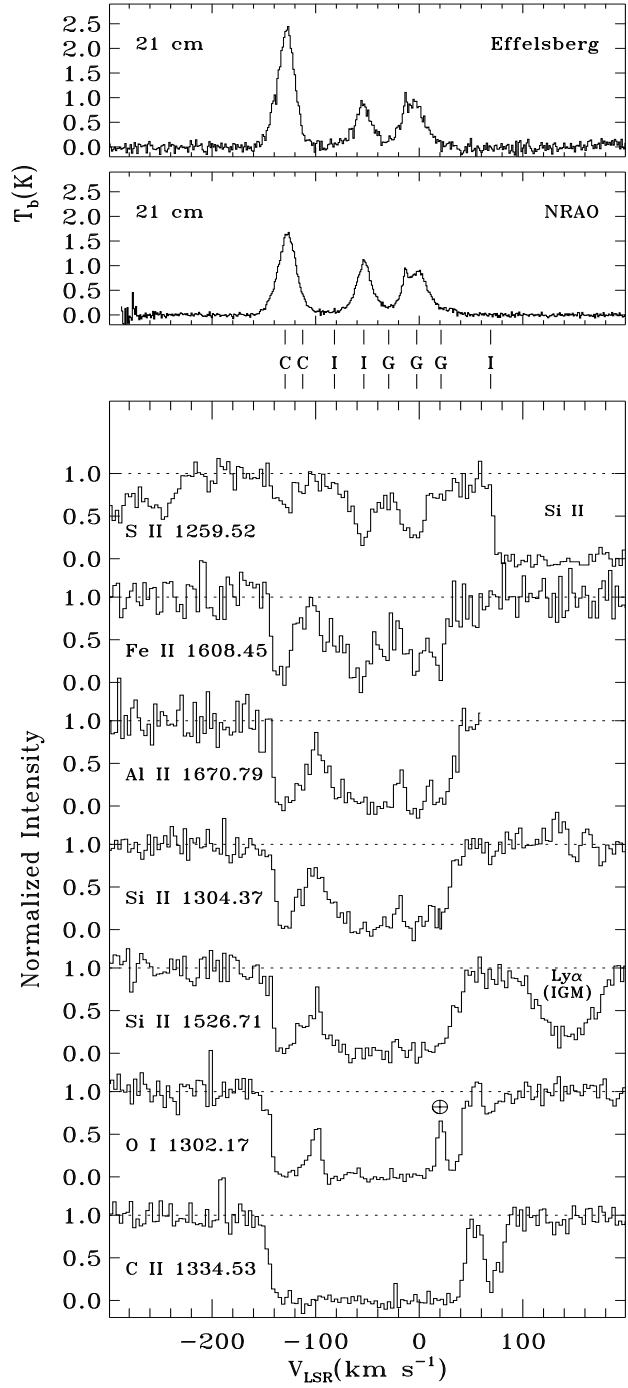}
\vspace{7.2in}
\caption{See caption on next page.}
\end{figure}

\clearpage
\newpage

\noindent
Fig. 2.---
{\it Top:} Brightness temperature versus LSR velocity for the 
\ion{H}{1} 21\,cm emission observed toward PG\,1259+593 with the Effelsberg
100-meter telescope and NRAO 140-foot telescope.  
The Effelsberg data have a beam size
of 9\farcm7 and a velocity resolution of 1.8 \kms\ after Hanning smoothing.  
The NRAO data have a beam
size of 21\arcmin\ and a velocity resolution of 1 \kms.
{\it Bottom:} Ultraviolet absorption-line profiles of various interstellar
species observed with HST/STIS.  The data have a resolution of $\sim7$ \kms\
(FWHM).  The velocities of the 8 absorption components along the 
sight line are indicated
above the top spectrum and are labeled C (Complex~C), G (Galactic ISM),
or I (intermediate velocity ISM). Note the multiple component nature of the 
Complex~C absorption between --150 and --100 \kms.  The weak 
intermediate-velocity feature at $\approx+69$ \kms\ is visible in the strong
\ion{O}{1} and \ion{C}{2} lines shown; it is also visible in the \ion{H}{1}
lines shown in Figure~1.  The \ion{Al}{2} line falls near the edge of a STIS
echelle order and extends only to +60 \kms. The broad absorption feature 
between +100 and +200 \kms\ next to \ion{Si}{2} $\lambda1526.707$ is an
intergalactic Ly$\alpha$ absorber at redshift $z = 0.2564$.

\clearpage
\newpage
\begin{figure}[ht!]
\includegraphics{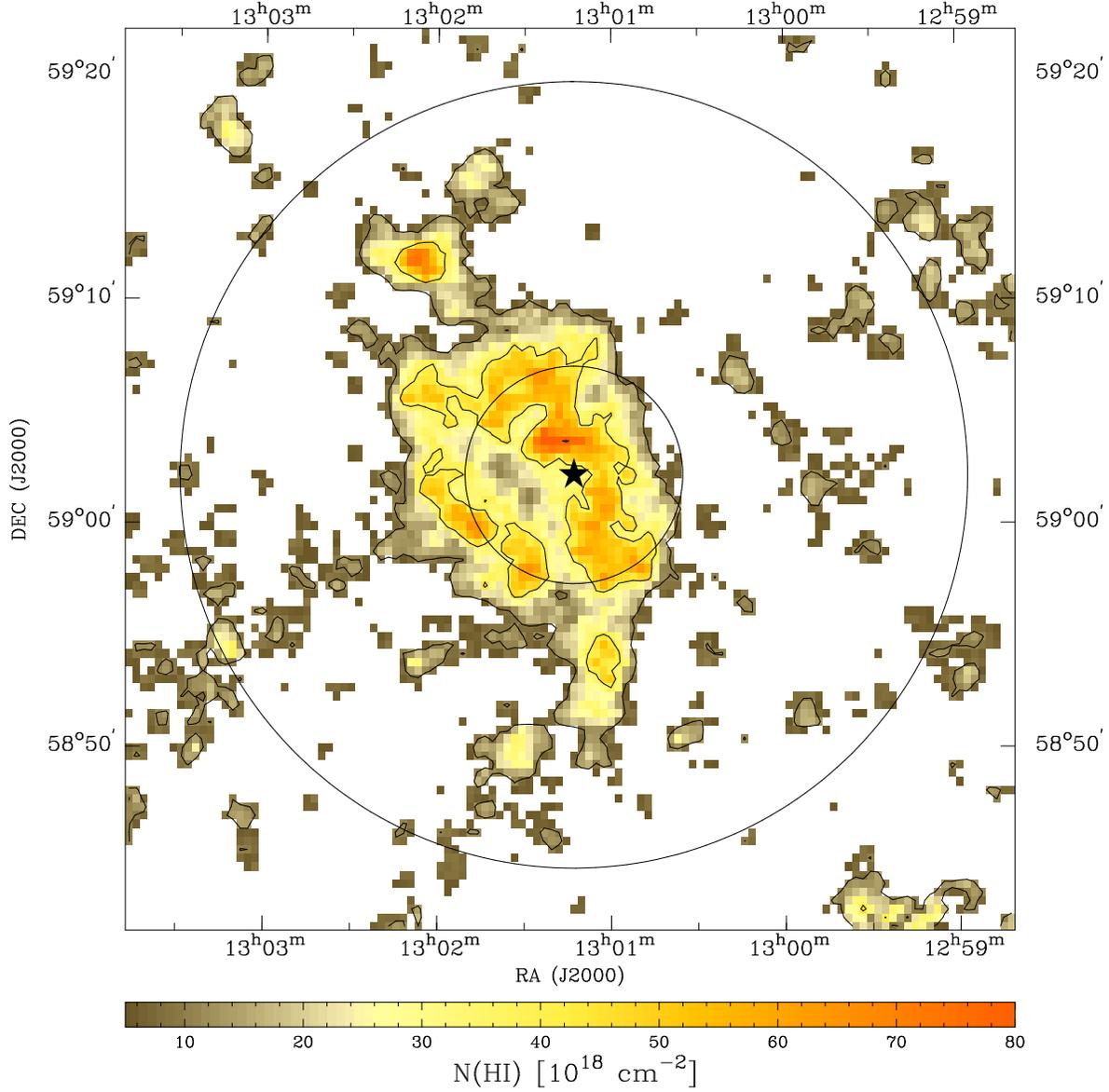}
\vspace{6.3in}
\caption{Filtered \ion{H}{1} column density of the WSRT data, integrated over 
the velocity range $-148$ to $-109$ \kms, which includes core CIII.  There
is a slowly varying offset of $(3.0-6.0)\times10^{19}$ cm$^{-2}$ across the 
field that is not included in this estimate because it has 
been filtered out by the interferometer.  At the position of 
PG\,1259+593, marked by the star at $\alpha$=13$^h$01$^m$13$^s$,
$\delta$=59$^\circ$02$^\prime$06$^{\prime\prime}$, 
this offset is $6\times10^{19}$ cm$^{-2}$.
The circles centered on PG\,1259+593 
show the FWHM of the Effelsberg and Westerbork primary beams.}
\end{figure}

\clearpage
\newpage
\begin{figure}[ht!]
\includegraphics{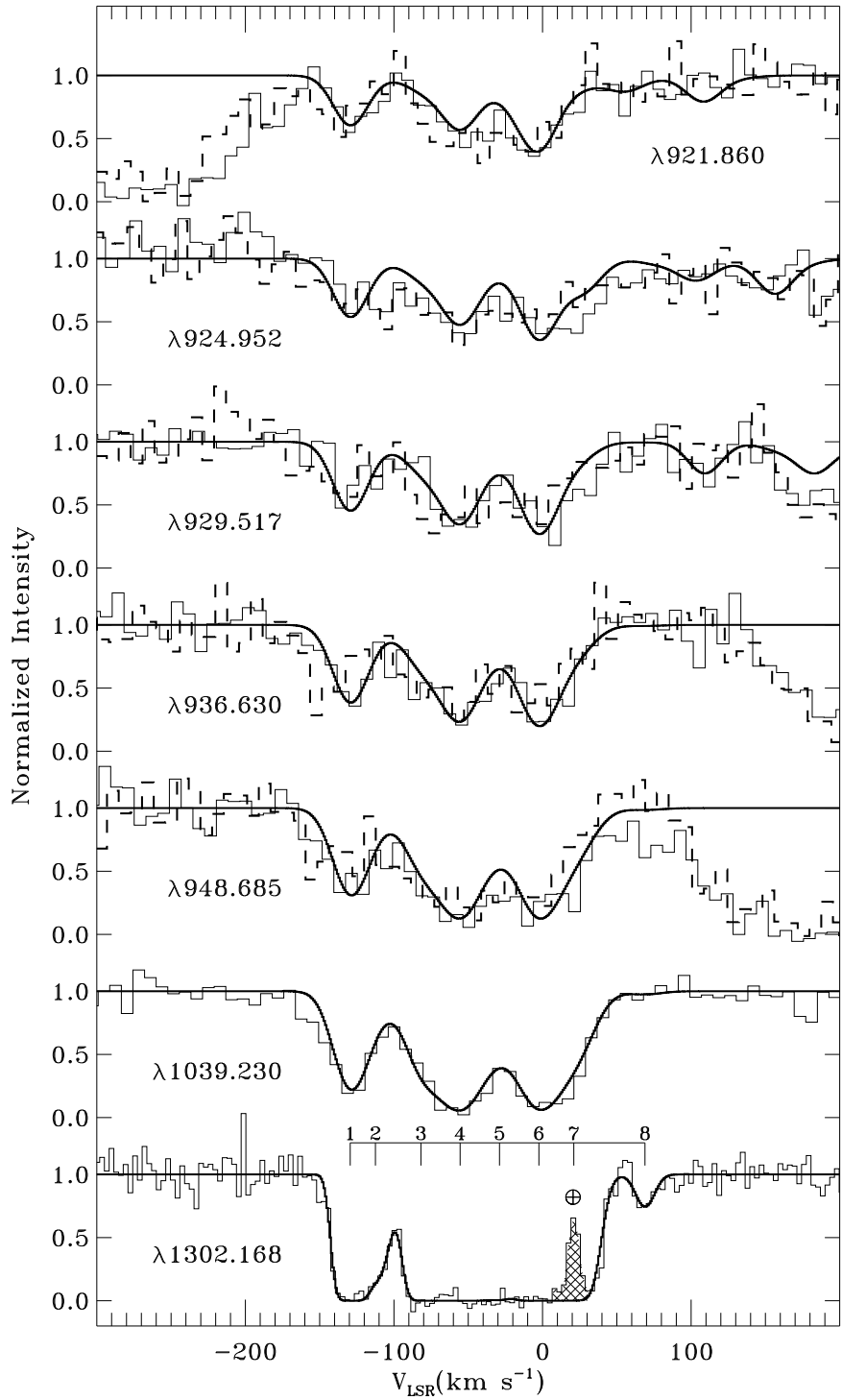}
\vspace{7.7in}
\caption{See caption on next page.}
\end{figure}

\clearpage
\newpage
\noindent
Fig. 4.--- A comparison of \ion{O}{1} lines toward PG\,1259+593
observed by 
FUSE and HST/STIS to our model for the \ion{O}{1} velocity distribution
along the sight line. 
The data are shown as histogrammed lines.  For 
$\lambda < 1000$\,\AA, the FUSE SiC2 data are shown as solid lines, and the 
SiC1 data are shown as dashed lines.  The \ion{O}{1} $\lambda1039.230$ 
line is from the FUSE LiF1 channel.  
In all cases, the model results are 
shown as smooth, heavy lines.  
Vertical tick marks above the STIS \ion{O}{1} $\lambda1302.168$ 
spectrum denote the velocities of the 8 components in the 
model. The deep broad absorption features near some of the \ion{O}{1} lines
are interstellar \ion{H}{1} Lyman-series lines.
The cross-hatched emission feature in the  \ion{O}{1}
$\lambda1302.168$ spectrum is terrestrial \ion{O}{1} airglow entering the 
STIS slit during the observation.  The FUSE profiles shown do not 
contain such features since we show only orbital night data for which the 
terrestrial \ion{O}{1} emission is negligible at these wavelengths.

\clearpage
\newpage
\begin{figure}[ht!]
\includegraphics{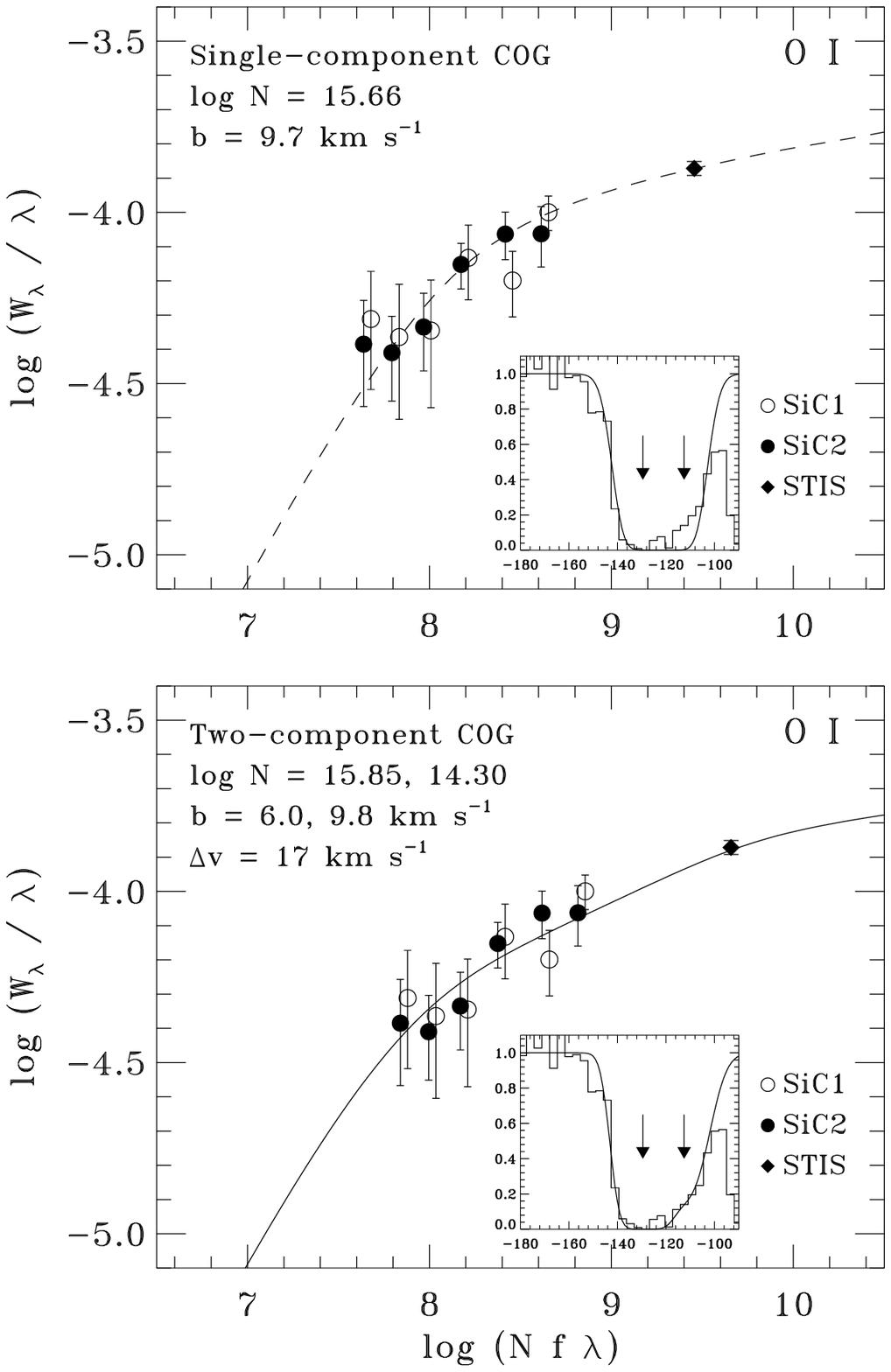}
\vspace{6.5in}
\caption{Curve of growth results for the \ion{O}{1} absorption in Complex~C
between $v_{\rm LSR} = -180$ \kms\ and $v_{\rm LSR} = -105$ \kms.  The 
x-axis has units of logarithmic inverse centimeters.  Filled 
circles are measurements based on data from the FUSE SiC2 
channel.  Open circles are measurements based on data from the 
FUSE SiC1 channel.  
The points for the 
two channels have been offset slightly (plus or minus 0.02 dex) 
from their nominal values
in the horizontal direction for 
clarity.  The filled diamond indicates the HST/STIS E140M measurement for the 
\ion{O}{1} $\lambda1302.168$ line.  Error bars are 
1$\sigma$ estimates. 
{\it Top}: Single-component COG fit with corresponding profile fit to the 
\ion{O}{1} $\lambda1302.168$ line (inset).  The two arrows in the inset panel
indicate the velocities (in \kms) of the two Complex~C components.
{\it Bottom}: Double-component COG fit based on the fitting parameters used
to construct the profiles shown in Figure~4 with corresponding profile fit to 
the \ion{O}{1} $\lambda1302.168$ line (inset). The two arrows in the inset 
panel indicate the velocities of the two Complex~C components.}
\end{figure}

\clearpage
\newpage
\begin{figure}[ht!]
\includegraphics{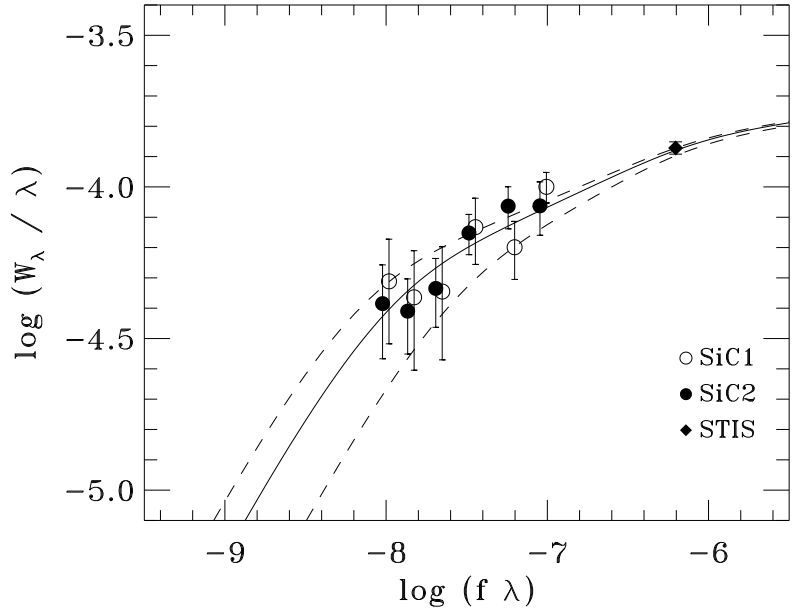}
\vspace{5.2in}
\caption{Double-component COG fit to the \ion{O}{1} lines 
based on the fitting parameters used
to construct the synthetic 
profiles shown in Figure~4.  The x-axis has units of 
logarithmic centimeters.  The symbols and best fit (solid)
curve are the same as those shown in the bottom panel of Figure~5.  The
dashed curves correspond to the $\pm2\sigma$ errors ($\pm4.2\times10^{15}$ cm$^{-2}$)
on the \ion{O}{1} column 
density.}
\end{figure}

\clearpage
\newpage
\begin{figure}[ht!]
\includegraphics{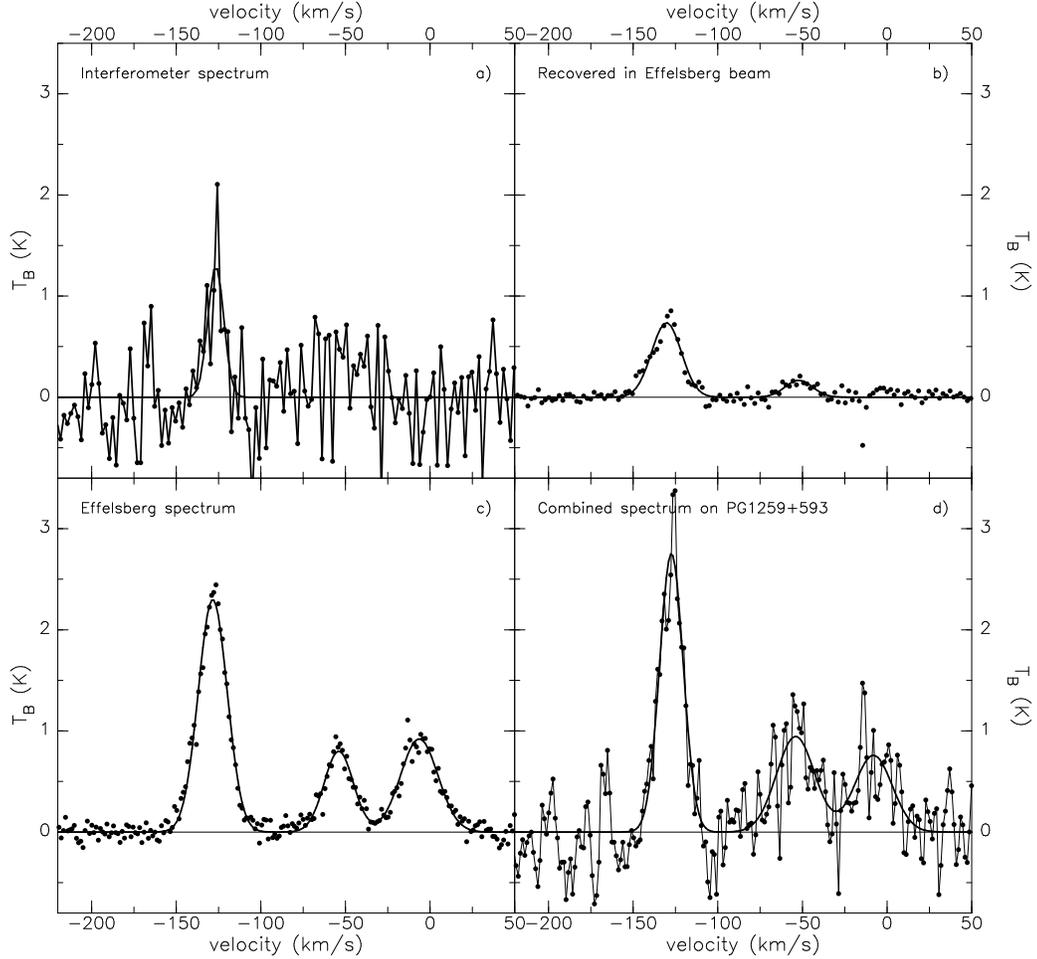}
\vspace{6.0in}
\caption{
These four panels show the derivation of the final
value for the \ion{H}{1} column density in the direction of PG\,1259+593. 
 For all spectra shown in the four panels, Gaussian fits 
are superimposed on the data points.
{\it a)}~WSRT-only data. {\it b)}~WSRT data ``observed'' 
by a 9\farcm7 beam.  The WSRT channel maps were multiplied by a Gaussian with 
FWHM = 9\farcm7, integrated, and converted back to brightness temperature. 
{\it c)}~Single-dish
spectrum taken with the Effelsberg 100-meter 
telescope. This clearly shows the three
components: Complex~C at $-128$ \kms, the IV-Arch at $\sim -54$ \kms, and the 
Milky Way ISM at $\sim-5$ \kms. {\it d)}~Final combined spectrum. }
\end{figure}

\clearpage
\newpage
\begin{figure}[ht!]
\includegraphics{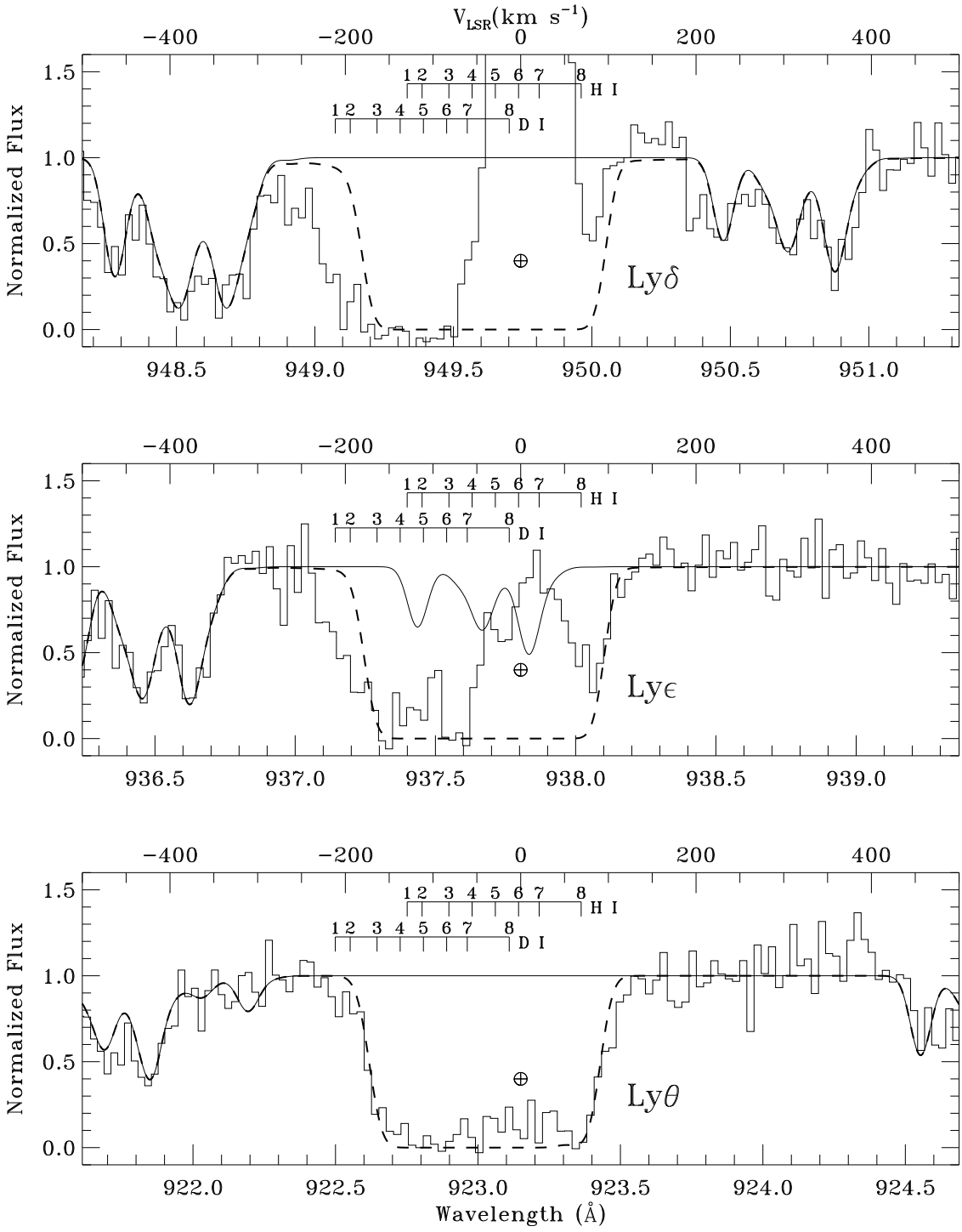}
\vspace{8.0in}
\caption{See caption on next page.}
\end{figure}
\clearpage
\newpage
\noindent
Fig. 8.--- Adopted model results for the \ion{H}{1}
Ly$\delta$, Ly$\epsilon$, and 
Ly$\theta$ lines, assuming (D/H)$_{\rm Complex~C}$ = 0.0.
The FUSE SiC2 data are shown as thin histogrammed lines.  The \ion{O}{1}
absorption in the vicinity of the \ion{H}{1} lines is represented by the 
smooth solid lines.  The total (\ion{O}{1} + \ion{H}{1} + \ion{D}{1}) 
absorption model
is shown as thick dashed curves.  Each panel spans a wavelength
range corresponding to a velocity range of $\pm500$ \kms\ centered on the 
\ion{H}{1} rest velocity, as indicated by the velocity scales 
above each panel.
The locations of the \ion{H}{1} and \ion{D}{1} components are denoted
by the vertical tick marks above each spectrum.  Terrestrial \ion{H}{1}
airglow occurs between --100 and +100 \kms\ in each panel and 
decreases in strength with increasingly higher lying 
lines in the Lyman series. 
The model reproduces
the observed absorption well, except at the velocities of the Complex~C
\ion{D}{1} components  ($\#1$ and $\#2$), for which N(\ion{D}{1}) was 
set to zero in this figure.  (Note the extra absorption in the 
blue wings of the observed absorption profiles.)
In all other components, D/H has been set to $1.5\times10^{-5}$.  
The source of the non-zero flux near 937.4\,\AA\ in the Ly$\epsilon$
profile is unknown.  It does not appear in the SiC1 data.  The steep
absorption walls at negative velocities in the higher order \ion{H}{1} 
Lyman-series transitions constrain the b-values of the Complex~C \ion{H}{1}
components (see text for details).

\clearpage
\newpage
\begin{figure}[ht!]
\includegraphics{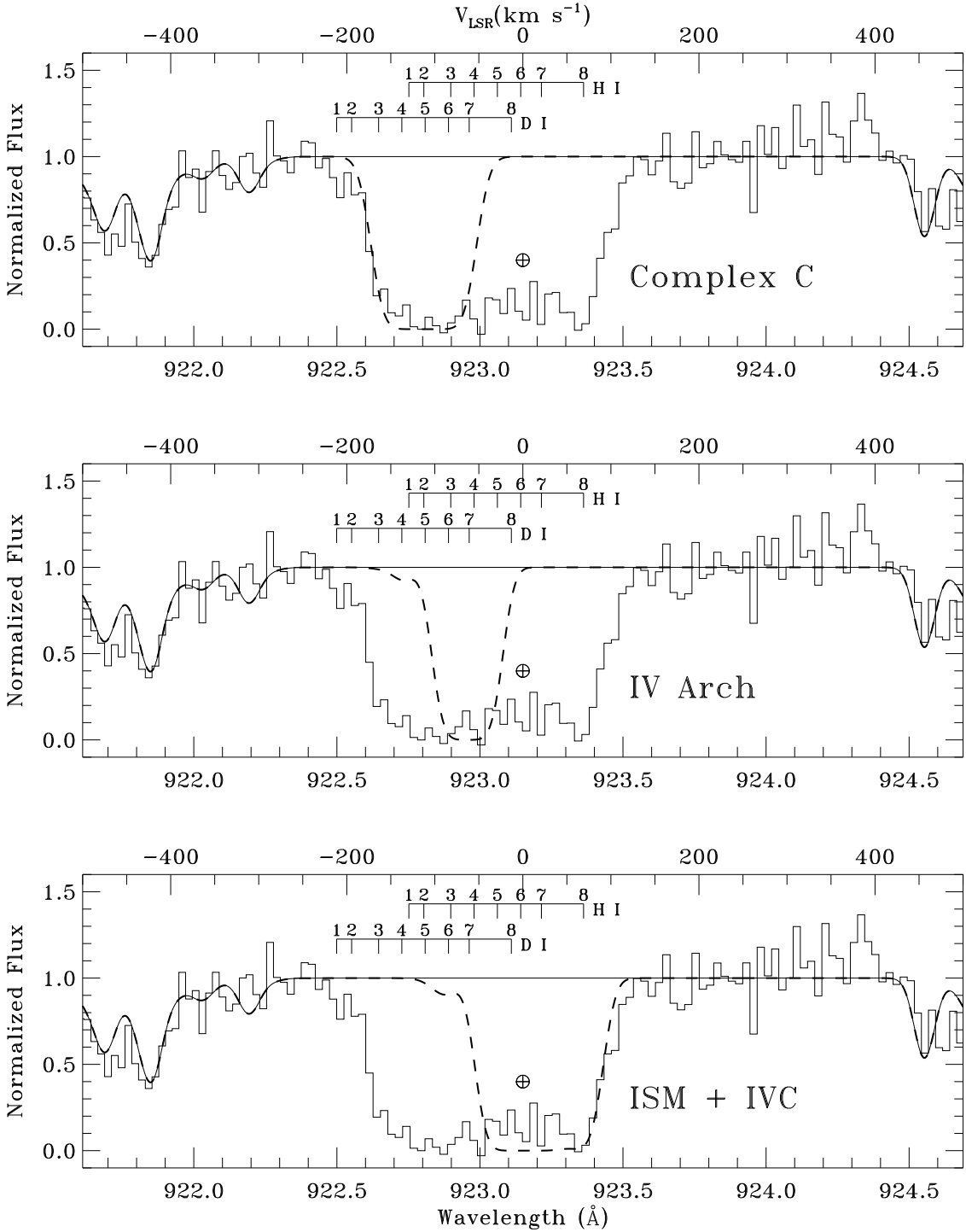}
\vspace{7.7in}
\caption{A breakdown of the component structure in the Ly$\theta$ line.  
The model results are shown as dashed lines. For Complex~C, D/H has been
set to zero. The smooth solid line indicates the fit to the
\ion{O}{1} absorption along the sight line.
{\it Top:} \ion{H}{1} absorption produced by Complex~C (components 1 and 2).
{\it Middle:} \ion{H}{1} and \ion{D}{1} 
absorption produced by the IV~Arch (components 3 and 4).
{\it Bottom:} \ion{H}{1} and \ion{D}{1} 
absorption produced by the Galactic ISM and positive velocity 
IVC (components $5-8$).  Note that the IV~Arch and Galactic ISM 
components contribute no absorption at the velocity of the \ion{D}{1}
lines in Complex~C.
  }
\end{figure}

\clearpage
\newpage
\begin{figure}[ht!]
\includegraphics{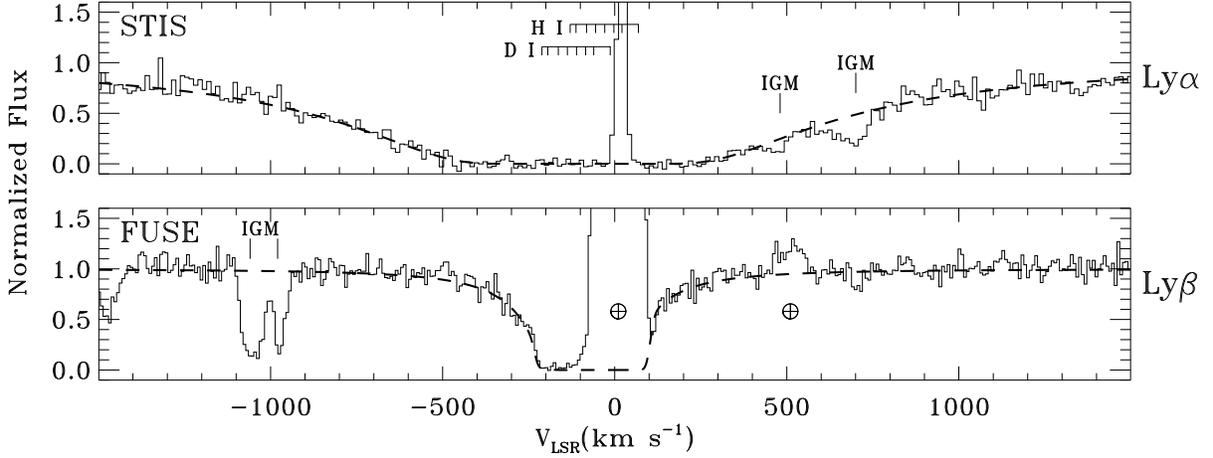}
\vspace{3.5in}
\caption{Model fits (dashed lines) to the \ion{H}{1} and \ion{D}{1}
Ly$\alpha$ (STIS) and Ly$\beta$ (FUSE) absorption 
lines.  The Ly$\beta$ profile shown is night-only data obtained with 
the FUSE LiF1 channel.
The velocities of the components in the model are indicated above
the Ly$\alpha$ spectrum.  
The \ion{H}{1} absorption is so strong that radiation damping
wings are visible in both lines; the \ion{D}{1} absorption is overwhelmed 
by the \ion{H}{1} absorption.  These lines were used as a consistency check
on the validity of the adopted \ion{H}{1} model for the sight line.
The absorption lines marked ``IGM'' are due to intervening intergalactic 
clouds along the sight line (see Richter et al. 2003). The emission 
feature near +500 \kms\ in the Ly$\beta$ spectrum is residual \ion{O}{1}
airglow in these night-only data; the emission from these lines is much 
stronger than the emission in the \ion{O}{1} lines used in our study 
of the \ion{O}{1} absorption along the sight line. The widths of the airglow
lines in the FUSE spectrum, $\Delta$v $\approx 100$ \kms, are broader than 
in the STIS spectrum because of the larger aperture used for the 
FUSE observation.}
\end{figure}

\clearpage
\newpage
\begin{figure}[ht!]
\includegraphics{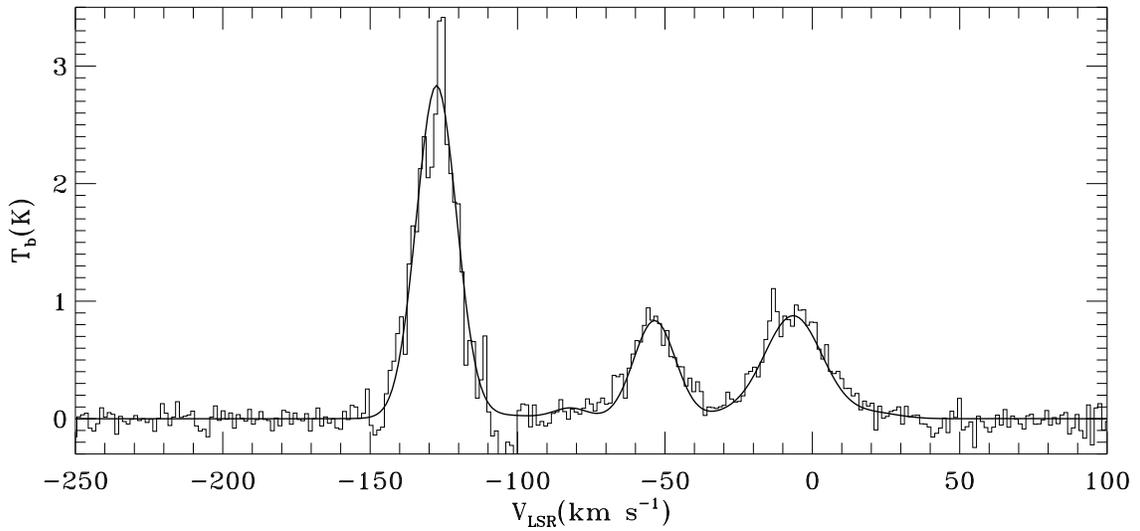}
\vspace{5.0in}
\caption{A comparison of the observed \ion{H}{1} 21\,cm emission in
the direction of PG\,1259+593 (thin histogrammed line) and the adopted 
\ion{H}{1} model described in the text (heavy smooth line).  
At velocities
$-150 < v_{LSR} < -100$ \kms\ (i.e., Complex~C velocities), 
the data are the interferometric
measurements described in \S2.3. At velocities outside this range, the
data are single-dish (9\farcm7 beam) Effelsberg measurements.  The model
has been scaled to brightness temperature with the assumption that the 
emission is optically thin.  }

\end{figure}

\clearpage
\newpage
\begin{figure}[ht!]
\includegraphics{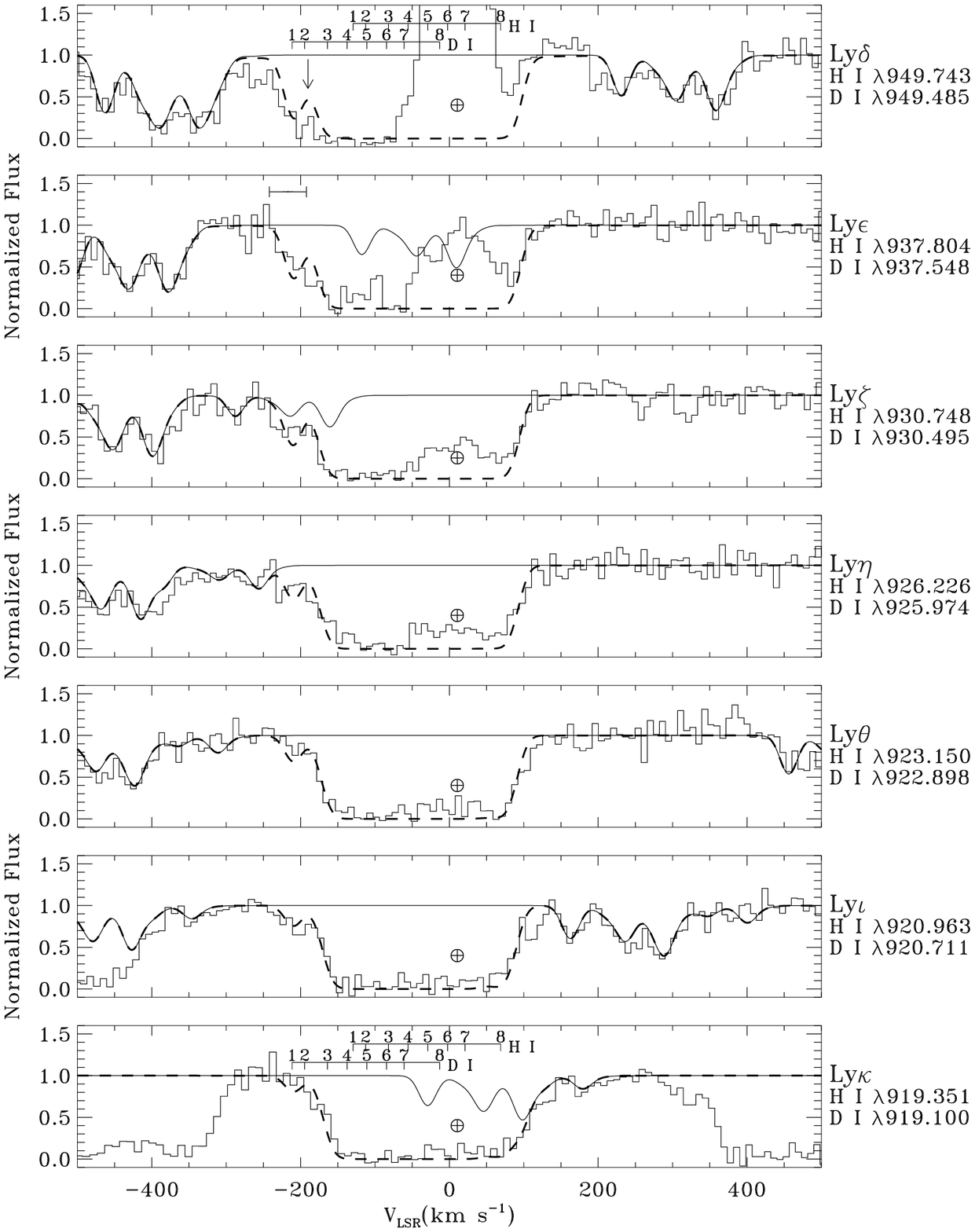}
\vspace{8.0in}
\caption{See caption on next page.}
\end{figure}

\clearpage
\newpage
\noindent
Fig.~12.--- Adopted model results for the \ion{H}{1}, \ion{D}{1}, and \ion{O}{1}
absorption along the sight line assuming (D/H)$_{\rm Complex~C} = 2.6\times10^{-5}$.  
The absorption features produced by the 
Lyman-series lines of Ly$\delta$--Ly$\kappa$ in the FUSE SiC2 channel
are shown as histogrammed lines.
The total (\ion{O}{1} + \ion{H}{1} + \ion{D}{1}) 
absorption model is shown as thick dashed curves.
The smooth solid line isolates the contribution to the fit produced by
\ion{O}{1} absorption along the sight line. The velocity scale applies to 
the \ion{H}{1} lines.  The velocities of the 8 \ion{H}{1} and \ion{D}{1} 
components discussed in the text are indicated above the Ly$\delta$
and Ly$\kappa$ lines.  The downward pointing arrow in the top panel
indicates the position of residual
absorption that is not reproduced by the model (see \S7.1). The horizontal
error bar above the Ly$\epsilon$ absorption in the second panel indicates
the absorption region used to calculate the equivalent widths in the \ion{D}{1}
curve-of-growth analyses discussed in the text.

\clearpage
\newpage
\begin{figure}[ht!]
\includegraphics{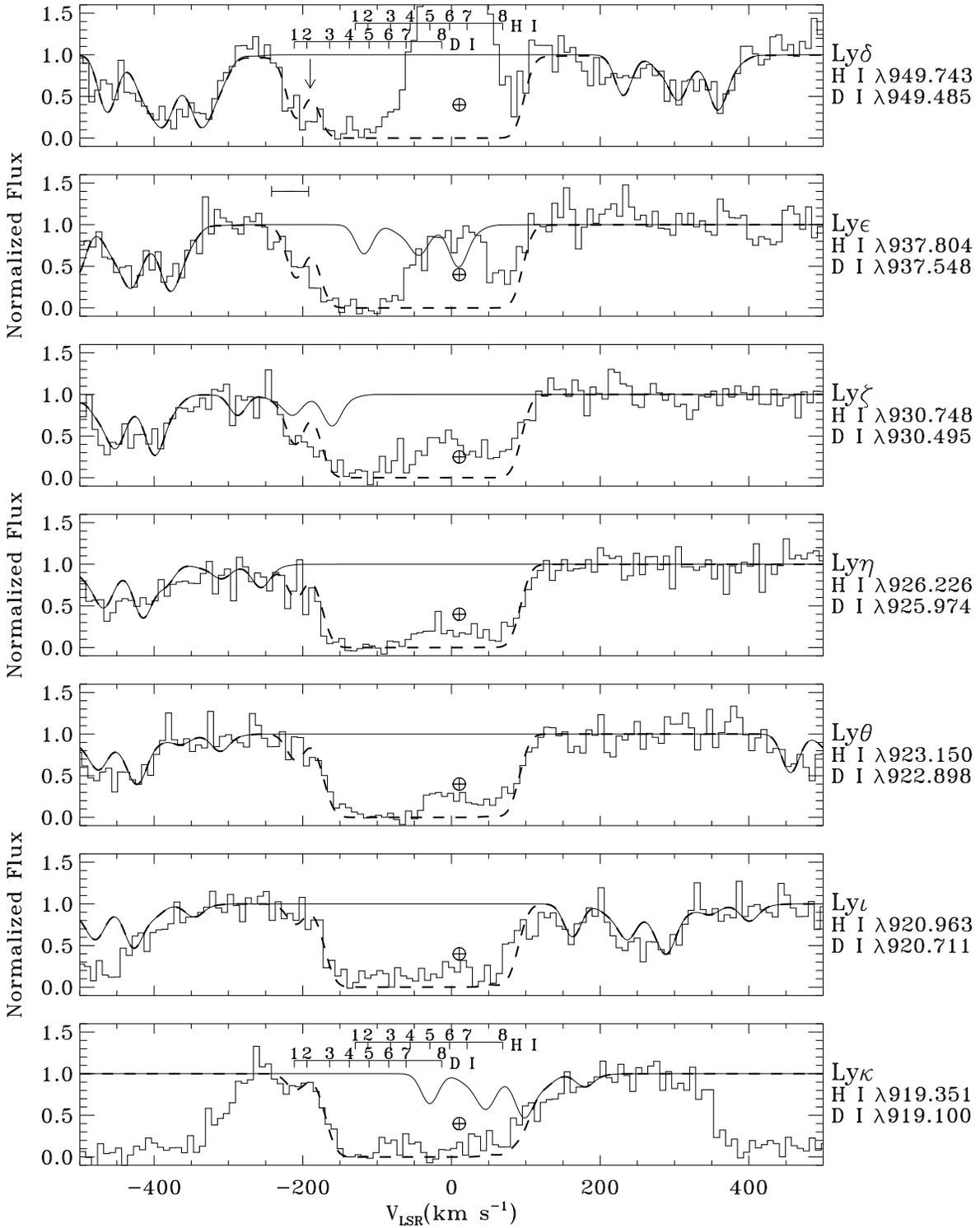}
\vspace{8.0in}
\caption{Same as Figure~12, except for data from the FUSE SiC1 channel. }
\end{figure}

\clearpage
\newpage
\begin{figure}[ht!]
\includegraphics{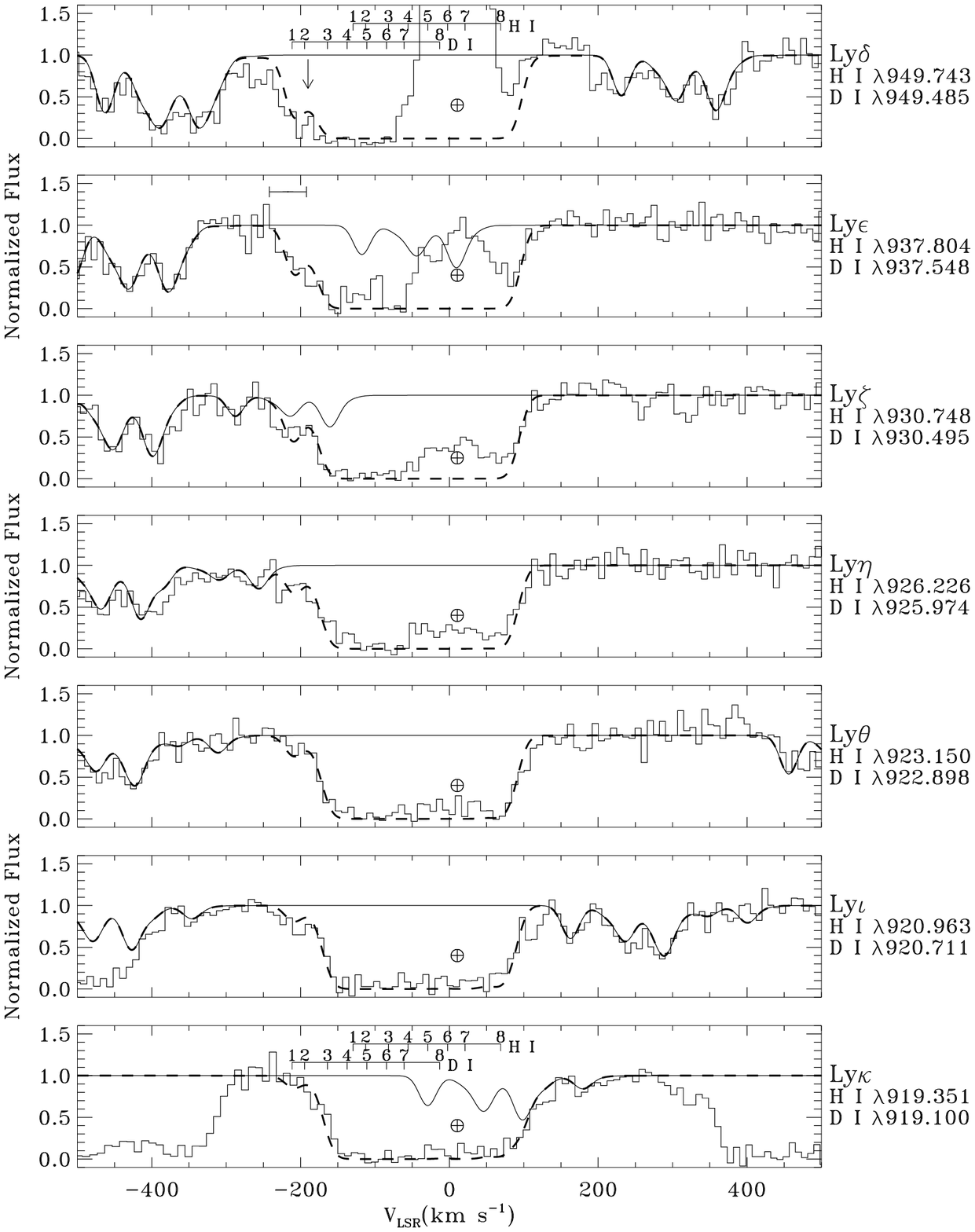}
\vspace{8.0in}
\caption{See caption on next page.}
\end{figure}

\clearpage
\newpage
\noindent
Fig.~14.--- Adopted model results for the \ion{H}{1}, \ion{D}{1}, and \ion{O}{1}
absorption along the sight line after incorporation of a weak \ion{H}{1}
absorption feature at --190 \kms.  This model has (D/H)$_{\rm Complex~C} = 1.8\times10^{-5}$.  
The absorption features produced by the 
Lyman-series lines of Ly$\delta$--Ly$\kappa$ in the FUSE SiC2 channel
are shown as histogrammed lines.
The total (\ion{O}{1} + \ion{H}{1} + \ion{D}{1}) 
absorption model is shown as thick dashed curves.
The smooth solid line isolates the contribution to the fit produced by
\ion{O}{1} absorption along the sight line. The velocity scale applies to 
the \ion{H}{1} lines.  The velocities of the 8 \ion{H}{1} and \ion{D}{1} 
components discussed in the text are indicated above the Ly$\delta$
and Ly$\kappa$ lines.  The downward pointing arrow in the top panel 
indicates the position of the residual
absorption modeled by the inclusion of a weak high-velocity \ion{H}{1} 
feature.

\clearpage
\newpage
\begin{figure}[ht!]
\includegraphics{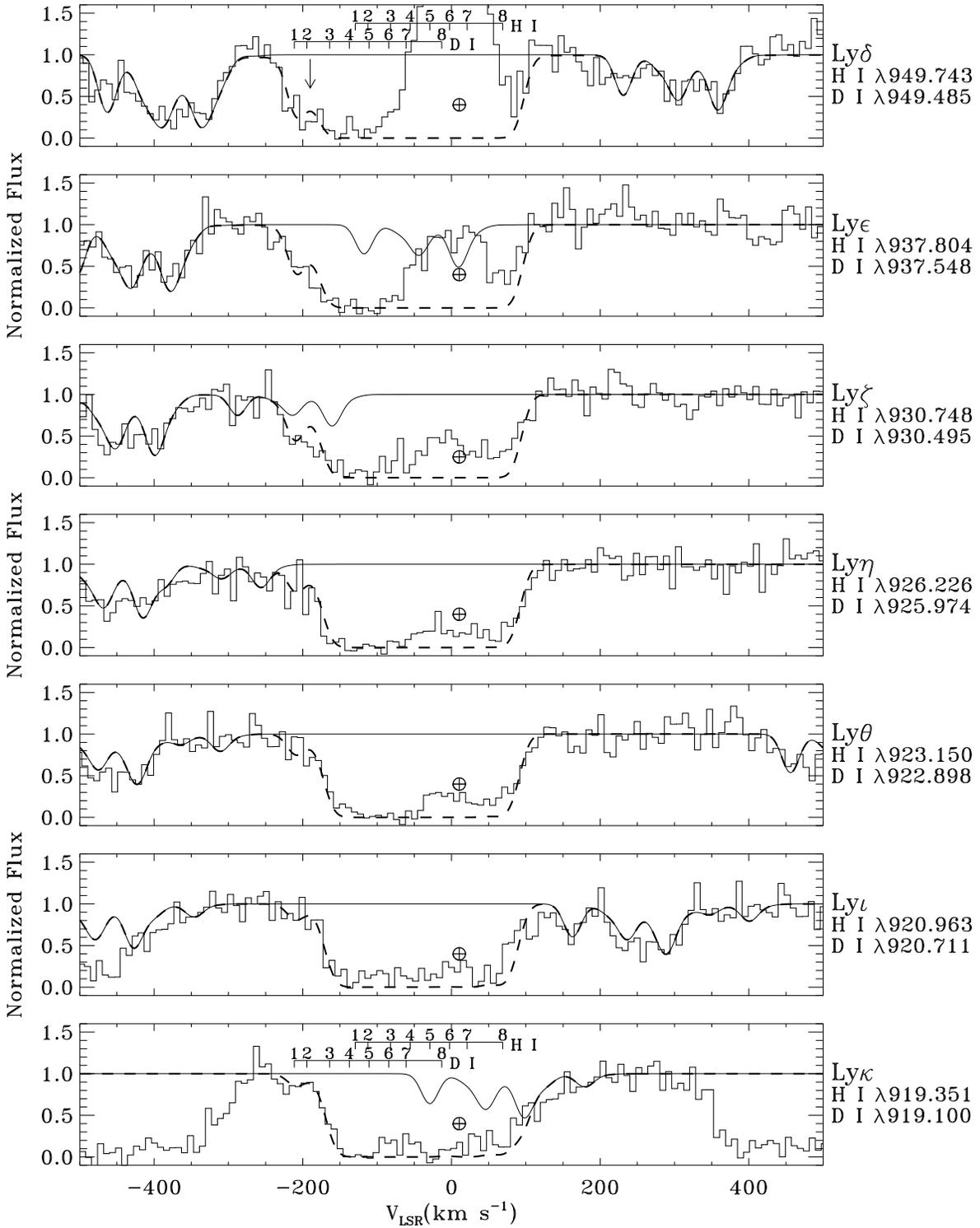}
\vspace{8.0in}
\caption{Same as Figure~14, except for data from the FUSE SiC1 channel. }
\end{figure}

\clearpage
\newpage
\begin{figure}[ht!]
\includegraphics{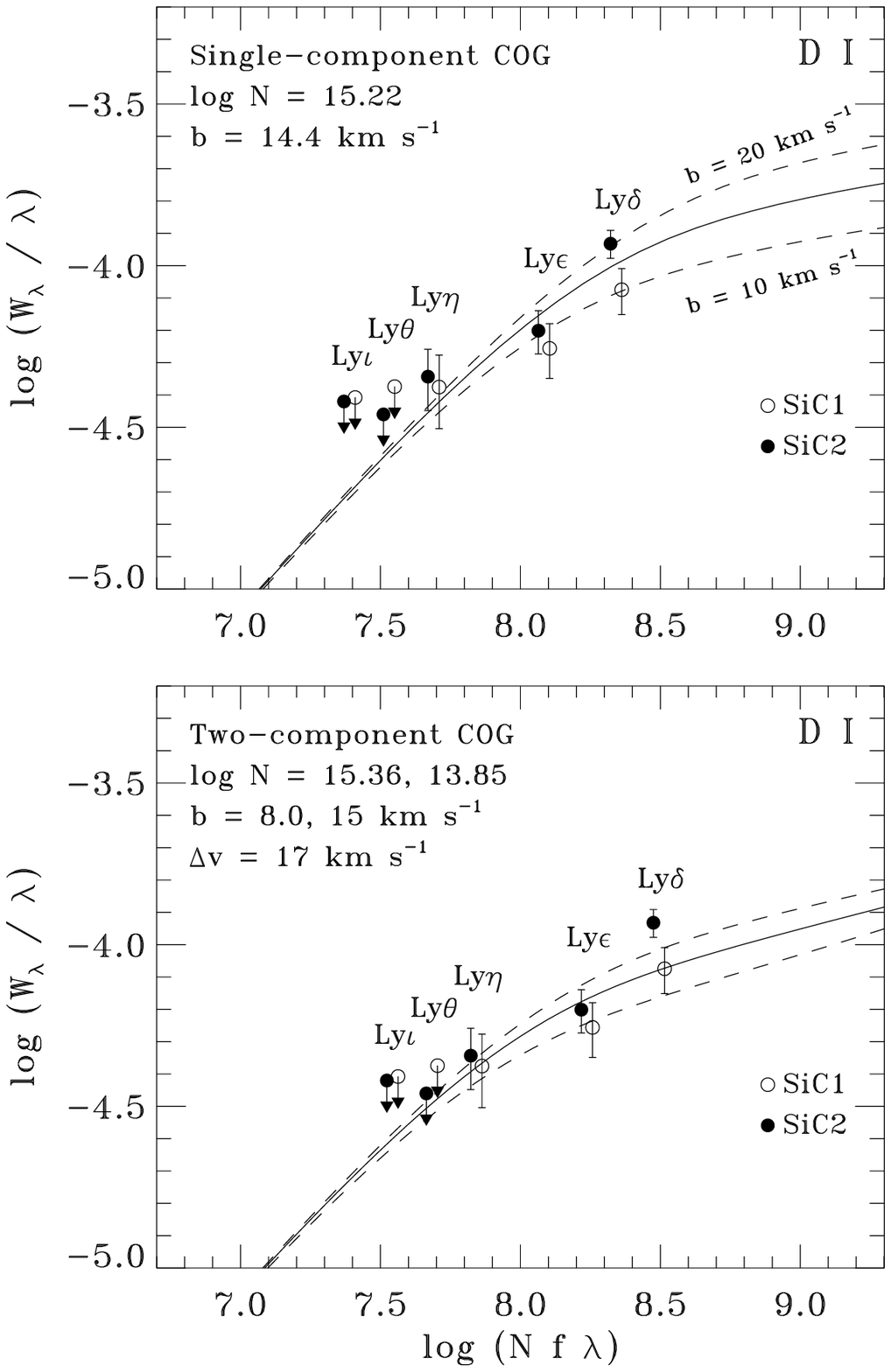}
\vspace{6.3in}
\caption{Curve of growth results for the \ion{D}{1} absorption in Complex~C
between $v_{\rm LSR} = -160$ \kms\ and $v_{\rm LSR} = -110$ \kms.  The 
x-axis has units of logarithmic inverse centimeters.  Filled 
circles are measurements based on data from the SiC2 
channel.  Open circles are measurements based on data from the SiC1 channel.  
The points for the 
two channels have been offset slightly (plus or minus 0.02 dex) from their 
nominal values
in the horizontal direction for 
clarity.  Error bars are 
1$\sigma$ estimates; limits are $3\sigma$ estimates. 
{\it Top}: Single-component COG fit derived from the equivalent widths 
listed in Table~8. Curves of growth with  
b-values of 10 and 20 \kms\ and log\,N = 15.22 are shown as dashed lines.
{\it Bottom}: Double-component COG model based on the fitting parameters in
Table~7 used
to construct the profiles shown in Figures~12 and 13. The parameters for the 
column densities, widths, and velocity separation of the two components
are listed in the top left corner. The model reproduces
the observed equivalent widths as well as the COG fit shown in the top
panel.  The dashed lines
indicate the curves corresponding to $\pm2$ \kms\ changes in the 
b-value of the stronger component.  The shape of the COG
is insensitive to the choice of b-value for the 
weaker component.}
\end{figure}

\clearpage
\newpage
\begin{figure}[ht!]
\includegraphics{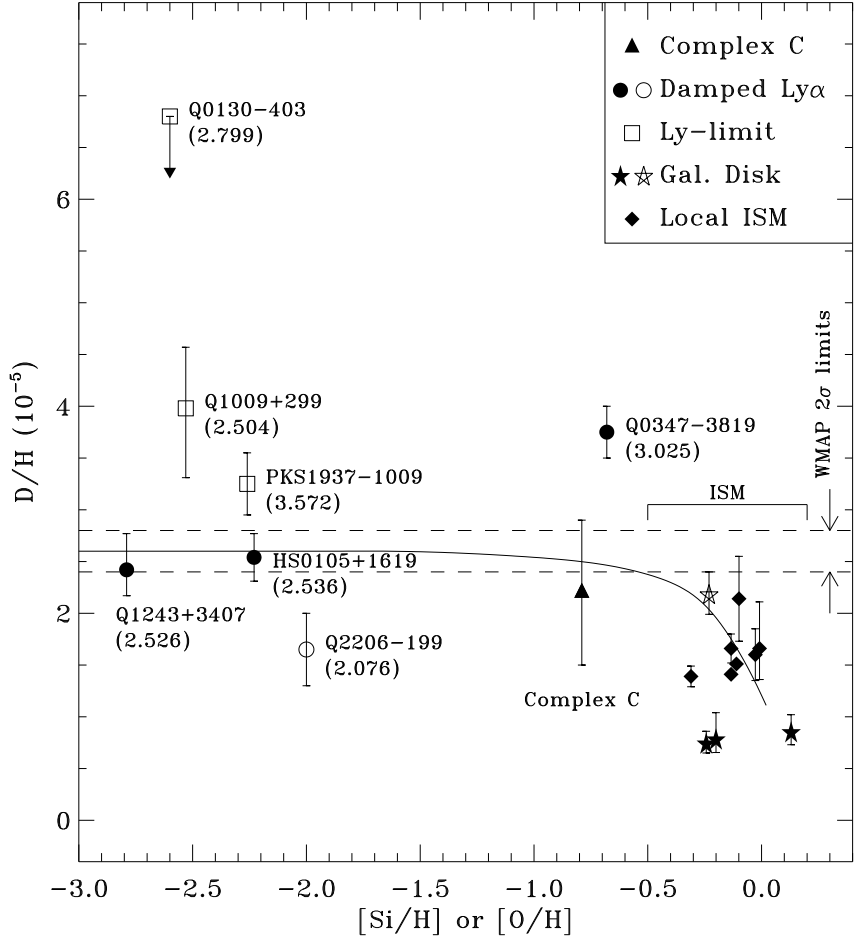}
\vspace{5.0in}
\caption{Values of D/H as a function of metallicity as measured by 
[Si/H] (open symbols) or [O/H] (filled symbols).  
The data for this plot are contained in Table~10.  The symbols are 
coded as follows:  Complex~C (triangle), high-redshift damped Ly$\alpha$ 
systems (circles), high-redshift Lyman-limit systems (squares),
local ISM (diamonds), and Galactic disk ISM (stars).
Redshifts are indicated underneath the name of 
 each extragalactic system.  Error bars
are $1\sigma$ estimates from the original sources.  The  region between
the horizontal dashed lines
indicates the range of primordial D/H values consistent with recent cosmic
microwave background measurements
made by WMAP.  The solid curve indicates the expected behavior
of D/H as a function of metallicity for a simple chemical evolution
model in which (D/H)$_p = 2.6\times10^{-5}$ (see text).}
\end{figure}

\begin{deluxetable}{ccccccccc}
\tablecolumns{8}
\tablewidth{0pt} 
\tablecaption{FUSE Observations of PG\,1259+593\tablenotemark{a}}
\tablehead{Dataset & Date & \# Exposures & \multicolumn{2}{c}{\underline{LiF1 / SiC1}}& \multicolumn{2}{c}{\underline{LiF2 / SiC2}} & Aperture & Note\tablenotemark{b}\\
		   & (U.T. Start) &   & t$_{\rm tot}$  & t$_{\rm ngt}$  & t$_{\rm tot}$  & t$_{\rm ngt}$\\
	           & &      & (ks) & (ks) & (ks) & (ks)}
\startdata
P1080101 & 2000-Feb-25 & 12 & \phn52.4 & 14.2  & \phn52.4 & 14.2 & $30\arcsec\times30\arcsec$\\
P1080102 & 2000-Dec-25 & 21 & \phn56.8  &  \phn0.0 & \phn\phn0.0 &  \phn0.0 & $30\arcsec\times30\arcsec$ & 1\\
P1080103 & 2001-Jan-29 & 31 &  \phn52.0  & 35.4 & \phn82.1 & 56.5 & $30\arcsec\times30\arcsec$ & 2\\
P1080104 & 2001-Mar-12 & 35 & 105.9 & 70.0 &103.2 & 69.5 & $30\arcsec\times30\arcsec$ \\
P1080105 & 2001-Mar-14 & 35 & 104.0 & 70.1 & \phn95.9 & 69.8 & $30\arcsec\times30\arcsec$ \\
P1080106 & 2001-Mar-17 & 35 & \phn63.6 & 43.8 & \phn66.7 & 45.9  & $30\arcsec\times30\arcsec$ & 3\\
P1080107 & 2001-Mar-19 & 38 & \phn95.7 & 70.2 & \phn96.1 & 70.2 & $30\arcsec\times30\arcsec$\\
P1080108 & 2001-Mar-22 & 13 & \phn33.5 & 24.1 & \phn33.5 & 24.1 & $30\arcsec\times30\arcsec$ \\
P1080109 & 2001-Mar-28 & 16 & \phn32.7 & 25.1 & \phn32.0 & 25.1 & $30\arcsec\times30\arcsec$\\
\enddata
\tablenotetext{a}{Entries in this table include the dataset identification, U.T. date
at the start of the observation, exposure
times (total and night-only) in kiloseconds 
for the LiF1/SiC1 channels and LiF2/SiC2 channels, and 
apertures used ($30\arcsec\times30\arcsec$ = LWRS apertures).  Exposure times are totals
after screening for valid data with event bursts removed.  The exposure times have 
been reduced when necessary to account for situations where the detector high voltage was down.
These exposure times do not 
necessarily reflect the exact amount of time that the light of PG\,1259+593 was in the 
aperture, except for the LiF1 channel used for guiding.}
\tablenotetext{b}{Notes:\\
1) SiC1, SiC2, and LiF2 were not aligned with LiF1, so no useful data were obtained in these 
channels. \\
2) Detector 1 high voltage was down during exposures 18--29.\\
3) High voltage was down on both detectors 
during exposures 6--17.  High voltage
was also down during exposure 22 on detector 1.}
\end{deluxetable}

\clearpage
\newpage
\begin{deluxetable}{ccccc}
\tablewidth{0pt} 
\tablecaption{HST/STIS Observations of PG\,1259+593\tablenotemark{a}}
\tablehead{Dataset & Date & t$_{\rm tot}$ & Grating & Aperture \\
		   & (U.T. Start)      & (ks)}
\startdata
O63G05010-O63G05060 & 2001-Jan-17 & 14.4 & E140M & $0.2\arcsec\times0.06\arcsec$ \\
O63G06010-O63G06060 & 2001-Jan-17 & 14.4 & E140M & $0.2\arcsec\times0.06\arcsec$ \\
O63G07010-O63G07060 & 2001-Jan-18 & 14.4 & E140M & $0.2\arcsec\times0.06\arcsec$ \\
O63G08010-O63G08060 & 2001-Jan-18 & 14.4 & E140M & $0.2\arcsec\times0.06\arcsec$ \\
O63G09010-O63G09060 & 2001-Jan-19 & 14.4 & E140M & $0.2\arcsec\times0.06\arcsec$ \\
O63G11010-O63G11040 & 2001-Jan-19 & \phn9.1 & E140M & $0.2\arcsec\times0.06\arcsec$ \\
O63G10010-O63G10060 & 2001-Dec-19 & 14.4 & E140M & $0.2\arcsec\times0.06\arcsec$ \\
\enddata
\tablenotetext{a}{Entries in this table include the dataset identifications, U.T. date
at the start of the observation, total exposure
time for the datasets listed, grating, and aperture used for the observation.}
\end{deluxetable}

\clearpage
\newpage
\begin{deluxetable}{cccccc}
\tabletypesize{\small}
\tablewidth{0pt} 
\tablecaption{\ion{O}{1} Lines\tablenotemark{a}}
\tablehead{Wavelength & $\log [f\lambda$(\AA)] & Instrument & S/N\tablenotemark{b} & Used in  & Note\tablenotemark{c} \\
(\AA) & & & ({\small $\Delta v$\,=\,20 \kms}) & ~~\ion{O}{1} Fit? }
\startdata
1302.168 & \phn1.796 & STIS/E140M & 21 & Yes & 1 \\
1039.230 & \phn0.974 & FUSE/LiF & 25 / 18  & Yes \\
\phn976.448     & \phn0.509 & FUSE/SiC & 13 / 11& Yes & 2 \\
\phn974.070     & --1.817 & FUSE/SiC & 13 / 11& No & 3 \\
\phn972.143	& --0.511 & FUSE/SiC & 13 / 11& No & 4\\
\phn972.142	& --1.521 & FUSE/SiC & 13 / 11 & No & 4 \\
\phn971.738	& \phn1.052 & FUSE/SiC & 13 / 11& No & 5 \\
\phn971.738	& \phn0.304 & FUSE/SiC & 13 / 11 & No & 5 \\
\phn971.737	& --0.872   & FUSE/SiC & 13 / 11 & No & 5 \\
\phn950.885     & \phn0.176 & FUSE/SiC & 13 / 10& Yes \\
\phn948.686     & \phn0.778 & FUSE/SiC & 13 / 10& Yes\\
\phn937.840	& --0.085 & FUSE/SiC & 13 / 10 & No & 6\\
\phn936.630     & \phn0.534 & FUSE/SiC & 13 / 10 & Yes\\
\phn930.257 	& --0.301 & FUSE/SiC & 13 / 11& No & 7\\
\phn929.517 	& \phn0.329 & FUSE/SiC & 13 / 11&  Yes\\
\phn925.446 	& --0.484 & FUSE/SiC & 13 / 11& Yes & 8\\
\phn924.950 	& \phn0.155 & FUSE/SiC & 13 / 11&  Yes\\
\phn922.200 	& --0.645 & FUSE/SiC & 13 / 11 & Yes \\
\phn921.857 	& --0.001 & FUSE/SiC & 13 / 11 & Yes\\
\phn919.917  	& --0.788 & FUSE/SiC & 13 / 11 & No & 9 \\
\phn919.658     & --0.137 & FUSE/SiC & 13 / 11 & No & 10 \\
\enddata
\tablenotetext{a}{This table lists prominent \ion{O}{1} lines in the FUSE and HST/STIS 
bandpasses.  Lines with strengths $\log f\lambda 
\le -2.0$ have been omitted from the table since they produce negligible
absorption features. Wavelengths and $f$-values are from the atomic data 
compilations by Morton (1991, 2003).}
\tablenotetext{b}{Signal-to-noise ratio per 20 \kms\ bin in the continuum near the 
line.  For the FUSE data, values are listed for either the LiF1 / LiF2 
channels ($\lambda > 1000$\,\AA) or the SiC1 / SiC2 channels
($\lambda<1000$\,\AA).}
\tablenotetext{c}{Notes: \\
\phn(1) High-resolution spectrum; \ion{P}{2} $\lambda1301.874$ 
at --68 \kms\ with respect to 
\ion{O}{1} features; \ion{O}{1} terrestrial airglow 
emission at $v_{LSR} \approx 20.5$ \kms. \\
\phn(2) Partial blend at low velocities
with high velocity \ion{C}{3} $\lambda977.020$.\\
\phn(3) Negligible absorption produced, blend with IGM line.\\
\phn(4) Blend with \ion{H}{1} and \ion{D}{1} Ly$\gamma$.  The published 
\ion{O}{1} $f$-values for the resonance lines in this multiplet (multiplet \#9)
are probably a factor of at least 5 times too high (see 
H\'ebrard \& Moos 2003).  We do not use these lines in our analysis.\\
\phn(5) Possible partial blend with \ion{D}{1} Ly$\gamma$.\\
\phn(6) Blend with \ion{H}{1} Ly$\epsilon$.\\
\phn(7) Blend with \ion{H}{1} and \ion{D}{1} Ly$\zeta$.\\
\phn(8) Close to, but does not blend with,  \ion{D}{1} Ly$\eta$.\\
\phn(9) Partial blend with \ion{H}{1} Ly$\kappa$. \\
(10) Blend with \ion{H}{1} Ly$\kappa$.}
\end{deluxetable}

\clearpage
\newpage
\begin{deluxetable}{ccccc}
\tablewidth{0pt} 
\tablecaption{\ion{O}{1} Velocity Model\tablenotemark{a}}
\tablehead{Component & $v_{\rm LSR}$ & b & log [N (cm$^{-2}$)] & Note\tablenotemark{b}\\
& (\kms) & (\kms) }
\startdata
\multicolumn{5}{c}{Complex C}\\
\tableline
1 & --129.5 & $6.0\pm1.0$ & $15.85\pm0.15$ & 1\\
2 & --112.5 & $9.8\pm^{1.0}_{3.0}$ & $14.30\pm0.04$ & 1\\
Total\,[1--2] & \nodata & \nodata & $15.86\pm0.15$ & 2\\
\tableline
\multicolumn{5}{c}{IV Arch}\\
\tableline
3 & \phn--81.9  & $\sim$8 & 15.08 & 3\\
4 & \phn--54.5  & $\sim$10 & 16.00 & 1\\
5 & \phn--29.0  & $\sim$10 & 14.78 & 3\\
Total\,[3--5] & \nodata & \nodata & $16.07\pm0.10$ & 2\\
\tableline
\multicolumn{5}{c}{Low Velocity ISM}\\
\tableline
6 & \phn\phn--2.5 & $\sim$8 & 16.08 & 1\\
7 & \phn+21.0 & $\sim$12 & 15.00 & 1\\
Total\,[6--7] & \nodata & \nodata & $16.11\pm0.10$ & 2\\
\tableline
\multicolumn{5}{c}{Positive IVC}\\
\tableline
8 & \phn+68.9 & $7.7\pm2.8$ & $13.43\pm0.08$ & 4 \\
\enddata
\tablenotetext{a}{Uncertainties on all quantities in this table are $1\sigma$ estimates.}
\tablenotetext{b}{Notes:\\
1) Obvious component in high-resolution HST/STIS E140M metal-line data. \\
2) Obvious component group in \ion{H}{1} 21\,cm emission. \\
3) Weak component required to improve fit to the \ion{O}{1} lines.\\
4) Weak component visible in the \ion{O}{1} $\lambda1302.168$ line.
Small inflection in \ion{O}{1} $\lambda1039.230$ line also present.}
\end{deluxetable}

\newpage
\begin{deluxetable}{ccccccc}
\tablewidth{0pt} 
\tablecaption{Equivalent Widths of \ion{O}{1} Lines in Complex~C }
\tablehead{$\lambda$ & log [$f\lambda$(\AA)] & \multicolumn{2}{c}{This Paper} &
\multicolumn{2}{c}{Previous Work}\\
& & $W_\lambda$(SiC1) & $W_\lambda$(SiC2)
& Richter et al.\ (2001b) & Collins et al.\ (2003)\\
(\AA) & & (m\AA) & (m\AA) & (m\AA) & (m\AA)}
\startdata
1302.168 & \phn1.796 & \multicolumn{2}{c}{$178\pm10$\tablenotemark{b}} & $214\pm20$\tablenotemark{c} & $203\pm8$\tablenotemark{c} \\
1039.230 & \phn0.974 & $90\pm18$  & $104\pm12$ & $100\pm13$ & $\phn97\pm6$\\
\phn948.685 & \phn0.778 & $60\pm13$  & $\phn82\pm13$ &  $\phn84\pm25$ & $\phn71\pm12$ \\
\phn936.630 & \phn0.534 & $69\pm17$  & $\phn66\pm10$ &  $\phn70\pm11$ & $\phn73\pm10$\\
\phn929.517 & \phn0.329 & $42\pm17$  & $\phn43\pm11$ &  $\le48$ & $\phn54\pm10$\\
\phn924.950 & \phn0.155 & $40\pm17$  & $\phn36\pm10$ &  \nodata & $\phn55\pm11$\\
\phn921.857 & --0.001 & $45\pm17$  & $\phn38\pm13$ &  \nodata & \nodata\\
\enddata
\tablenotetext{a}{Errors are $1\sigma$ uncertainties.  Limits are 
$3\sigma$ estimates. The integration range in all cases was 
$-160 \le v_{\rm LSR} \le -110$ \kms, which includes both Complex~C 
components.}
\tablenotetext{b}{Equivalent width measurement is from HST/STIS E140M data.}
\tablenotetext{c}{Equivalent width measurement is from partial set of HST/STIS E140M data.}
\end{deluxetable}

\clearpage
\newpage
\begin{deluxetable}{ccccc}
\tablewidth{0pt} 
\tablecaption{\ion{H}{1} Velocity Model\tablenotemark{a}}
\tablehead{Component & $v_{\rm LSR}$ & b & log [N (cm$^{-2}$)] & Note\tablenotemark{b}\\
& (\kms) & (\kms) }
\startdata
\multicolumn{5}{c}{Complex C}\\
\tableline
1 & --127.4 & $9.6\pm1.2$ & $19.94\pm0.06$ & 1\\
2 & --112.5 & $24\pm2$ & $18.43\pm0.30$ & 2\\
Total\,[1--2] & \nodata & \nodata & $19.95\pm0.06$ & 3\\
\tableline
\multicolumn{5}{c}{IV Arch}\\
\tableline
3 & \phn--81.9  & $\sim$8 & 18.34 &4\\
4 & \phn--53.5  & $\sim$10 & 19.43 &4\\
5 & \phn--29.0  & $\sim$10 & 18.04 &4\\
Total\,[3--5] & \nodata & \nodata & $19.48\pm0.01$ &5\\
\tableline
\multicolumn{5}{c}{Low Velocity ISM}\\
\tableline
6 & \phn\phn--6.5 & $\sim$14 & 19.63 & 4\\
7 & \phn+21.0 & $\sim$12 & 18.56 & 4\\
Total\,[6--7] & \nodata & \nodata & $19.67\pm0.02$ & 5\\
\tableline
\multicolumn{5}{c}{Positive IVC}\\
\tableline
8 & \phn+68.9 & $13\pm2$ & $16.69\pm0.08$ & 6, 7 \\
\enddata
\tablenotetext{a}{Uncertainties on all quantities in this table are 
$1\sigma$ estimates.}
\tablenotetext{b}{Notes:\\
1) Component width and velocity set by interferometric 21\,cm emission data. \\
2) Component width constrained by the negative velocity wings
 of the \ion{H}{1} absorption profiles.  The large width implies that 
this component is likely to be a 
blend of several narrower unresolved sub-components.\\
3) Total \ion{H}{1} column density of Complex~C
set by interferometric 21\,cm data.\\
4) Component width and velocity similar to the values for \ion{O}{1}. 
Values constrained by 21\,cm data.\\
5) Total \ion{H}{1} column density of IV~Arch and ISM set by single-dish
(9\farcm7 beam) Effelsberg 21\,cm emission data. \\
6) Component width set by the positive velocity wings of 
the \ion{H}{1} absorption profiles. Component velocity set by \ion{O}{1} 
$\lambda1302.168$ line.  \\
7) Column density of positive IVC set by value of N(\ion{O}{1})
assuming (O/H) = (O/H)$_\odot$; no useful constraint on N(\ion{H}{1}) is 
set by the \ion{H}{1} Lyman-series lines.}
\end{deluxetable}

\clearpage
\newpage
\begin{deluxetable}{ccccc}
\tablewidth{0pt} 
\tablecaption{\ion{D}{1} Velocity Model\tablenotemark{a,b}}
\tablehead{Component & $v_{\rm LSR}$ & b & log [N (cm$^{-2}$)]  & Note\tablenotemark{c}\\
& (\kms) & (\kms) }
\startdata
\multicolumn{5}{c}{Complex C}\\
\multicolumn{5}{c}{(without additional residual \ion{H}{1} at --190 \kms)}\\
\tableline
1 & --127.4 & $8\pm1$ & 15.36 & 1\\
2 & --112.5 & 15 & 13.85 & 2\\
Total\,[1--2] & \nodata & \nodata &  $15.37\pm0.10$& 3\\
\tableline
\multicolumn{5}{c}{Complex C} \\
\multicolumn{5}{c}{(with additional residual \ion{H}{1} at --190 \kms)}\\
\tableline
1 & --127.4 & $8\pm1$ & 15.20 & 1\\
2 & --112.5 & 15 & 13.69 & 2\\
Total\,[1--2] & \nodata & \nodata &  $15.21\pm0.12$& 4\\
\tableline
\multicolumn{5}{c}{IV Arch}\\
\tableline
3 & \phn--81.9  & \phn8 & 13.52 &5\\
4 & \phn--53.5  & 10 & 14.61 &5\\
5 & \phn--29.0  & 10 & 13.22 &5\\
Total\,[3--5] & \nodata & \nodata & 14.66 & 5\\
\tableline
\multicolumn{5}{c}{Low Velocity ISM}\\
\tableline
6 & \phn\phn--6.5 & 14 & 14.81 & 5\\
7 & \phn+21.0 & 12 & 13.74 & 5\\
Total\,[6--7] & \nodata & \nodata &  14.85 & 5\\
\tableline
\multicolumn{5}{c}{Positive IVC}\\
\tableline
8 & \phn+68.9 & 13 & 11.87 & 5 \\
\enddata
\tablenotetext{a}{Velocities and widths of the \ion{D}{1} absorption
components were set equal to the values for \ion{H}{1} in Table~6
unless noted otherwise.}
\tablenotetext{b}{Two sets of parameters are given for the Complex~C
absorption.  The first set is the fit without additional, weak
high-velocity \ion{H}{1} that could account for the 
discrepancy in the fits to the Ly$\delta$ and Ly$\epsilon$ lines shown in 
Figures~12 and 13.  The second
set includes a weak \ion{H}{1} feature with N(\ion{H}{1}) = $3\times10^{14}$
cm$^{-2}$ and b$_{\rm HI} = 10$ \kms\ in the fit, as shown in 
Figures~14 and 15.}
\tablenotetext{c}{Notes appear on next page.}
\end{deluxetable}

\clearpage
\newpage

\indent $^c$Notes:\\
1) Component width allowed to vary between b$_{\rm O\,I} = 6$ \kms\ 
and b$_{\rm H\,I} = 10$ \kms. \\
2) Component width poorly constrained by data, but does not affect 
total column density of \ion{D}{1} in Complex~C. \\
3) D/H ratio allowed to vary such that (D/H)$_{\rm Comp\,1}$ = (D/H)$_{\rm Comp\,2}$.
Best fit is (D/H) = $2.6\times10^{-5}$.\\
4) D/H ratio allowed to vary such that (D/H)$_{\rm Comp\,1}$ = (D/H)$_{\rm Comp\,2}$.
Best fit is (D/H) = $1.8\times10^{-5}$.\\
5) D/H ratio set equal to $1.5\times10^{-5}$.\\

\clearpage
\newpage
\begin{deluxetable}{ccccc}
\tablewidth{0pt} 
\tablecaption{Equivalent Widths of \ion{D}{1} Lines in Complex~C }
\tablehead{Line\tablenotemark{a} & $\lambda$ & log [$f\lambda(\AA)]$ & $W_\lambda$(SiC1)\tablenotemark{b} & $W_\lambda$(SiC2)\tablenotemark{b} \\
& (\AA) & & (m\AA) & (m\AA)}
\startdata
Ly$\delta$ 	& 949.485 & 1.122 & $80\pm13$ & $111\pm11$ \\
Ly$\epsilon$	& 937.548 & 0.865 & $52\pm10$ & $\phn59\pm09$ \\
Ly$\eta$	& 925.974 & 0.470 & $39\pm10$ & $\phn42\pm09$ \\
Ly$\theta$	& 922.898 & 0.311 & $<39$     & $<32$ \\
Ly$\iota$ 	& 920.711 & 0.170 & $<36$     & $<35$\\
\enddata
\tablenotetext{a}{The Ly$\zeta$ line is omitted from this table because
of substantial blending with \ion{O}{1} absorption (see Figures~12 and 13).}
\tablenotetext{b}{Errors are $1\sigma$ uncertainties.  Limits are 
$3\sigma$ estimates. The integration range in all cases was 
$-160 \le v_{\rm LSR} \le -110$ \kms, which includes both Complex~C 
components.}
\end{deluxetable}

\clearpage
\newpage
\begin{deluxetable}{lcc}
\tablewidth{0pt} 
\tablecaption{Summary of Complex~C Results}
\tablehead{Quantity & Value~$\pm~1\sigma$ & Value~$\pm~2\sigma$ \\
	& (68\% Conf.) & (95\% Conf.)}
\startdata
N(\ion{H}{1}) (cm$^{-2}$) & $(9.0\pm1.0)\times10^{19}$ & $(9.0\pm1.6)\times10^{19}$\\
N(\ion{D}{1}) (cm$^{-2}$) & $(2.0\pm0.6)\times10^{15}$ & $(2.0\pm0.9)\times10^{15}$\\
N(\ion{O}{1}) (cm$^{-2}$) & $(7.2\pm2.1)\times10^{15}$ & $(7.2\pm4.2)\times10^{15}$\\
\\
D/H  		 &  $(2.2\pm0.7)\times10^{-5}$ &  $(2.2\pm1.1)\times10^{-5}$ \\
O/H 		 &  $(8.0\pm2.5)\times10^{-5}$ &  $(8.0\pm4.2)\times10^{-5}$ \\
D/O 		 &  $0.28\pm0.12$ &  $0.28\pm0.20$ \\
\enddata
\tablenotetext{~}{Note -- Errors and confidence limits include statistical
uncertainties as well as systematic uncertainties discussed in the text.}

\end{deluxetable}

\clearpage
\newpage

\begin{deluxetable}{lcccccl}
\tabletypesize{\footnotesize}
\tablewidth{0pt}
\tablecaption{Light Element Abundance Ratios in Different Environments\tablenotemark{a}}
\tablehead{System & $z$ or & log N(H\,I [cm$^{-2}$]) & [O/H]\tablenotemark{b}&  D/H	& D/O & Reference \\
& [$d$ (pc)] &  & & (10$^{-5}$) }
\startdata
\multicolumn{7}{c}{High Velocity Gas}\\
\tableline
\medskip
Complex~C	  & [$>5000$] & 19.95 & $-0.79\pm^{0.12}_{0.16}$ & $2.2\pm0.7$   & $0.28\pm0.12$ & This paper \\
\tableline
\multicolumn{7}{c}{Milky Way ISM}\\
\tableline
\medskip
LISM (7 l.o.s.)   & [$\lesssim180$] &  $17.93-20.14$ & $-0.21\pm^{0.03}_{0.03}$ & $1.52\pm0.08$	& $0.040\pm0.002$	& Moos et al. (2002)\tablenotemark{c} \\
\medskip
$\gamma^2$ Vel & [258] & 19.71 & (--0.23) & $2.18\pm^{0.22}_{0.19}$ & \nodata & Sonneborn et al. (2000)\tablenotemark{d,e}\\
\medskip
$\zeta$ Pup    & [430] & 19.96 & \nodata & $1.42\pm^{0.15}_{0.14}$ & \nodata & Sonneborn et al. (2000)\tablenotemark{e}\\
\medskip
$\delta$ Ori~A & [500] & 20.19 & $-0.24\pm^{0.06}_{0.07}$ & $0.74\pm^{0.12}_{0.09}$ & $0.026\pm^{0.005}_{0.002}$ & Jenkins et al. (1999)\tablenotemark{e,f}\\
\medskip
HD\,195965     & [794] & 20.95 & $+0.13\pm^{0.03}_{0.04}$ & $0.85\pm^{0.17}_{0.12}$  & $0.013\pm^{0.003}_{0.002}$ & Hoopes et al. (2003)\\
\medskip
HD\,191877     & [2200] & 21.05& $-0.20\pm^{0.08}_{0.12}$  & $0.78\pm^{0.26}_{0.13}$ & $0.025\pm^{0.011}_{0.004}$ & Hoopes et al. (2003)\\
\tableline
\multicolumn{7}{c}{Damped Ly$\alpha$ Systems} \\
\tableline
\medskip
HS\,0105+1619	& 2.536 & 19.42  & $-1.73\pm0.02$ & $2.54\pm0.23$  & $2.8\pm0.3$ & O'Meara et al. (2001) \\
\medskip
Q\,0347--3819 	& 3.025 & 20.63 & $-0.68\pm0.06$ & $3.75\pm0.25$ & $0.37\pm0.03$ & Levshakov et al. (2002) \\
\medskip
Q\,2206--199	& 2.076 & 20.44 & (--2.23)  & $1.65\pm0.35$ & \nodata & Pettini \& Bowen (2001)\\
\medskip
Q\,1243+3407	& 2.526 & 19.73 & $-2.79\pm0.05$ & $2.42\pm^{0.35}_{0.25}$ & $30\pm3$ & Kirkman et al. (2003)\\
\tableline
\multicolumn{7}{c}{Lyman Limit Systems} \\
\tableline
\medskip
PKS\,1937--1009 & 3.572 & 17.86 & (--2.26) & $3.25\pm0.30$ & \nodata &  O'Meara et al. (2001) \\
\medskip
Q\,0130--403 & 2.799 & 16.66 & (--2.6)  & $<6.8$  & \nodata& O'Meara et al. (2001) \\
\medskip
Q\,1009+299 & 2.504 & 17.39 & (--2.53)  & $3.98\pm^{0.59}_{0.67}$ & \nodata& O'Meara et al. (2001) \\
\enddata
\tablenotetext{a}{Uncertainties are $1\sigma$ estimates.}
\tablenotetext{b}{[O/H] = log N(\ion{O}{1})/N(\ion{H}{1}) -- log (O/H)$_\odot$,
where (O/H)$_\odot = 4.90\times10^{-4}$ (Allende Prieto et al. 2001).  All
values of [O/H] have been scaled to this reference abundance.
Items in parentheses
are estimates of [Si/H] when no estimate of [O/H] is available.  These values 
of [Si/H] should only be considered approximate substitutions for [O/H] since
ionization corrections for Si may be very large, especially for the 
Lyman-limit systems.}
\tablenotetext{c}{Values are [O/H], D/H, and D/O are weighted averages for the seven
sight lines considered.  Measurements for each sight line can be found in 
Kruk et al. (2002), Friedman et al. (2002), Sonneborn et al. (2002), 
Lemoine et al. (2002), Lehner et al. (2002), Wood et al. (2002), and
H\'ebrard et al. (2002).  Additional measurements of D/O can be found in
H\'ebrard \& Moos (2003), who measure a local ISM value of D/O = $0.038\pm0.002$.}
\tablenotetext{d}{N(\ion{Si}{2}) is from Fitzpatrick \& Spitzer (1994).}
\tablenotetext{e}{Errors on D/H have been converted from 90\% confidence to $1\sigma$
estimates (see Moos et al. 2002).}
\tablenotetext{f}{N(\ion{O}{1}) is from Meyer et al. (1998).}
\end{deluxetable}

\clearpage
\newpage
\begin{deluxetable}{lcccl}
\tablewidth{0pt} 
\tablecaption{Searching for Deuterium in Other Complex~C Sight Lines}
\tablehead{Sight Line & $l$ & $b$ & $\theta$\tablenotemark{a} & Comment \\
	& (\degr) & (\degr) & (\degr)}
\startdata
PG\,1626+554 & \phn84.51 & 42.19 & 27.7 & Velocity structure similar to PG\,1259+593, but flux is lower and\\
	&	&	& & Complex~C component is less pronounced than Milky Way ISM.\\
\\
3C\,351 & \phn90.08 & 36.38 & 29.5 & Ly-limit system at $z\approx0.22$ attenuates quasar flux shortward \\
	&	&	&	 &  of $\approx1115$ \AA. \\
\\
Mrk~290 & \phn91.49 & 47.95 & 20.0 & Potentially good velocity structure, but quasar is faint.\\
\\
Mrk~876 & \phn98.27 & 40.38 & 22.7 & Good Complex~C structure, but \ion{D}{1} absorption blends with \ion{H}{1}.\\
\\
Mrk~817 & 100.30 & 53.48 & 12.2 & Complex~C at relatively low velocity (--109 \kms),\\
	&	&	&	& leading to blending of \ion{D}{1} and \ion{H}{1}.\\
\\
PG\,1351+640 & 111.89 & 52.02 & \phn7.8 & Strong intermediate velocity structure confuses Complex~C \\
	&	&	&	& \ion{D}{1} absorption.\\	
\\
Mrk~279 & 115.04 & 56.86 & \phn3.2 & Complicated structure at intermediate velocities blends with\\
	&	 &	 &	& Complex~C absorption. \\
\enddata
\tablenotetext{a}{Angular distance from PG\,1259+593 sight line.}
\end{deluxetable}

\end{document}